\documentclass[fleqn,a4paper,oneside,openany]{book}
\usepackage[graphicx]{realboxes}
\usepackage{graphicx,xspace,amssymb,amsmath,amsfonts}
\usepackage{enumitem,wrapfig,xcolor}
\usepackage[font=small,labelfont=bf,labelsep=space]{caption}
\usepackage{subcaption}
\usepackage{mathtools}

\DeclarePairedDelimiter\floor{\lfloor}{\rfloor}

\usepackage{booktabs}
\usepackage{placeins}

\usepackage{hyperref}
\hypersetup{hidelinks}

\usepackage[british]{babel}
\usepackage{booktabs}
\usepackage{multirow}
\usepackage{subcaption}

\usepackage{graphicx}
\usepackage{marginnote}

\usepackage{xtab}
\usepackage{pdflscape}
\usepackage{placeins}
\usepackage{longtable}
\usepackage{pifont}
\usepackage{csquotes}

\usepackage[numbers, sort&compress, super]{natbib}
\newcommand\id[1]{{\hfill\normalsize{\texttt #1}}}
\newcommand\textid[1]{{\normalsize{\texttt #1}}}

\usepackage{caption}
\DeclareCaptionLabelSeparator{upright}{ | }
\usepackage{geometry}
\usepackage{setspace}
\usepackage{hyperref}
\hypersetup{hidelinks}

\usepackage{titlesec}
\titlespacing\subsubsection{0pt}{8pt plus 2pt minus 2pt}{0pt plus 2pt minus 2pt}

\usepackage{enumitem}

\usepackage{tcolorbox}
\begin{document}
\emergencystretch 3em
\renewcommand{\tablename}{Table}

\frontmatter
\title{Standardised convolutional filtering for Radiomics\\
\vspace{2mm}
\large Image Biomarker Standardisation Initiative (IBSI) \\
\vspace{10mm}}
\author{Adrien Depeursinge$^{1,2}$, 
Vincent Andrearczyk$^1$, 
Philip Whybra$^3$,\\
Joost van Griethuysen$^{4,5,6}$,
Henning M\"{u}ller$^{1,7}$, 
Roger Schaer$^1$,\\
Martin Valli\`{e}res$^{8,9,*}$,
Alex Zwanenburg$^{10,11,*}$}
\date{
\small
$^1$Institute of Informatics, University of Applied Sciences Western Switzerland (HES-SO), Switzerland\\
$^2$Service of Nuclear Medicine and Molecular Imaging, Centre Hospitalier Universitaire Vaudois (CHUV), Lausanne, Switzerland\\
$^3$School of Engineering, Cardiff University, Cardiff, United Kingdom\\
$^4$Department of Radiology, the Netherlands Cancer Institute (NKI), Amsterdam, the Netherlands\\
$^5$GROW-School for Oncology and Developmental Biology, Maastricht University Medical Center, Maastricht, The Netherlands\\
$^6$Department of Radiation Oncology, Dana-Farber Cancer Institute, Brigham and Women’s Hospital, Harvard Medical School, Boston, MA\\
$^{7}$University of Geneva, Geneva, Switzerland\\
$^8$Department of Computer Science, University of Sherbrooke, Sherbrooke, Qu\'{e}bec, Canada\\
$^9$GRIIS, University of Sherbrooke, Sherbrooke, Qu\'{e}bec, Canada\\
$^{10}$OncoRay -- National Center for Radiation Research in Oncology, Faculty of Medicine and University Hospital Carl Gustav Carus, Technische Universität Dresden, Helmholtz-Zentrum Dresden - Rossendorf, Dresden, Germany\\
$^{11}$National Center for Tumour Diseases (NCT), Partner Site Dresden, Germany: German Cancer Research Center (DKFZ), Heidelberg, Germany; Faculty of Medicine, University Hospital Carl Gustav Carus, Technische Universität Dresden, Dresden, Germany, and; Helmholtz Association / Helmholtz-Zentrum Dresden - Rossendorf (HZDR), Dresden, Germany\\
\vspace{5mm}
$^{*}$These authors contributed equally to this work\\
\vspace{10mm}
\large
%\today
}
\maketitle

\newpage
\chapter*{The image biomarker standardisation initiative}
The Image Biomarker Standardisation Initiative (IBSI) is an independent international collaboration that works towards standardising the extraction of image biomarkers from acquired imaging for the purpose of high-throughput quantitative image analysis (radiomics). Lack of reproducibility and validation of radiomic studies is considered to be a major challenge for the field. Part of this challenge lies in the scantiness of consensus-based guidelines and definitions for the process of translating acquired imaging into high-throughput image biomarkers. The IBSI therefore seeks to provide standardised image biomarker nomenclature and definitions, a standardised general image processing workflow, tools for verifying radiomics software implementations and reporting guidelines for radiomic studies.

Additional information can be found on the official IBSI website\footnote{\texttt{\url{https://theibsi.github.io/}}, as of February 2024.}.

\subsubsection*{Related work}
Results related to this reference manual are presented in Whybra and Zwanenburg \emph{et al.} \cite{Whybra2024-yb}.

\subsubsection*{Permanent identifiers}
The IBSI uses permanent identifiers for image biomarker definitions and important related concepts such as image processing. These consist of four-character codes and may be used for reference. Please do not use page numbers or section numbers as references, as these are subject to change.

\subsubsection*{Copyright}
This work is licensed under the Creative Commons Attribution 4.0 International License (CC-BY).

Copyright information regarding the benchmark data sets may be found on GitHub\footnote{\texttt{\url{https://github.com/theibsi/data\_sets}}, as of February 2024.}.

\subsection*{Change notes}

\noindent Changes in version 9:
\begin{itemize}
    \item Added link to repository with reference consensus response maps.
    \item Added link to related publication in Radiology \cite{Whybra2024-yb}.
\end{itemize}

\noindent Changes in version 8:
\begin{itemize}
    \item Added permanent identifiers throughout the document.
    \item Added and updated reporting guidelines in chapter \ref{sec:guidelinesDetails}.
    \item Removed some details regarding comparison between teams from chapter \ref{sec:benchmarking}.
    \item Added reference feature values to chapter \ref{sec:reference_values}.
\end{itemize}

\noindent Changes in version 7:
\begin{itemize}
    \item Reworked the filter configurations for the validation phase to be more realistic, see section \ref{sec:validationPhase}.
\end{itemize}

\noindent Changes in version 6:
\begin{itemize}
    \item Updated list of participants.
    \item Updated description of the spatial extent of Gabor filters to produce the most consistent results (section \ref{sec:Gabor}).
    \item Added implementation troubleshooting to clarify how undecimated wavelet decomposition should be implemented (section \ref{sec:separableWavelets}).
    \item The description of how to align Riesz kernels was expanded and is now considerably more detailed (section \ref{sec:RieszAlignment}).
    \item Added boundary condition for Simoncelli filters in phase 1 (configurations 8.a.1 - 3), see Table \ref{tab:benchmark_filter_settings}.
    \item Added a new section for the validation phase (section \ref{sec:validationPhase}).
\end{itemize}

\noindent Changes in version 5:
\begin{itemize}
    \item Updated list of participants.
    \item Clarified the intended use of the "mirror" boundary condition (section \ref{sec:boundaryConditions}).
    \item The 2D tests in phase 2 are now without interpolation (section \ref{sec:featureBenchmarkLungCT}).
\end{itemize}

\noindent Changes in version 4:
\begin{itemize}
    \item A list of participants was added.
    \item Moved embedding of convolutional filtering in radiomics workflow to section \ref{sec:overallWorkflow}.
    \item Added several implementation troubleshooting boxes (chapter \ref{sec:filtersDescription}).
    \item Clarified units of $\sigma$ and $\lambda$ parameters of Gabor filters (section \ref{sec:Gabor}).
    \item Clarified spatial extent of Gabor filter support (section \ref{sec:Gabor}).
    \item Clarified implementation of the \emph{\`{a} trous} algorithm (section \ref{sec:undecimatedWT}).
    \item Extended description of non-separable wavelets (section \ref{sec:nonseparableWavelets}).
    \item Added recommendation to not perform image decomposition beyond the first level using separable wavelets (section \ref{sec:waveletConsiderations}).
    \item Extended description of the Riesz transform (section \ref{sec:Riesz})
    \item Removed chapter 5 \emph{Filtering for Radiomics in Practice}.
\end{itemize}

\subsection*{Participants}
\marginnote{\footnotesize v4: Added list of participants. v5: updated list. v6: updated list. v8: updated list to reflect manuscript}
The following people contributed to this reference manual and the project as a whole.

% TODO: Add list of participants
\small
\begin{longtable}{p{4cm}p{10cm}}

\toprule
\textbf{Name} & \textbf{Affiliation} \\
\midrule
\endfirsthead

\toprule
\textbf{Name} & \textbf{Affiliation} \\
\midrule
\endhead

\bottomrule
\multicolumn{2}{r}{\textit{Continued on next page}}
\endfoot

\\
\endlastfoot
Vincent Andrearczyk & Institute of Informatics, University of Applied Sciences and Arts Western Switzerland (HES-SO), Sierre, Switzerland\\
Aditya P Apte & Department of Medical Physics, Memorial Sloan Kettering Cancer Center, New York, NY, USA\\
Alexandre Ayotte & Department of Computer Science, Université de Sherbrooke, Sherbrooke, QC, Canada\\
Bhakti Baheti & Center for Artificial Intelligence and Data Science for Integrated Diagnostics (AI2D) and Center for Biomedical Image Computing and Analytics (CBICA), University of Pennsylvania, Philadelphia, PA, USA; Department of Pathology and Laboratory Medicine, Perelman School of Medicine, University of Pennsylvania, Philadelphia, PA, USA; and Department of Radiology, Perelman School of Medicine, University of Pennsylvania, Philadelphia, PA, USA\\
Spyridon Bakas & Center for Artificial Intelligence and Data Science for Integrated Diagnostics (AI2D) and Center for Biomedical Image Computing and Analytics (CBICA), University of Pennsylvania, Philadelphia, PA, USA; Department of Pathology and Laboratory Medicine, Perelman School of Medicine, University of Pennsylvania, Philadelphia, PA, USA; and Department of Radiology, Perelman School of Medicine, University of Pennsylvania, Philadelphia, PA, USA\\
Andrea Bettinelli & Medical Physics Department, Veneto Institute of Oncology IOV - IRCCS, Padua, Italy\\
Ronald Boellaard & Radiology and Nuclear Medicine, Amsterdam UMC, Amsterdam, the Netherlands\\
Luca Boldrini & Fondazione Policlinico Universitario “A. Gemelli” IRCCS, Rome, Italy\\
Irène Buvat & Institut Curie, Université PSL, Inserm U1288, Laboratoire d’Imagerie Translationnelle en Oncologie, Orsay, France\\
Gary J R Cook & Cancer Imaging, School of Biomedical Engineering and Imaging Sciences, King's College London, London, United Kingdom\\
Adrien Depeursinge & Institute of Informatics, University of Applied Sciences and Arts Western Switzerland (HES-SO), Sierre, Switzerland; and Department of Nuclear Medicine and Molecular Imaging, Lausanne University Hospital (CHUV), Lausanne, Switzerland\\
Florian Dietsche & Department of Radiation Oncology, University Hospital Zurich, University of Zurich, Zurich, Switzerland\\
Nicola Dinapoli & Fondazione Policlinico Universitario “A. Gemelli” IRCCS, Rome, Italy\\
Hubert S Gabryś & Department of Radiation Oncology, University Hospital Zurich, University of Zurich, Zurich, Switzerland\\
Vicky Goh & Cancer Imaging, School of Biomedical Engineering and Imaging Sciences, King's College London, London, United Kingdom; and Department of Radiology, Guy's \& St Thomas' NHS Foundation Trust, London, United Kingdom\\
Matthias Guckenberger & Department of Radiation Oncology, University Hospital Zurich, University of Zurich, Zurich, Switzerland\\
Mathieu Hatt & LaTIM, INSERM, UMR 1101, Université de Bretagne-Occidentale, Brest, France\\
Mahdi Hosseinzadeh & Technological Virtual Collaboration (TECVICO Corp.), Vancouver, BC, Canada; and Department of Electrical and Computer Engineering, Tarbiat Modares University, Tehran, Iran\\
Aditi Iyer & Department of Medical Physics, Memorial Sloan Kettering Cancer Center, New York, NY, USA\\
Vincent Jaouen & IMT-Atlantique, Plouzané, France\\
Jacopo Lenkowicz & Fondazione Policlinico Universitario “A. Gemelli” IRCCS, Rome, Italy\\
Mahdi A L Loutfi & Department of Computer Science, Université de Sherbrooke, Sherbrooke, QC, Canada\\
Steffen Löck & OncoRay – National Center for Radiation Research in Oncology, Faculty of Medicine and University Hospital Carl Gustav Carus, Technische Universität Dresden, Helmholtz-Zentrum Dresden - Rossendorf, Dresden, Germany\\
Francesca Marturano & Medical Physics Department, Veneto Institute of Oncology IOV - IRCCS, Padua, Italy\\
Olivier Morin & Department of Radiation Oncology, University of California San Francisco, San Francisco, CA, USA\\
Henning Müller & Institute of Informatics, University of Applied Sciences and Arts Western Switzerland (HES-SO), Sierre, Switzerland\\
Christophe Nioche & Institut Curie, Université PSL, Inserm U1288, Laboratoire d’Imagerie Translationnelle en Oncologie, Orsay, France\\
Fanny Orlhac & Institut Curie, Université PSL, Inserm U1288, Laboratoire d’Imagerie Translationnelle en Oncologie, Orsay, France\\
Sarthak Pati & Center for Artificial Intelligence and Data Science for Integrated Diagnostics (AI2D) and Center for Biomedical Image Computing and Analytics (CBICA), University of Pennsylvania, Philadelphia, PA, USA; Department of Pathology and Laboratory Medicine, Perelman School of Medicine, University of Pennsylvania, Philadelphia, PA, USA; and Department of Radiology, Perelman School of Medicine, University of Pennsylvania, Philadelphia, PA, USA\\
Arman Rahmim & Departments of Radiology and Physics, University of British Columbia, Vancouver, BC, Canada\\
Seyed Masoud Rezaeijo & Department of Medical Physics, Faculty of Medicine, Ahvaz Jundishapur University of Medical Sciences, Ahvaz, Iran.\\
Christopher G Rookyard & Cancer Imaging, School of Biomedical Engineering and Imaging Sciences, King's College London, London, United Kingdom; and Repository Unit, Cancer Research UK National Cancer Imaging Translational Accelerator, United Kingdom\\
Mohammad R Salmanpour & Department of Integrative Oncology, BC Cancer Research Institute, Vancouver, BC, Canada; and Technological Virtual Collaboration (TECVICO Corp.), Vancouver, BC, Canada\\
Roger Schaer & Institute of Informatics, University of Applied Sciences and Arts Western Switzerland (HES-SO), Sierre, Switzerland\\
Andreas Schindele & Department of Nuclear Medicine, Universitätsklinikum Augsburg, Augsburg, Germany\\
Isaac Shiri & Division of Nuclear Medicine and Molecular Imaging, Geneva University Hospital, Geneva, Switzerland\\
Emiliano Spezi & School of Engineering, Cardiff University, Cardiff, United Kingdom\\
Stephanie Tanadini-Lang & Department of Radiation Oncology, University Hospital Zurich, University of Zurich, Zurich, Switzerland\\
Florent Tixier & LaTIM, INSERM, UMR 1101, Université de Bretagne-Occidentale, Brest, France\\
Taman Upadhaya & Department of Radiation Oncology, University of California San Francisco, San Francisco, CA, USA\\
Vincenzo Valentini & Dipartimento Radiodiagnostica, Radioterapia ed Ematologia, Fondazione Policlinico Universitario “A. Gemelli” IRCCS, Rome, Italy; and Professore ordinario di Radioterapia, Università Cattolica del Sacro Cuore - Milano, Milan, Italy\\
Martin Vallières & Department of Computer Science, Université de Sherbrooke, Sherbrooke, QC, Canada; and Centre de recherche du Centre hospitalier universitaire de Sherbrooke (CHUS), Sherbrooke, QC, Canada\\
Joost J M van Griethuysen & Department of Radiology, The Netherlands Cancer Institute, Amsterdam, the Netherlands; and Department of Radiology, UMC Utrecht, Utrecht, the Netherlands\\
Philip Whybra & School of Engineering, Cardiff University, Cardiff, United Kingdom\\
Fereshteh Yousefirizi & Department of Integrative Oncology, BC Cancer Research Institute, Vancouver, BC, Canada\\
Habib Zaidi & Division of Nuclear Medicine and Molecular Imaging, Geneva University Hospital, Geneva, Switzerland\\
Alex Zwanenburg & OncoRay – National Center for Radiation Research in Oncology, Faculty of Medicine and University Hospital Carl Gustav Carus, Technische Universität Dresden, Helmholtz-Zentrum Dresden - Rossendorf, Dresden, Germany; and National Center for Tumor Diseases (NCT), Partner Site Dresden, Germany: German Cancer Research Center (DKFZ), Heidelberg, Germany, Faculty of Medicine and University Hospital Carl Gustav Carus, Technische Universität Dresden, Dresden, Germany, and Helmholtz Association / Helmholtz-Zentrum Dresden - Rossendorf (HZDR), Dresden, Germany\\
 
\bottomrule
\caption{Alphabetical list of IBSI2 contributors.\label{participantList}}
\end{longtable}
\normalsize

\newpage
\setcounter{tocdepth}{1}
\tableofcontents

\mainmatter

\chapter{Introduction}
Medical imaging is often used to support clinical decision-making, but currently only through visual inspection or simple measures.
Additional relevant information concerning \textit{e.g.}, disease phenotypes, may be present in medical images but remains unassessed~\cite{Lambin2017}.
Radiomics characterises regions of interest in medical images using quantitative measures for \textit{e.g.} morphology, image intensities and texture.
Such characteristics are called image biomarkers or image features.
Since 2012, an exponentially growing number of radiomics-related publications are demonstrating the relevance of artificial intelligence and medical image analysis in the context of various diseases and imaging modalities.
However, the standardisation and reproducibility of image biomarkers is a major remaining challenge~\cite{Zwanenburg2019-ky}.
In particular, the interoperability of radiomics software is hindered by the lack of consensus concerning the exact calculation of image biomarkers.
As a result, considerable variations in biomarker values have been reported, even if they are computed from the same image~\cite{Hatt2017-zp,Bogowicz2017-kh,Foy2018-mx}.
To improve interoperability, the Image Biomarker Standardisation Initiative (IBSI), an independent international collaboration, proposes (i) reference manuals documenting mathematical definitions of image analysis processes and (ii) reference values for image biomarkers based on phantoms with controlled image input and well-defined configurations of the image processing parameters.

This reference manual draws on and extends the first IBSI reference manual~\cite{ZLV2017} by defining and standardising image biomarkers based on so-called convolutional filters, \textit{e.g.}, wavelets, Laws kernels and Laplacians of Gaussian (LoG).
Here, we specifically address several so-called convolutional filters.
Filters such as entropy, median, max and morphological filtering are not covered.

This document is structured as follows.
The first section of this chapter describes the place of convolutional filtering in the overall radiomics workflow.
The second section of this chapter introduces notations and conventions (Section~\ref{sec:notations}).
The final section provides a brief overview of the image filtering process for image biomarker extraction (Section~\ref{sec:overview}).
Chapter~\ref{sec:convolution} details all aspects of convolutional operations, the fundamental process for linear image filtering.
Chapter~\ref{sec:linearFilterProperties} lists important properties of convolutional filters.
%Aggregation operations are a fundamental step to transform filter responses (called response maps) into scalar-valued quantitative image biomarkers and is described in Chapter~\ref{sec:aggregation}.
Commonly used convolutional filtering methods in radiomics are identified and formally defined in Chapter~\ref{sec:filtersDescription}.
Chapter~\ref{sec:guidelinesDetails} contains technical reporting guidelines and an overview of parameters for the convolutional filtering methods detailed in the previous chapter.
Chapter~\ref{sec:qualitativeComparison} qualitatively compares these filtering methods in terms of the properties introduced in Chapter~\ref{sec:linearFilterProperties}.
Finally, benchmarking of software implementations of the filters defined in Chapter~\ref{sec:filtersDescription} is proposed in Chapter~\ref{sec:benchmarking}, where phantoms are defined and the benchmarking procedure, including response map comparisons, is detailed. 

\section{Convolutional Filtering within the Radiomics Workflow}\label{sec:overallWorkflow}
\begin{figure}
\centering
\includegraphics[]{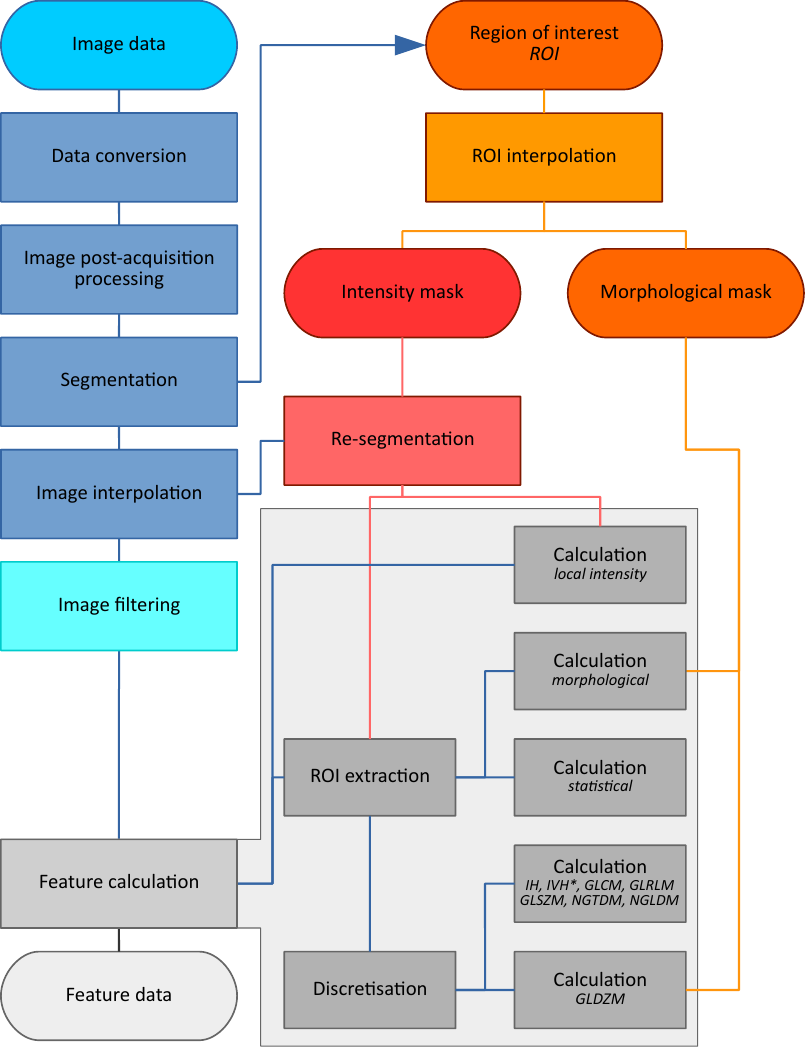}
\caption{Overall image processing scheme with image filtering (adapted from Zwanenburg \textit{et al.}~\cite{ZLV2017}). After loading a medical image, the image data are optionally converted (\textit{e.g.} SUV normalisation). The image may then undergo further post-processing (\textit{e.g.} noise reduction, bias field correction). Subsequently, a segmentation mask is created or loaded to identify the ROI in the image. The image is subsequently interpolated to ensure that image voxels are isometric, after which image filters may be applied to the image. The ROI mask is interpolated to the same grid as the image, prior to forming intensity and morphological ROI masks. The intensity (but not morphological) mask is optionally re-segmented based on image intensities of the unfiltered image. Subsequently, features are computed from the filtered image and the applicable ROI masks. The intensities in ROI intensity mask may undergo discretisation prior to computing features from \textit{e.g.} texture families. IH: intensity histogram; IVH: intensity-volume histogram; GLCM: grey level cooccurrence matrix; GLRLM: grey level run length matrix; GLSZM: grey level size zone matrix; NGTDM: neighbourhood grey tone difference matrix; NGLDM: Neighbouring grey level dependence matrix; GLDZM: grey level distance zone matrix; *A different discretisation scheme is usually used for computing IVH features.}
\label{fig:imageProcessing}
\end{figure}

 \marginnote{\footnotesize v4: Moved embedding of convolutional filtering in radiomics workflow to section \ref{sec:overallWorkflow}.}
Convolutional filtering is part of image processing for radiomics. Within the image processing workflow, filtering is conducted after image interpolation, see Figure \ref{fig:imageProcessing}. Filtering an image creates a response map that for the filters discussed in this document has the same dimensions as the input image. Radiomic features, such as those detailed in the first IBSI reference manual~\cite{ZLV2017}, may then be computed from the response map as well.

\section{Notations and Conventions}\label{sec:notations}
We follow the notations of Depeursinge \textit{et al.}~\cite{DFA2017} and Unser~\cite{Uns2012}.
The imaginary symbol for complex numbers is denoted by $\mathrm{j}$.
A continuous image is modeled as a $D$-dimensional function of the spatial coordinates $\boldsymbol{x} = (x_1, \dots , x_D) \in \mathbb{R}^D$, taking values $f(\boldsymbol{x}) \in \mathbb{R}$. 
The Fourier transform of an integrable $f(\boldsymbol{x})$ is noted $\hat{f}(\boldsymbol{\omega})\in\mathbb{C}$ and is defined as
\begin{equation}\label{eq:continuousFourierTransform}
f(\boldsymbol{x})\overset{\mathcal{F}}{\longleftrightarrow}\hat{f}(\boldsymbol{\omega})=
\int_{\boldsymbol{x}\in\mathbb{R}^D} f(\boldsymbol{x}) \,\, e^{-\mathrm{j}\langle\boldsymbol{\omega},\boldsymbol{x}\rangle} \mathrm{d}\boldsymbol{x},
\end{equation}
where $\boldsymbol{\omega}=(\omega_1,\dots,\omega_D) \in \mathbb{R}^D$ is the frequency coordinate vector and $\langle\cdot,\cdot\rangle$ denotes the scalar product.

A discrete image is modeled as a $D$-dimensional function of the variable $\boldsymbol{k} = (k_1, \dots , k_D) \in \mathbb{Z}^D$, taking values $f[\boldsymbol{k}] \in \mathbb{R}$. 
A subset of $\mathbb{Z}^D$ is considered in practice for the image domain with dimensions $N_1\times\cdots\times N_D$ as possible values for the index vector $\boldsymbol{k}$.
The discrete Fourier transform\footnote{In this discrete setting, the normalised Nyquist frequency is denoted by $\nu_B=\pi$.} of a summable $f[\boldsymbol{k}]$ is noted $\hat{f}[\boldsymbol{\nu}]\in\mathbb{C}$, with $\boldsymbol{\nu}=(\nu_1,\dots,\nu_D) \in \mathbb{Z}^D$ the discrete frequency coordinate vector, and is defined as
\begin{equation}\label{eq:discreteFourierTransform}
f[\boldsymbol{k}]\overset{\mathcal{F}}{\longleftrightarrow}\hat{f}[\boldsymbol{\nu}]=
\sum_{\boldsymbol{k}\in\mathbb{Z}^D} f[\boldsymbol{k}] \,\, e^{-\mathrm{j}\langle\boldsymbol{\nu},\boldsymbol{k}\rangle}.
\end{equation}

\subsection[Image Directions]{Image Directions\id{XKP5}}\label{sec:imageDirections} 
Two reference frames are important for image filters\footnote{For further information, please see \texttt{\url{https://www.slicer.org/wiki/Coordinate_systems}}, as of November 2019.}: (i) the patient frame of reference and (ii) the image frame of reference. The patient frame of reference determines how the image itself is oriented with regard to anatomical directions. Filters are computed in the image frame of reference, which does not need to exactly match the patient frame of reference. However, to use filters consistently, all images in a given study should have the same orientation relative to the patient frame of reference. This may require the rotation of image sets (\textit{e.g.} scans) to a common orientation relative to the patient frame of reference, \textit{e.g.} in a study with mixed sagittal and axial images. Besides, if data stored in different image formats are used in the same study, special care needs to be taken to ensure that the image reference frame of all image sets have the same orientation relative to the patient frame of reference (\textit{e.g.} this may require flipping and/or transposing images along different directions). For example, image file types such as Digital Imaging and COmmunications in Medicine (DICOM) and Neuroimaging Informatics Technology Initiative (NIfTI) formats may store an image using a different orientation relative to the patient frame of reference. Overall, to maximize the reproducibility potential of a given study, two important parameters should be reported: (i) the common orientation of the image frame of reference of all image sets relative to the patient frame of reference (\textit{e.g.} the Image Orientation Patient (\texttt{0020 0037}) field for the DICOM format) and (ii) if applicable, the (potentially different) rotation operations applied to each image set separately to ensure a common orientation of the image frame of reference relative to the patient frame of reference.  

The image frame of reference determines how the stored image grid is oriented and thus where image voxels are located. Within the image reference frame, we define filter directions as follows:
\begin{itemize}
    \item $k_1$ (\textit{i.e.} $x$) goes from left to right with increasing grid index values,\hfill \textid{1F0N}
    \item $k_2$ (\textit{i.e.} $y$) goes from top to bottom  with increasing grid index values,\hfill \textid{M2EZ}
    \item $k_3$ (\textit{i.e.} $z$) goes from front to back  with increasing grid index values.\hfill \textid{R83H}
\end{itemize}

\vspace{2mm}
\begin{tcolorbox}[width=150mm, halign=left, colframe=black, colback=white, boxsep=0mm, arc=3mm, colframe=black!50!white,
title=Implementation Troubleshooting, title filled=true, fonttitle=\bfseries]
\begin{itemize}
\item The orientation phantom (section \ref{sec:benchmarkingMethodo}) can be used to assess whether the orientation used in an implementation conforms to the expected orientation.
\item The orientation phantom has a dimension of $(32, 48, 64)$ voxels along $k_1$ ($x$), $k_2$ ($y$) and $k_3$ ($z$) axes, respectively. The pixel intensity increases with the distance from the origin, which has an intensity of $0$. The most distal voxel has an intensity of $141$.
\end{itemize}
\end{tcolorbox}

\section{Steps in convolutional filtering}\label{sec:overview}
\begin{figure}
\centering
\includegraphics[trim = 0 0 0 0, clip, width=\linewidth]{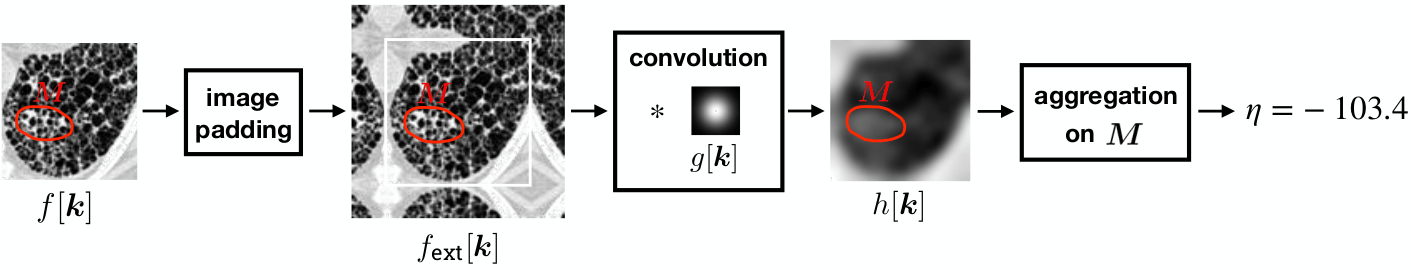}\\
\caption{Overview of the filter-based image biomarker extraction process.}
  \label{fig:overview}
\end{figure}

This section provides an overview of the filter-based image biomarker extraction process, which is shown in Fig.~\ref{fig:overview}.
It consists of the following steps:
\subsection[Image Padding]{Image Padding \id{GBYQ}}
The input image $f[\boldsymbol{k}]$ is first padded with an appropriate boundary condition to form an extended image $f_{\text{ext}}[\boldsymbol{k}]$. This step is further required to calculate the response of a filter close to the boundaries of $f$ and is detailed in Section~\ref{sec:boundaryConditions}.
\subsection[Convolution]{Convolution \id{Y4GJ}}
$f_{\text{ext}}[\boldsymbol{k}]$ is convolved with a filter $g[\boldsymbol{k}]$, which yields a response map $h[\boldsymbol{k}]$ (\textid{5P3T}) with the same dimension as $f[\boldsymbol{k}]$.
This process is described in Section~\ref{sec:convolution} and consists of applying the filter at all possible locations $\boldsymbol{k}$ in the input image.
Local or global rotation invariance operations may be applied during this step to minimise the variation of  $h[\boldsymbol{k}]$ with respect to rotations of the input image, as well as variations in orientations of local image patterns (\textit{e.g.} tumor walls or vessels, see Chapter~\ref{sec:linearFilterProperties}).
\subsection[Aggregation]{Aggregation \id{YHT3}}
$h[\boldsymbol{k}]$ is not a directly usable quantitative image feature and needs to be aggregated (\textit{i.e.} summarised) over a given Region Of Interest (ROI) $\boldsymbol{R}$ using an aggregation function~\cite{DFA2017}.
The latter will transform the response map $h$ into a scalar measurement $\eta$, which can be further used in a statistical or machine learning model.
$h$ can be seen as a new image derived from $f$, on which any feature defined in the first IBSI reference manual can be computed~\cite{ZLV2017}.
However,interpretation of the latter would be difficult (unless the filter is simply used to \textit{e.g.} remove high frequency noise), and the large number of possible combinations of filters and features increases the risk of fortuitous correlations between features and clinical endpoints.
%%% RECOMMENDATION NOT TO USE 2ND ORDER on top of response maps
In practice, the most obvious aggregation functions are the first four statistical moments of the voxel value distribution in $\boldsymbol{R}$, the first one being the average:
\begin{equation}\label{eq:aggregationAverage}
\eta=
\frac{1}{|\boldsymbol{R}|}
\sum_{\boldsymbol{k}\in \boldsymbol{R}} h[\boldsymbol{k}],
\end{equation}
where $|\boldsymbol{R}|$ denotes the number of voxels in the ROI.
It is worth noting that the influence of the aggregation function is as important as the filter and it should be considered carefully~\cite{Dep2017}.
For the particular case of average-based aggregation in Eq.~\eqref{eq:aggregationAverage}, one must be careful when using zero-mean filters (\textit{e.g.} LoG, some Laws kernels, Gabor, wavelets, Riesz) leading to $\eta\approx 0$. In this case, taking the energy $h^2[\boldsymbol{k}]$ or the absolute value $|h[\boldsymbol{k}]|$ of the response map before the aggregation is recommended.
%%% RECOMMENDATION for zero mean filters
%
\chapter{Convolution}\label{sec:convolution}
The $D$-dimensional continuous convolution of a filter $g(\boldsymbol{x})$ and an image $f(\boldsymbol{x})$ yields a response map $h(\boldsymbol{x})$ with the same dimension $N^D$ as $f$ following
\begin{equation}\label{eq:continuousConvolution}
h(\boldsymbol{x}_0)=(g\ast f)(\boldsymbol{x}_0)=
\int_{\boldsymbol{x}\in\mathbb{R}^D}
g(\boldsymbol{x})f(\boldsymbol{x}_0-\boldsymbol{x})
\mathrm{d}\boldsymbol{x}.
\end{equation}
The continuous convolution is only explained here for reference. 
In practice, images are discretised and the corresponding discrete convolution\footnote{We consider here filters and images with square ($D=2$) or cubic ($D=3$) dimensions for simplicity.} (\textid{Y4GJ}) of a $M^D=M\times\cdots\times M$ filter $g[\boldsymbol{k}]$ and a $N^D=N\times\cdots\times N$  image $f[\boldsymbol{k}]$ is
\begin{equation}\label{eq:discreteConvolution}
h[\boldsymbol{k}_0]=(g\ast f)[\boldsymbol{k}_0]=
\sum_{\boldsymbol{k}\in M^D}
g[\boldsymbol{k}]f[\boldsymbol{k}_0-\boldsymbol{k}].
\end{equation}
(\ref{eq:continuousConvolution}) and~(\ref{eq:discreteConvolution}) can also be computed in the Fourier domain and the convolution becomes in this case just the product
\begin{equation}\label{eq:convolutionFFT}
(g \ast f) (\boldsymbol{x})  \overset{\mathcal{F}}{\longleftrightarrow}   \hat{g}(\boldsymbol{\omega}) \hat{f}(\boldsymbol{\omega}),
\end{equation}
which remains valid for the discrete case via a Hadamard product between the arrays of the image and the filter.

The latter implementation becomes efficient when the filter size is large as the computational cost of the Fast Fourier Transform (FFT) becomes small when compared to implementing the $D$ interleaved sums in~(\ref{eq:discreteConvolution}).
As a rule of thumb in 2D\footnote{This estimation only counts the number of operations. In practice, the efficiency will strongly depend on software language (complex data storage and manipulation) and hardware.}, for a $N\times N$ image and a $M\times M$ filter, the computational cost to convolve them in the spatial domain is $N^2 \cdot M^2$.
In the Fourier domain, the cost is $N^2\cdot(1+2\log_2(N^2))$ when both the filter and the image are already defined in the Fourier domain and the filter is interpolated to the dimension $N$ to allow pointwise multiplication.
For instance, with a $N^2=512\times 512$ image, FFT filtering becomes more efficient than spatial filtering when the filter is larger than $M^2=7\times 7$.

At a fixed position $\boldsymbol{x}_0$, the value of the convolution response $(g\ast f)(\boldsymbol{x}_0)$, \textit{i.e.}, the application of the filter to the image, is the scalar product 
\begin{equation}\label{eq:scalProd}
(g\ast f)(\boldsymbol{x}_0)=
\langle f(\boldsymbol{x}_0 -\cdot) , g( \cdot) \rangle,
\end{equation}
with $f(x_0 - \cdot)$ is the function $x \mapsto f(x_0 - x)$.
\section{Separability of the Convolution}\label{sec:separableConv}
Specific filter subtypes are \emph{separable}, which means that their $D$-dimensional kernel $g_{123\dots D}[\boldsymbol{k}]$ can be obtained from the outer product of $D$ 1-dimensional kernels $g_i[k_i]$ as
\begin{equation}\label{eq:separability}
g_{123\dots D}[\boldsymbol{k}]=\prod_{i=1}^D g_i[k_i].
\end{equation}
An example of a separable 2$D$ unit-norm $3\times 3$ smoother $g_{ss}[\boldsymbol{k}]$ is
\begin{equation}\label{eq:separabilityExample}
g_{ss}[\boldsymbol{k}]= g_{s}[k]\otimes g_{s}[k]=
g_{s}[k]\cdot g_{s}^T[k]=
\frac{1}{\sqrt{6}}\begin{bmatrix}1 \\ 2 \\ 1\end{bmatrix} \cdot
\frac{1}{\sqrt{6}}\begin{bmatrix}1 & 2 & 1\end{bmatrix}=
\frac{1}{6}\begin{bmatrix}1 & 2 & 1 \\ 2 & 4 & 2 \\ 1 & 2 & 1\end{bmatrix}.   
\end{equation}

\begin{figure}
\centering
\includegraphics[trim = 0 0 0 0, clip, width=\linewidth]{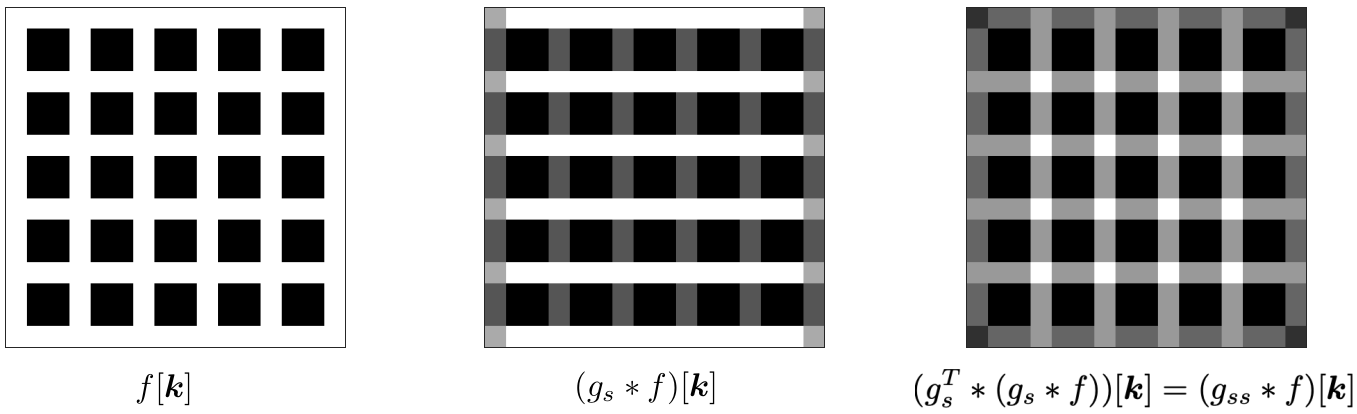}\\
\caption{A $16\times 16$ image $f[\boldsymbol{k}]$ (left) is filtered by $g_{ss}$ defined in Eq.~\eqref{eq:separabilityExample} using a separable convolution.
The intermediate image $(g_{s}\ast f)[\boldsymbol{k}]$ is shown (centre), where the convolution is performed with $g_s$ aligned along the lines (\textit{i.e.} $k_1$) of $f$.
After convolving this intermediate image with $g_{s}^T$ (\textit{i.e.} aligned along the columns $k_2$), the result (right) is equivalent to a 2$D$ convolution of $f$ and $g_{ss}$.}
  \label{fig:separableConv}
\end{figure}

The steps of a separable convolution process of a $16\times 16$ image $f[\boldsymbol{k}]$ with $g_{ss}[\boldsymbol{k}]$ are illustrated in Fig.~\ref{fig:separableConv}.
%
%
%A filter is separable if and only if all its rows are multiples of each other. Then you can pick one, call it $f$, make a column of the multiplicative factors, call it $g$, and find that $h=f\ast g$. This can also be checked by looking at the rank of the kernel (considered as a matrix). If rank is one, than (by SVD decomposition) you can find two vectors whose outer product is the kernel.
Thanks to the associativity of convolution, separable filters can be implemented efficiently with $D$ successive convolutions with 1$D$ filters when compared to more computationally expensive convolutions with  $D$-dimensional filters.
For instance, we have the composition
\begin{equation}\label{eq:separability2Dexample}
(g_2\ast(g_1\ast f))[\boldsymbol{k}]=
(g_1\ast(g_2\ast f))[\boldsymbol{k}]=
(g\ast f)[\boldsymbol{k}].
\end{equation}
%Where $g[\boldsymbol{k}]=g_1[k]\otimes g_2[k]=g_2[k]\otimes g_1[k]$, thanks in turn to the associativity of the outer product.

However, satisfying Eq.~(\ref{eq:separability}) strongly constrains the design of the filter. Moreover, because image axes are analysed separately, separable convolutions yield directional image analyses biased along $k_1$, $k_2$ and $k_3$ with no rotation invariance by default (the importance of the latter is detailed in Section~\ref{sec:geomInvariances}). The only separable and directionally insensitive kernel is the Gaussian.

As a convention, the filter responses are noted $h_{123}[\boldsymbol{k}]$ when resulting from a convolution with the three 1$D$ filters as follows: $g_1$ along $k_1$, $g_2$ along $k_2$, and $g_3$ along $k_3$ (see section \ref{sec:imageDirections}).
The corresponding 3$D$ kernel is noted $g_{123}[\boldsymbol{k}]$.

% $$f[\boldsymbol{k}]$$
% $$(g_s\ast f)[\boldsymbol{k}]$$
% $$(g_s^T\ast (g_s\ast f)[\boldsymbol{k}]=(g_{ss}\ast f)[\boldsymbol{k}]$$

\section[Boundary Conditions]{Boundary Conditions \id{GBYQ}}\label{sec:boundaryConditions}
\begin{figure}
\centering
   \begin{minipage}[b]{150pt}
     \centering
     \includegraphics[trim = 0 0 0 0, clip, scale=0.57]{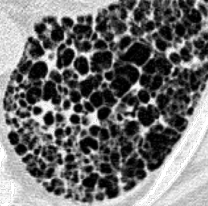}
     \subcaption{$f[\boldsymbol{k}]$}\label{fig:boundaryConditions1_a}
     \hspace{100pt}
   \end{minipage}
   \begin{minipage}[b]{150pt}
     \centering
     \includegraphics[trim = 0 0 0 0, clip, scale=0.57]{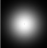}
     \subcaption{$g[\boldsymbol{k}]$}\label{fig:boundaryConditions1_b}
     \hspace{100pt}
   \end{minipage}
   \begin{minipage}[b]{150pt}
     \centering
     \includegraphics[trim = 0 0 0 0, clip, scale=0.57]{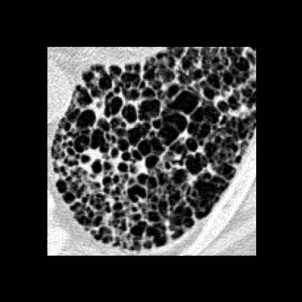}
     \subcaption{$f_{\text{ext}}^{\text{padding}}[\boldsymbol{k}]$}\label{fig:boundaryConditions1_c}
     \hspace{100pt}
   \end{minipage}
   \begin{minipage}[b]{150pt}
     \centering
     \includegraphics[trim = 0 -34 0 0, clip, scale=0.57]{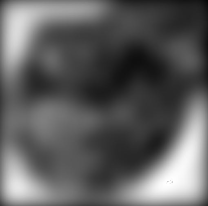}
     \subcaption{$(g\ast f_{\text{ext}}^{\text{padding}})[\boldsymbol{k}]$}\label{fig:boundaryConditions1_d}
     \hspace{100pt}
   \end{minipage}
   \begin{minipage}[b]{150pt}
     \centering
     \includegraphics[trim = 0 0 0 0, clip, scale=0.57]{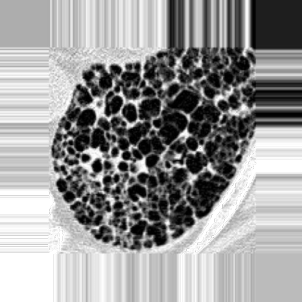}
     \subcaption{$f_{\text{ext}}^{\text{nearest}}[\boldsymbol{k}]$}\label{fig:boundaryConditions1_e}
   \end{minipage}
   \begin{minipage}[b]{150pt}
     \centering
     \includegraphics[trim = 0 -34 0 0, clip, scale=0.57]{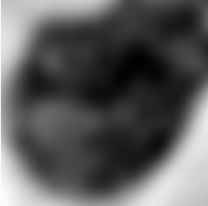}
     \subcaption{$(g\ast f_{\text{ext}}^{\text{nearest}})[\boldsymbol{k}]$}\label{fig:boundaryConditions1_f}
   \end{minipage}
  \caption{Qualitative comparison of various methods for imputing image values at the boundaries.
  The image $f[\boldsymbol{k}]$ is smoothed by convolution with the Gaussian filter $g[\boldsymbol{k}]$.
  The response maps using either zero padding or nearest methods are compared.
  }
  \label{fig:boundaryConditions1}
\end{figure}

Computing the discrete convolution as in Eq.~(\ref{eq:discreteConvolution}) when the distance between the center $\boldsymbol{k}_0$ of the filter and a boundary of the image is smaller than half of the support $M$ of the filter requires accessing pixel values outside the support of the image.
Therefore, a method must be used to impute pixel values in the vicinity of the image boundaries, \emph{i.e.}, to generate an extended image $f_{\text{ext}}[\boldsymbol{k}]$ that includes a margin with a width/height/depth of $\lfloor M/2-1 \rfloor$.
Four common padding methods are described in the following subsections and are compared qualitatively in Fig.~\ref{fig:boundaryConditions1} and~\ref{fig:boundaryConditions2}, where a CT image of honeycombing lung parenchyma (Fig.~\ref{fig:boundaryConditions1_a}, $N=301$) is smoothed by convolution with an isotropic Gaussian filter (Fig.~\ref{fig:boundaryConditions1_b}, $M=70$, $\sigma = 15$).

Because all padding methods make arbitrary assumptions concerning image content beyond the boundaries, a method should be chosen based on the expected image background.
However, general advantages and disadvantages exist and are discussed in the next sections.
%While separable convolutions do not need to access locations beyond the corners of image $f$, the importance of boundary conditions remains fundamental.

\subsection[Constant Value Padding]{Constant Value Padding \id{Z3VE}}\label{sec:constantValuePadding}
The simplest method to extend the support of $f[\boldsymbol{k}]$ is to pad it with a constant value $C$ as
$$
\begin{cases}
f_{\text{ext}}^{\text{constant}}[\boldsymbol{k}]=f[\boldsymbol{k}] \quad\forall\quad (k_1,\dots,k_D)\in \{0,\dots,N-1\}^D \quad \text{and},\\
f_{\text{ext}}^{\text{constant}}[\boldsymbol{k}]=C \quad \text{otherwise}. \\
\end{cases}
$$
Thus, all pixels outside the original image are assigned the constant value $C$. This is illustrated in Fig.~\ref{fig:boundaryConditions1_c}.
Though constant value padding is simple to implement, potentially sharp transitions between $C$ and the image value at a boundary may yield inconsistent filter responses.
Distortions may appear in the boundary region of the response image as a consequence.
Such artefacts can be observed in all boundaries of the filtered image in Fig.~\ref{fig:boundaryConditions1_d}.
With $C=0$, this method is also called \textit{zero padding}.
\subsection[Nearest Value Padding]{Nearest Value Padding \id{SIJG}}
The nearest value padding method consists of repeating the values of the image at the boundary.
We have 
$$
\begin{cases}
f_{\text{ext}}^{\text{nearest}}[\boldsymbol{k}]=f[\boldsymbol{k}] \quad\forall\quad (k_1,\dots,k_D)\in \{0,\dots,N-1\}^D \quad \text{and},\\
f_{\text{ext}}^{\text{nearest}}[\boldsymbol{k}]=f[\boldsymbol{k}_{\text{nearest}}] \quad \text{otherwise},\\
\end{cases}
$$
where $\boldsymbol{k}_{\text{nearest}}:=\underset{\boldsymbol{k}_{\text{nearest}}}{\arg\min} \biggl(||\boldsymbol{k}-\boldsymbol{k}_{\text{nearest}}|| \biggr)$ is the closest neighbour in $f$. Thus, all pixels outside the original image get the intensity value of the closest pixel in the original image.
This is illustrated in Fig.~\ref{fig:boundaryConditions1_e}.
The advantage of this method is that the transitions at the boundary are usually relatively small.
However, it introduces a nonexistent pattern in the response image.
This method is also called \textit{replicate}.
\subsection[Periodisation]{Periodisation \id{Z7YO}}
Another straightforward method is to repeat the image along every dimension, \emph{i.e.}, so to periodise the image content.
The extended image is therefore equivalent to the original image modulo its support as
$$f_{\text{ext}}^{\text{periodise}}[\boldsymbol{k}]=f[k_1\mod\,\,\, N,\,\,\, \dots,\,\,\, k_D\mod\,\,\, N]$$
for indices $k_i = 0,\dots, N-1$.
This is illustrated in Fig.~\ref{fig:boundaryConditions2_a}.
The introduced patterns are consistent with the actual image content. 
However, strong transitions are possible at the boundaries.
The subsequent artefacts can be observed in the upper part of the left boundary of the example response image (see Fig.~\ref{fig:boundaryConditions2_b}).
It is worth noting that this periodisation is implicit if the convolution operation is performed in the Fourier domain using Eq.~(\ref{eq:convolutionFFT}).
This method is also called \textit{wrapping}, \textit{circular} or \textit{tiling}.
\subsection[Mirror]{Mirror \id{ZDTV}}
\marginnote{\footnotesize v5: Clarified use of mirror boundary condition by the IBSI.}
The mirror folding method consists of symmetrising the image at the boundaries. The extended image is
$f_{\text{ext}}^{\text{mirror}}[\boldsymbol{k}]=f[\boldsymbol{k'}]$, with $\boldsymbol{k'} = (k'_1,\cdots,k'_D) \in \mathbb{Z}^D$, where
% $$
% \begin{cases}
% k_i'= k_i \quad\forall\quad k_i\in \{1,\dots,N\} \quad\text{and},\\
% k_i'=-k_i+1 \quad\forall\quad k_i < 1 \quad \text{and},\\
% k_i'= 2N - k_i  \quad\forall\quad k_i > N.\\

% \end{cases}
% $$
$$
k_i'=
\begin{cases}
  k_i \,\mod\,\, N \quad\text{if}\quad \lfloor \frac{k_i}{N}\rfloor\quad\text{is even},\\
  N - (k_i \,\mod\,\, N+1) \quad \text{otherwise},\\
 \end{cases}$$
for indices $k_i = 0,\dots N-1$.

\begin{figure}
\centering
   \begin{minipage}[b]{150pt}
     \centering
     \includegraphics[trim = 0 0 0 0, clip, scale=0.57]{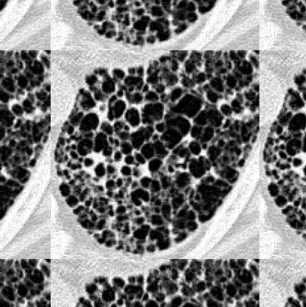}
     \subcaption{$f_{\text{ext}}^{\text{periodise}}[\boldsymbol{k}]$}\label{fig:boundaryConditions2_a}
     \hspace{100pt}
   \end{minipage}
   \begin{minipage}[b]{150pt}
     \centering
     \includegraphics[trim = 0 -34 0 0, clip, scale=0.57]{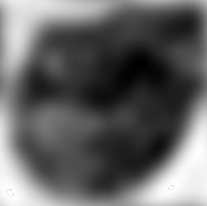}
     \subcaption{$(g\ast f_{\text{ext}}^{\text{periodise}})[\boldsymbol{k}]$}\label{fig:boundaryConditions2_b}
     \hspace{100pt}
   \end{minipage}
   \begin{minipage}[b]{150pt}
     \centering
     \includegraphics[trim = 0 0 0 0, clip, scale=0.57]{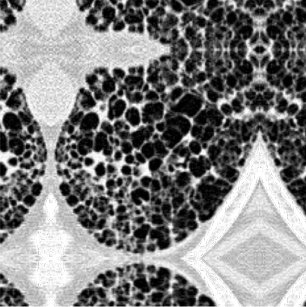}
     \subcaption{$f_{\text{ext}}^{\text{mirror}}[\boldsymbol{k}]$}\label{fig:boundaryConditions2_c}
   \end{minipage}
   \begin{minipage}[b]{150pt}
     \centering
     \includegraphics[trim = 0 -34 0 0, clip, scale=0.57]{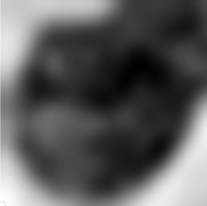}
     \subcaption{$(g\ast f_{\text{ext}}^{\text{mirror}})[\boldsymbol{k}]$}\label{fig:boundaryConditions2_d}
   \end{minipage}
  \caption{Qualitative comparison of various methods for imputing image values at the boundaries.  The image $f[\boldsymbol{k}]$ is smoothed by convolution with the Gaussian filter $g[\boldsymbol{k}]$ (see Fig.~\ref{fig:boundaryConditions1}).
  The response maps using either periodisation or mirroring methods are compared.}
  \label{fig:boundaryConditions2}
\end{figure}

This is illustrated in Fig.~\ref{fig:boundaryConditions2_c}.
The introduced patterns are consistent with the actual image content and the transitions at the boundaries are relatively smooth, minimising convolution artefacts (see Fig.~\ref{fig:boundaryConditions2_d}).
Depending on the software implementation, the extended image may include or exclude the boundary pixels of the original image.
This method is also called \textit{symmetric}.

For consistency reasons, we assume that boundary pixels are included.

\subsection{Considerations for Radiomics}
In many radiomics applications, the choice of boundary extension method is generally not important, as long as the ROI is sufficiently far from the image boundary, \textit{i.e.} at least by more than half of the convolution filter width/height/depth. In the uncommon case where the ROI is close to the border, the \textit{mirror} method may generally be recommended because it avoids sharp transitions and artificial patterns.

\chapter{Filtering Properties}\label{sec:linearFilterProperties}
Due to the characteristics common to medical images, several filter properties are important for image analysis methods to be optimal.
In this section, we list these properties and suggest how they can be fulfilled with filtering techniques.
\section{Geometric Invariances for Medical Image Analysis}\label{sec:geomInvariances}
Analysis of medical imaging requires invariance or equivariance to various geometric transformations.
%In medical imaging, the data structure and acquisition result in particular needs for invariances or equivariances to geometric transformations.
As in Depeursinge \textit{et al.}~\cite{DFA2017}, we define the invariance of a function $g(\cdot)$ to a transformation $\mathcal{T}$ of the image $f(x)$ as $g(\mathcal{T}\{f\})(x) = g(f)(x)$. In the case of filtering, this means that the response map obtained by the convolution of the image with a filter is unchanged under the effect of $\mathcal{T}$. The equivariance is defined as $g(\mathcal{T}\{f\})(x) = \mathcal{T}\{g(f)\}(x)$, \textit{i.e.} the response map undergoes the same transformation $\mathcal{T}$ as the image. 
\begin{enumerate}
\item \textbf{Translation}: The response maps of linear filters are equivariant to translation by construction, thanks to the convolution operation. For a translation transformation $\mathcal{T}_{t,\boldsymbol{x}_0}$ by $\boldsymbol{x}_0$, $g(\mathcal{T}_{t,\boldsymbol{x}_0}\{f\})(x) = \mathcal{T}_{t,\boldsymbol{x}_0}\{g(f)\}(x)$, meaning that the filtering process does not differ according to position, \textit{e.g.} if an input $(0,1,2,1,0)$ leads to response $(0,0,1,0,0)$, then input $(0,0,1,2,1)$ should lead to response $(0,0,0,1,0)$.
Equivariance to translations is required in medical imaging since we want to keep track of the positions where the filter responded to create full response maps, where patterns of interest may appear at random positions.
\item \textbf{Scaling}: In most cases, scale is a discriminative property in medical images. Thus, filters should not be scale invariant. 
\item \textbf{Rotation}: Response maps should be equivariant to global rotations and invariant to local rotations as defined in Depeursinge \textit{et al.}~\cite{DFA2017}.
We consider a global rotation when applied around the origin $\boldsymbol{0}$ of the image.
In this case, equivariance to global rotations is required for the same reasons as translation equivariance, where the positions where the filter responded should rotate in the same fashion as the image itself.
A local rotation around a given position $\boldsymbol{x}_0$ requires the following:
$\mathcal{T}_{r,\boldsymbol{x}_0}=\mathcal{T}_{t,\boldsymbol{x}_0}\mathcal{T}_{r}\mathcal{T}_{t,-\boldsymbol{x}_0}$.
For $\mathcal{T}_{r,\boldsymbol{x}_0}$ a transformation that locally rotates patterns, we seek the invariance $g(\mathcal{T}_{r,\boldsymbol{x}_0}\{f\})(x) = g(f)(x)$. 
Invariance to local rotations is required to equivalent obtain filter responses to any orientation of a local pattern.
For example, patterns such as vessels may have arbitrary orientations and all must be equivalently characterised (see Section 3.3.1. of Depeursinge \textit{et al.}~\cite{DFA2017}).
In general, we are not interested in the local orientation of the pattern itself, but rather to its presence only, hence justifying the invariance to local rotations.
\end{enumerate}
\section{Directional Sensitivity}
Most structures of interest in medical images are composed of directional components such as edges and corners. Filters may be sensitive to image directions, but not necessarily. This has a one-to-one relation with the circular/spherical symmetry of the filter~\cite{Depeursinge2018}.
We adopt the definition of directionality introduced in Tamura \textit{et al.}~\cite{Tamura1978}.
Formally, a non-directionally sensitive filter consists of a circular/spherical averaging, \textit{i.e.} $g[\boldsymbol{k}]$ only depends on the radius $||\boldsymbol{k}||$.
Directional sensitivity is important: such circularly/spherically symmetric filters cannot differentiate between blob and tubular structures (\textit{e.g.} a small nodule and a vessel).
An example of a non-directionally sensitive filter is the Laplacian of Gaussian (LoG) defined in Section~\ref{sec:LaplacianOfGaussian} and in particular Eq.~\eqref{eq:LoG} (see Fig.~\ref{fig:LoG}).
An example of a directionally sensitive filter is a Gabor filter defined in Section~\ref{sec:Gabor} (see Fig.~\ref{fig:Gabor}).

\section{Combining Directional Sensitivity and Invariance to Local Rotations}\label{sec:combiningDSandRotInv}
Ideally, filters should combine directional sensitivity with invariance to local rotations.
A simple convolutional filter is invariant to local rotations (as defined in Section~\ref{sec:geomInvariances}) if and only if the filter is directionally-insensitive, \textit{i.e.} circularly symmetric (see Proposition 1 of Andrearczyk \emph{et al.}~\cite{AFO2019b}).
Therefore, slightly more complex and non-linear filtering designs are needed to combine the two properties.

A suitable strategy to achieve invariance to local rotations with a directionally-sensitive filter is to (i) compute a pseudo rotation equivariant representation via a collection of rotated filter responses at uniformly sampled orientations and (ii) voxelwise orientation pooling: taking either the average or the max of the oriented responses, yielding the locally rotation invariant response maps $h_{\text{avg}}$ or $h_{\text{max}}$, respectively.

\begin{figure}
\centering
\includegraphics[trim = 0 0 0 0, clip, width=\linewidth]{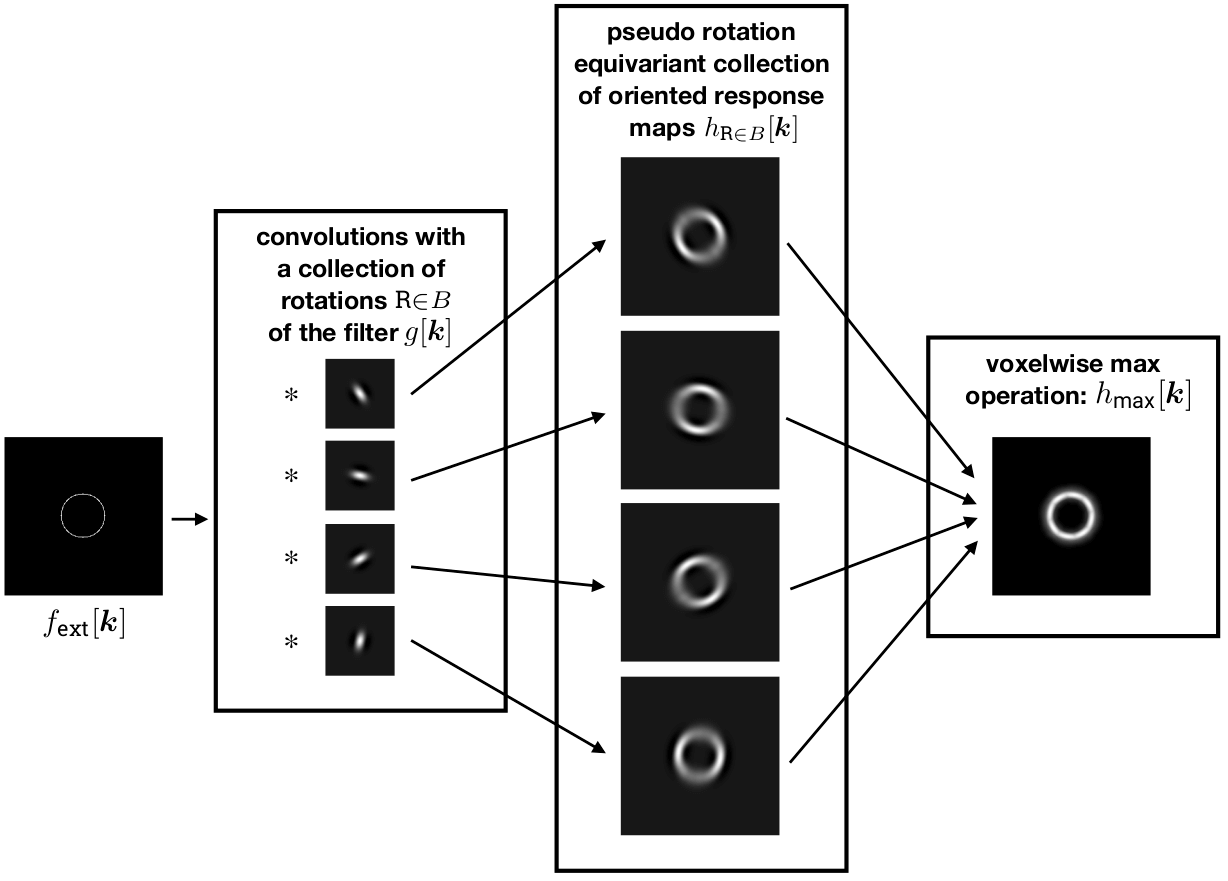}\\
\caption{Overview of the process to achieve filtering operations combining directional sensitivity with invariance to rotations.}
  \label{fig:LRI_overview}
\end{figure}

The pseudo equivariant representation (i) is (\textid{O1AQ})
\begin{equation}\label{eq:rotationEquivariantRepresentation}
    \{ f\ast g[\mathrm{R}\cdot]\}_{\mathrm{R}\in B},
\end{equation}
where $\mathrm{R}$ is a rotation matrix and $B$ is a uniformly sampled set of rotations (\textit{e.g.} the set of right-angle rotations in 2$D$).
The process is illustrated in 2$D$ in Fig.~\ref{fig:LRI_overview}.
It is worth noting that Eq.~\eqref{eq:rotationEquivariantRepresentation} can be efficiently computed using steerable filters, which obviates the need to reconvolve the image with rotated versions of the filters~\cite{WHS2018,AFO2019}. 
In the particular case of separable filters and right angle rotations, the equivariant representation can be efficiently obtained using unidimensional filter flipping and permutation. This is described for the 2$D$ and 3$D$ cases in Appendix~\ref{app:separableConvRightAngleEquivariant}.

Alternatively, an equivariant representation can be obtained by permuting the image orientation in the same manner, convolving the rotated image with the separable filters, and then rotating the response map back to the original orientation.
This method is useful when transforms are performed using a standard software implementation, e.g. of discrete wavelet transforms, instead of using filter kernels directly.

\begin{figure}
    \centering
    \includegraphics[trim= 0 0 0 0, clip, width=\linewidth]{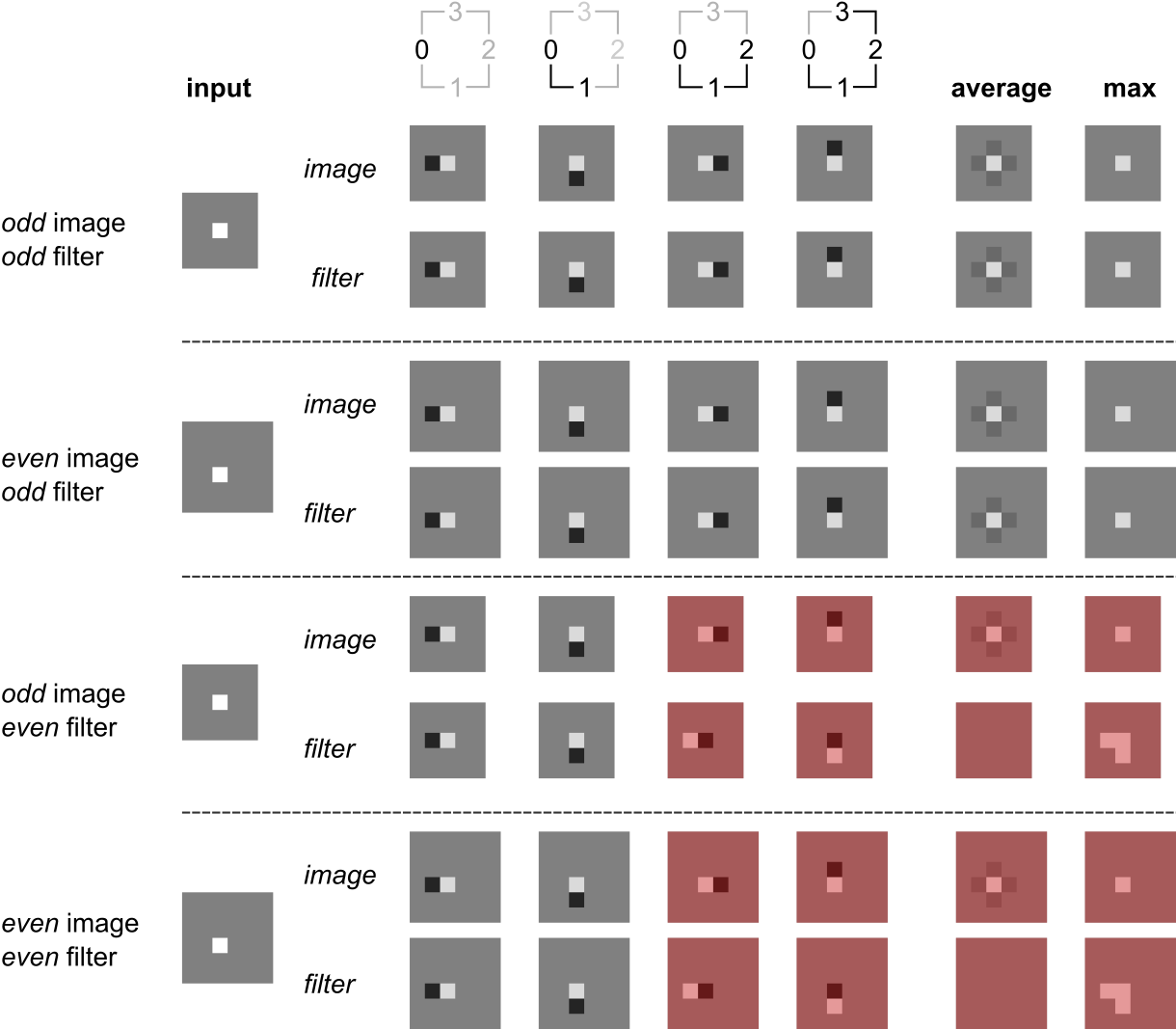}\\
    \caption{Comparison of methods for achieving rotation equivariance for filtering. Two methods can be used: 1) \textit{image}: rotating the image, convolution with the filter, and then rotating the response map back to the original orientation; 2) \textit{filter}: through filter flipping and permutation. In this example, the image is either $5 \times 5$ pixels (\textit{odd} image) or $6 \times 6$ pixels (\textit{even} image). A single pixel with intensity $1.0$ is located at $(3,3)$. Both 1$D$ filter kernels in this example are based on the high-pass filter of the Haar wavelet. The \textit{even} kernel is $\left[-1 / \sqrt{2}, 1 / \sqrt{2} \right]$, whereas the \textit{odd} kernel is $\left[-1 / \sqrt{2}, 1 / \sqrt{2}, 0 \right]$. The \textit{even} and \textit{odd} kernels are completely equivalent except for their length, and produce the same response maps and pooled images for the \textit{rotation} method. For the \textit{odd} kernel both methods to achieve rotational invariance are equivalent. However, for the \textit{even} kernel several response maps (red overlay) can be seen to differ between \textit{rotation} and \textit{filter} methods. Hence we recommend that, when rotational invariance is desired, any even-sized kernel should be extended by appending a $0$ to create an odd-sized kernel for the \textit{filter} method. This behaviour moreover generalises to 2$D$ and 3$D$ filters, and is independent of whether the input image has an even or odd size.}
    \label{fig:rotation_invariance_comparison}
\end{figure}

As illustrated in Fig.~\ref{fig:rotation_invariance_comparison}, both methods are equivalent if and only if the filter kernels have an odd length, or odd dimensions for 2$D$ and 3$D$ filters. This can be achieved by appending a $0$ to extend the kernel prior to filter flipping and permutation. Note that some implementations of convolution might require that a $0$ is prepended instead, but we did not encounter this for MATLAB and scipy standard implementations. 
Filter kernels do not need to be altered when permuting the image orientation instead.

Voxelwise orientation pooling (ii) can be done using either the average or the max over the elements of the equivariant representation provided by Eq.~\eqref{eq:rotationEquivariantRepresentation} (\textid{SVKW}).
Although commonly used, average orientation pooling strongly deteriorates the directional sensitivity of the filtering operation. For a large set of rotations, averaging is equivalent to filtering with a single circularly symmetric filter. 
The max response preserves the directional sensitivity of the filters and achieves invariance to local rotations~\cite{CoW2016b,AFO2019}.
Taking the max response can be  interpreted as ``aligning'' the filter locally to seek for the best match between the filter and the local image pattern (see Fig.~\ref{fig:LRI_overview}).

It is worth noting that particular filtering methods allow combining directional sensitivity with invariance to rotations via the calculation of invariants. Notable examples are circular and spherical harmonics~\cite{KaM2010,eickenberg2017solid,Depeursinge2018}.
\section{Spectral Coverage}\label{sec:spectalCoverage}
The type of filters considered here change the frequency contents of an image in the Fourier domain, thereby altering its appearance.
Each filter has a frequency profile that can be characterised in the Fourier domain. Filters are usually split into one of four categories based on their frequency profiles:
\begin{enumerate}
\item \textbf{All-pass filter}: An all-pass filter does not change the frequency content of the image. Such filters are rarely used. Examples are the identity filter, where the response image is identical to the input image, and a translation filter, where the response image is identical to the input, except for a shift.
\item \textbf{Low-pass filter}: A low-pass filter attenuates the high-frequency content of the image. Examples are mean and Gaussian filters, which produce a smoothed version of the input image.
\item \textbf{High-pass filter}: A high-pass filter attenuates the low-frequency content of the image. High-pass filters may be used to sharpen an image, at the cost of amplifying noise. 
\item \textbf{Band-pass filter}: A band-pass filter attenuates both high- and low-frequency content in an image in specific ranges.
Band-pass filters can capture specific spectral signatures of patterns at relevant scales that are related to the problem at hand (\textit{e.g.}, fibrosis, necrosis).
\end{enumerate}

Examples of low-, high- and band-pass filtered images are shown in Fig.~\ref{fig:nonseparableWT}, respectively $h_L^3$, $h_H^1$ and $h_H^2$.
\chapter{Common Convolutional Filters in Radiomics}\label{sec:filtersDescription}
This section details the definitions of common convolutional filters used for radiomics. Guidance is also provided on how to set their parameters and to obtain appropriate features via matching aggregation functions.
\section{Identification of Common Approaches}
As a starting point, we listed the filtering techniques used in common radiomics software libraries. 
The considered libraries are TexRAD\footnote{\texttt{\url{http://texrad.com}}, as of September 2019.}, Definiens~\cite{BGW2014}, PyRadiomics~\cite{Van_Griethuysen2017-qp} , CGITA~\cite{FLS2014}, IBEX~\cite{ZFF2015}, CERR~\cite{Apte2018-sf}, MAZDA~\cite{SzK2017}, QIFE~\cite{EBR2017}, LIFEx~\cite{Nioche2018-jr}, and QuantImage~\cite{DCS2017}.
\section[Mean Filter]{Mean Filter \id{S60F}} \label{subsec:meanFilter}
One of the simplest existing kernels is the mean filter that computes the average intensity over its $M\times\dots\times M$ spatial support\footnote{We restrict the definition to odd values of $M$.} (\textid{YNOF}) as
\begin{equation}\label{eq:meanFilter}
g[\boldsymbol{k}]= 
\begin{cases}
\frac{1}{M^D} \quad\text{if}\quad k_1,\dots,k_D \in  \Big[ -\floor*{\frac{M}{2}},\floor*{\frac{M}{2}} \Big],\\
0\quad\text{otherwise}.
\end{cases}
\end{equation}

\begin{figure}[!t] 
\centering
   \begin{minipage}[b]{0.45\textwidth} 
     \centering
     \includegraphics[trim = 0 0 0 0, clip,scale=0.40]{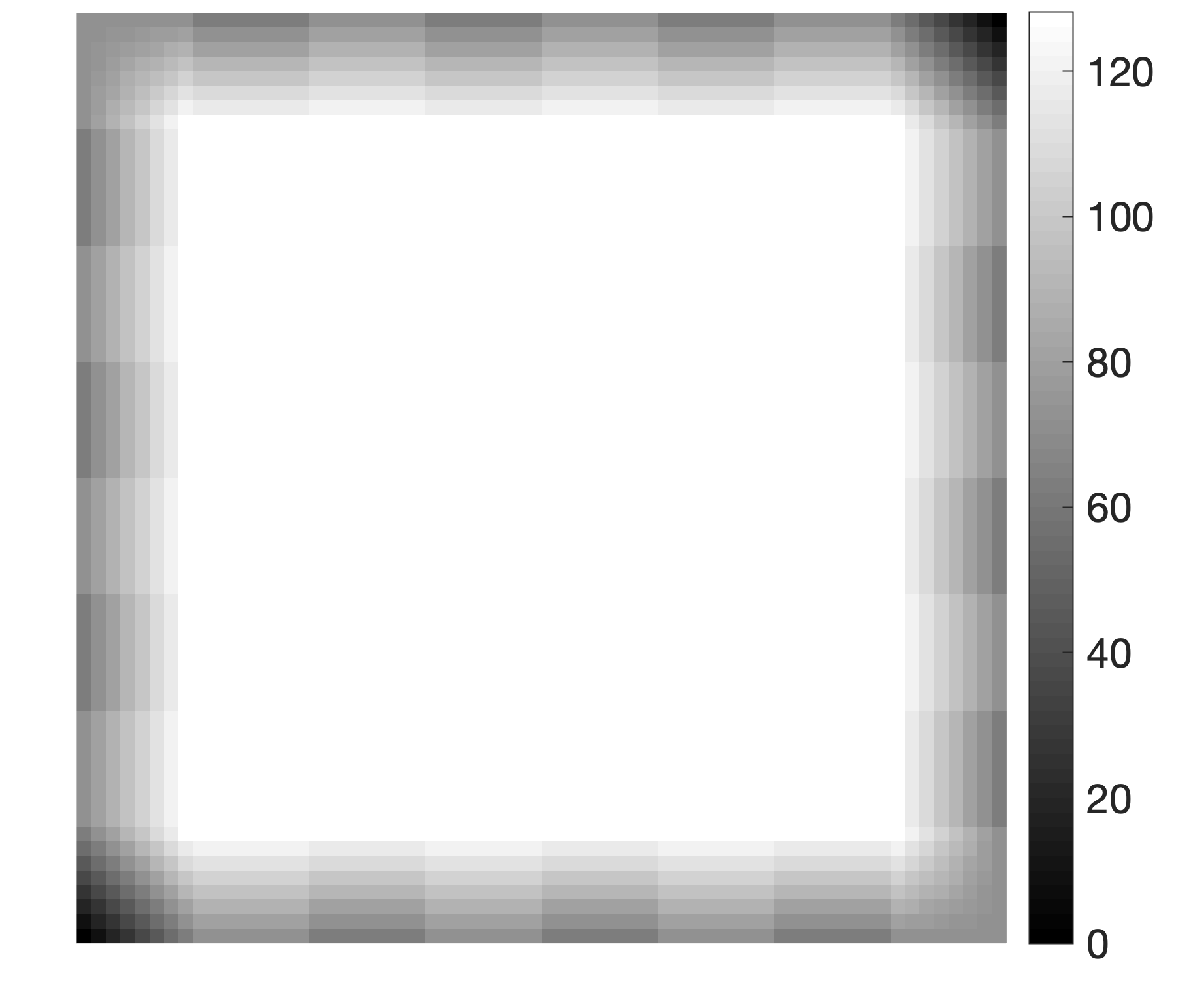}
     \subcaption{\footnotesize Outlier submission.}
     %\hspace{50pt}
   \end{minipage}
   \begin{minipage}[b]{0.45\textwidth}
     \centering
     \includegraphics[trim = 0 0 0 0, clip, scale=0.40]{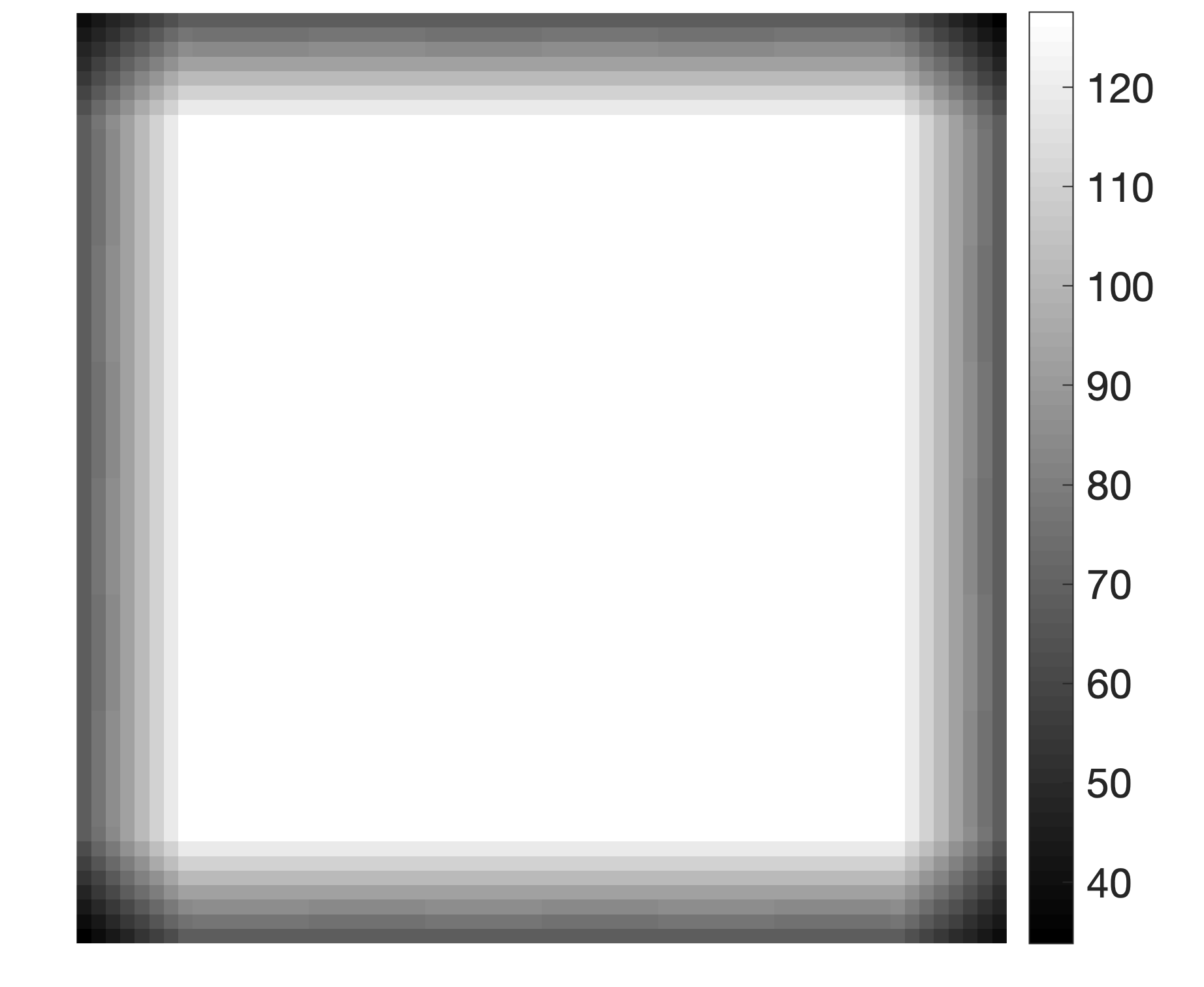}
     \subcaption{\footnotesize Valid CRM.}
     %\hspace{50pt}
   \end{minipage}
      \begin{minipage}[b]{0.45\textwidth}
     \centering
     \includegraphics[trim = 0 0 0 0, clip, scale=0.40]{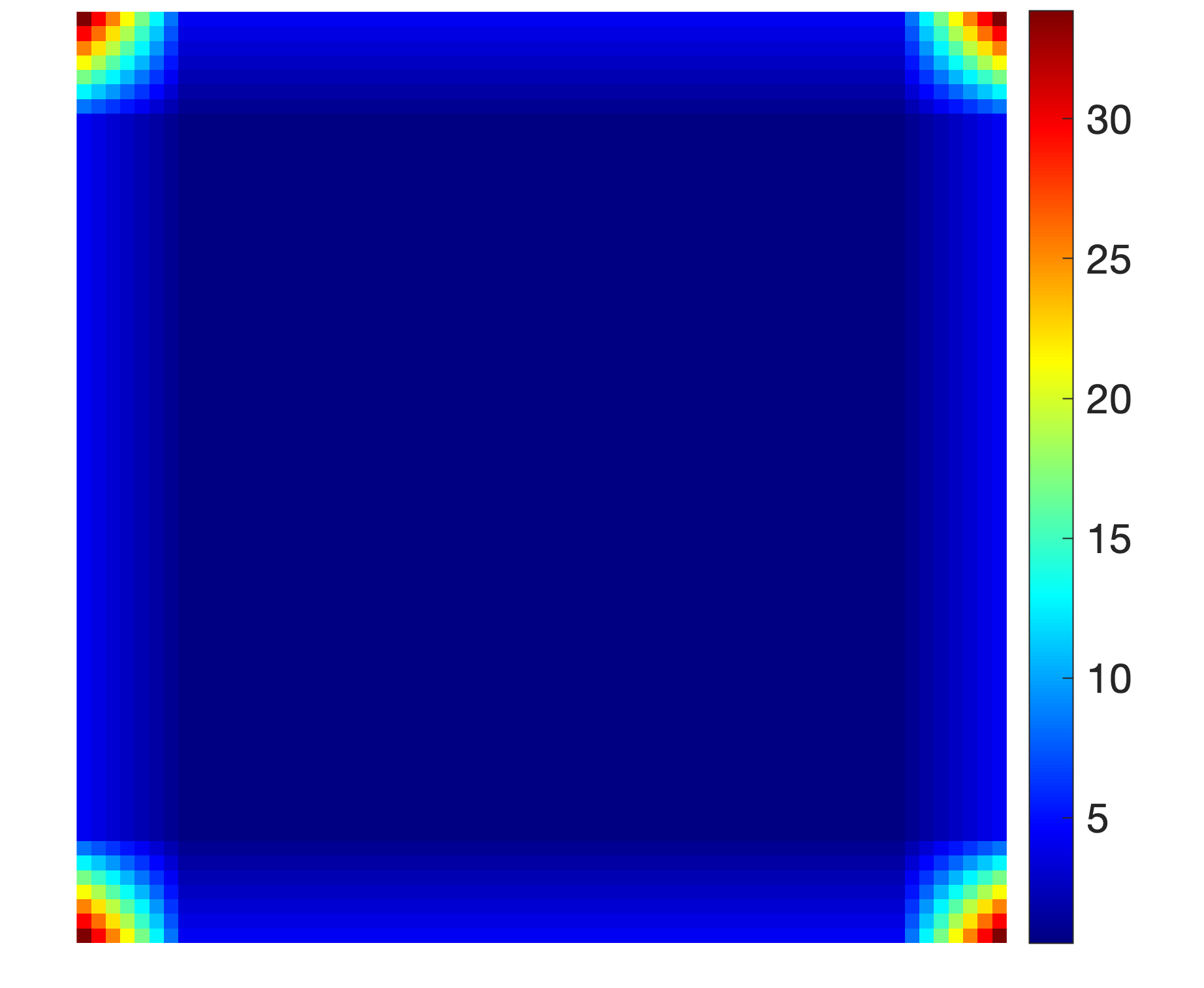}
     \subcaption{\footnotesize Absolute difference.}
     %\hspace{50pt}
   \end{minipage}
   \begin{minipage}[b]{0.45\textwidth}
     \centering
     \includegraphics[trim = 0 0 0 0, clip, scale=0.40]{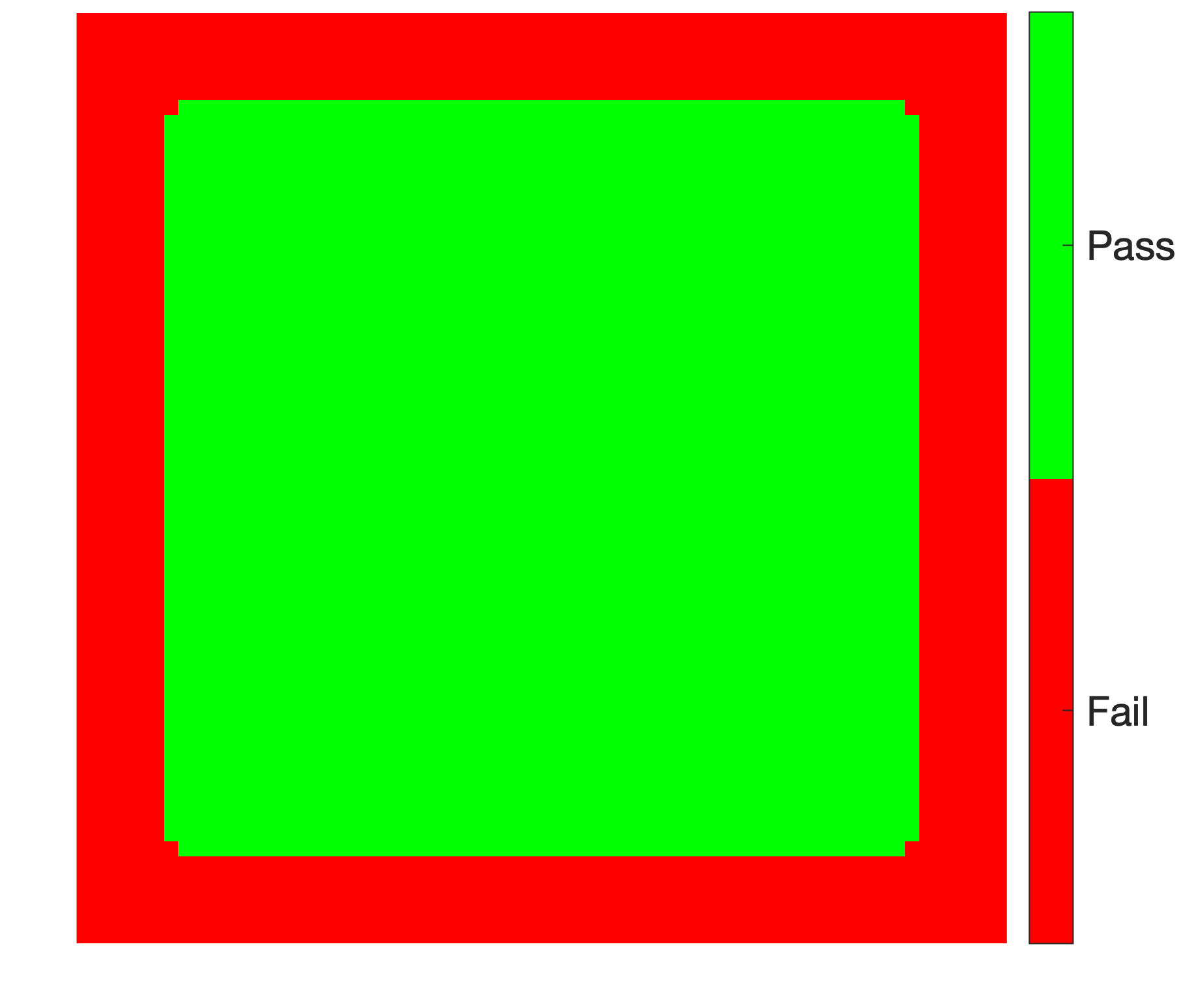}
     \subcaption{\footnotesize Voxel-wise \textit{passing map}.}
   \end{minipage} 
   \caption{Example of outlier discrepancy for mean filter test 1.a.1 (Table \ref{tab:benchmark_filter_settings}) caused by padding differences. The central 2D slice of the 3D volumes are visualised for: \textbf{(a)} an outlier submission, \textbf{(b)} the valid \textit{consensus response map} (CRM) found, \textbf{(c)} the absolute difference between the outlier and consensus,  \textbf{(d)} and voxel-wise passing map of this difference. }
   \label{fig:1a1_mean_example}
\end{figure}

\vspace{2mm}
\begin{tcolorbox}[width=150mm, halign=left, colframe=black, colback=white, boxsep=0mm, arc=3mm, colframe=black!50!white,
title=Implementation Troubleshooting, title filled=true, fonttitle=\bfseries]
\begin{itemize}
\itemsep0em 
\item Ensure that filter support $M$ is defined in voxel units. 
\item Check that the padding method is correctly applied (particularly mirror). An example of an outlier submission due to padding is shown in Fig.\ \ref{fig:1a1_mean_example}.
\end{itemize}
\end{tcolorbox}

\section[Laplacian of Gaussian]{Laplacian of Gaussian \id{L6PA}}\label{sec:LaplacianOfGaussian}
The Laplacian of Gaussian (LoG) is a band-pass and circularly/spherically symmetric convolutional operator.
It is therefore invariant to local rotations but also directionally insensitive.
Its profile $g[\boldsymbol{k}]$ only depends on the 1$D$ radius $||\boldsymbol{k}||$ and corresponds to a radial second-order derivative of a $D$-dimensional Gaussian filter as
\begin{equation}\label{eq:LoG}
g_{\sigma}[\boldsymbol{k}]=
-\frac{1}{\sigma^2}
\left(\frac{1}{\sqrt{2\pi}\sigma}\right)^D
\left(D-\frac{||\boldsymbol{k}||^2}{\sigma^2}\right)
e^{-\frac{||\boldsymbol{k}||^2}{2\sigma^2}},
\end{equation}
where the standard deviation of the Gaussian $\sigma$ controls the scale of the operator (\textid{41LN}). Note that \(\sigma\) is implied to be in voxel units, i.e. $\sigma = \sigma^* / s$, with $s$ the voxel spacing.

LoG filtering cannot be implemented with separable convolution and requires a full $D$-dimensional convolution.
However, it can be approximated using a Difference of two Gaussians (DoG) when the ratio between their respective standard deviations $\sigma_1$ and $\sigma_2$ is $\sigma_1 = \frac{\sigma_2}{\sqrt{2}}$.
Because Gaussian kernels are separable, DoG filtering can also be efficiently implemented using separable convolutions.
With an adequate sequence of $\sigma_i$, a collection of LoGs can cover the entire image spectrum. In this case, they form wavelets and are often called Mexican hat, Ricker, or Marr wavelet.

The spatial support of the LoG is \((-\infty, \infty\)) (\emph{i.e.} not compact). 
Because this would require a filter with infinite spatial support, the LoG filter is cropped in practice, usually based on the $\sigma$ parameter (\textid{WGPM}). The 1$D$ filter size is then
$$M=1 + 2\lfloor d\, \sigma + 0.5\rfloor,$$
with \(d\) the truncation parameter, \(\sigma\)  in voxel units (\emph{e.g.}, \(\sigma=2.5\) voxels when it parameterised as \(\sigma^* = 5\) mm with voxel spacing of \(2\) mm). As a consequence, the size of a (1$D$) LoG filter is at least \(M=1\) and will have an odd, integer value. \(d=4\) is commonly used for truncation.
%%% RECOMMENDATION ???

\begin{figure}
\centering
\includegraphics[trim = 0 0 0 0, clip, scale=0.8]{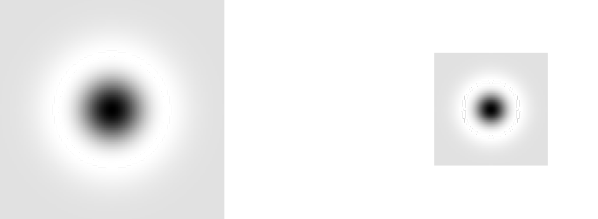}\\
\caption{Examples of 2$D$ LoG filters with $\sigma=16$, $M=129$ (left) and $\sigma=8$, $M=65$ (right).}
  \label{fig:LoG}
\end{figure}

\begin{figure}
\centering
\includegraphics[trim = 0 0 0 0, clip, width=\linewidth]{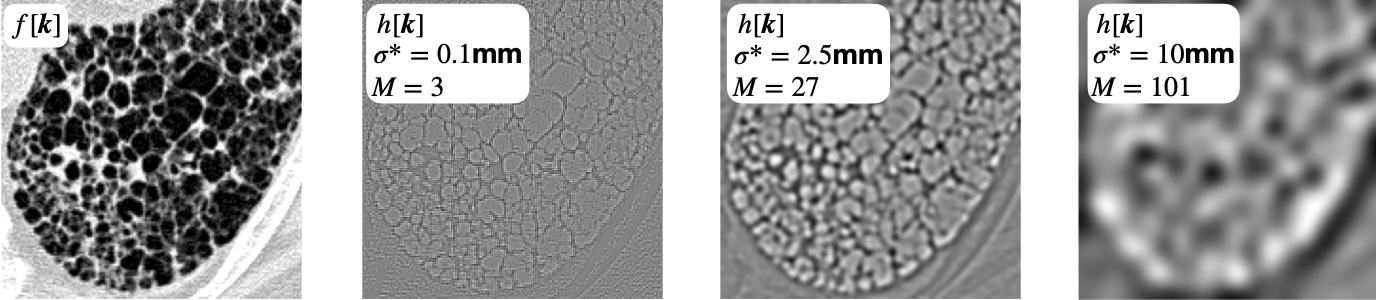}\\
% \hspace{0.9cm}
% $f[\boldsymbol{k}]$
% \hspace{2.1cm}
% \tiny $h[\boldsymbol{k}]$, $\sigma=0.1$mm, $M=27$
% \hspace{1.1cm}
% $h[\boldsymbol{k}]$, $\sigma=2.5$mm, $M=27$
% \hspace{1.1cm}
% $h[\boldsymbol{k}]$, $\sigma=10$mm, $M=101$
\caption{Examples of image filtering with a LoG filter (the pixel spacing is 0.8mm, mirror boundary conditions used for the convolution) at various scales.
The small scale ($\sigma^*=0.1$mm) highlights tiny collagen fibers, whereas the larger scale ($\sigma^*=10$mm) highlights larger image blobs or clusters present in the diseased lung tissue (honeycombing).}
  \label{fig:LoG_examples}
\end{figure}

The profile of a 2$D$ LoG is illustrated in Fig.~\ref{fig:LoG}.
The LoG filter can be used to enhance image blobs and ridges at a specific scale, controlled by $\sigma^*$.
This is illustrated in Fig.~\ref{fig:LoG_examples}.

\vspace{2mm}
\begin{tcolorbox}[width=150mm, halign=left, colframe=black, colback=white, boxsep=0mm, arc=3mm, colframe=black!50!white,
title=Implementation Troubleshooting, title filled=true, fonttitle=\bfseries]
\begin{itemize}
\item The scale parameter $\sigma^*$ for the filter tests (Table \ref{tab:benchmark_filter_settings}) is defined in physical distance. Voxel dimensions for the digital phantoms are stored in millimeters in NifTI or DICOM headers. Many standard implementations assume that $\sigma$ is defined in voxel units, so a conversion between physical and voxel units is required (e.g. \textit{fspecial3} in Matlab).
\item To test implementation of Eq.\ \ref{eq:LoG}, check that the LoG kernel sums to approximately 0.
\end{itemize}
\end{tcolorbox}

\section[Laws Kernels]{Laws Kernels \id{JTXT}}
Laws kernels are a collection of five types of small 1$D$ filters $g[k]$~\cite{Law1980}.
They are combined using outer products to obtain 2$D$ and 3$D$ filters (\textid{JVAD}).

The first type is a low-pass kernel called \emph{Level} for grey level averaging, which is available in two scales with a spatial support of 3 (\textid{B5BZ}) or 5 (\textid{6HRH}) pixels:
$$g_{L3}[k] = \frac{1}{\sqrt{6}}\cdot\left[ 1,2,1 \right],\quad g_{L5}[k] = \frac{1}{\sqrt{70}}\cdot\left[ 1,4,6,4,1 \right].$$
The next kernels are all zero mean, which makes them insensitive to the average grey level. They will solely focus on spatial transitions between the values (\textit{i.e.} texture). The four types of transitions are:
\begin{itemize}
    \item edges:
    \begin{itemize}
        \item \(g_{E3}[k] = \frac{1}{\sqrt{2}}\cdot\left[-1, 0, 1 \right]\) \hfill \textid{LJ4T}
        \item \(g_{E5}[k] = \frac{1}{\sqrt{10}}\cdot\left[-1, -2, 0, 2, 1 \right]\) \hfill \textid{2WPV}
    \end{itemize}
    \item spots:
    \begin{itemize}
        \item \(g_{S3}[k] = \frac{1}{\sqrt{6}}\cdot\left[-1, 2, -1 \right]\) \hfill \textid{MK5Z}
        \item \(g_{S5}[k] = \frac{1}{\sqrt{6}}\cdot\left[-1, 0, 2, 0, -1 \right]\) \hfill \textid{RXA1}
    \end{itemize}
    \item wave:
    \begin{itemize}
        \item \(g_{W5}[k] = \frac{1}{\sqrt{10}}\cdot\left[-1, 2, 0, -2, 1 \right]\) \hfill \textid{4ENO}
    \end{itemize}
    \item ripple:
    \begin{itemize}
        \item \(g_{R5}[k] = \frac{1}{\sqrt{70}}\cdot\left[ 1, -4, 6, -4, 1 \right]\) \hfill \textid{3A1W}
    \end{itemize}
\end{itemize}

\begin{figure}
\centering
\includegraphics[trim = 0 0 0 0, clip, scale=0.6]{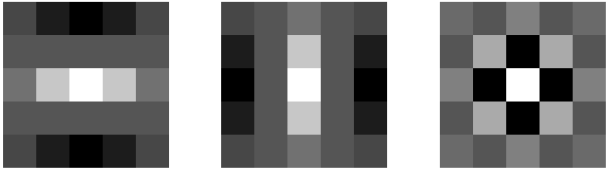}\\
$g_{L5S5}[\boldsymbol{k}]$
\hspace{1.9cm}
$g_{S5L5}[\boldsymbol{k}]$
\hspace{1.9cm}
$g_{R5R5}[\boldsymbol{k}]$
\caption{Example of 2$D$ Laws kernels.}
  \label{fig:Laws}
\end{figure}

Laws 2$D$ and 3$D$ kernels are separable by design, and response maps are created by 1$D$ kernel convolution along each image direction. Such response maps are referred to by their 1-$D$ kernel names. For instance, a 2$D$ response map called $h_{L5S5}$ is obtained after first convolving the image with the $g_{L5}$ kernel along the image lines (\textit{i.e.} $k_1$) and a subsequent convolution of the image with the $g_{S5}$ kernel along the image columns (\textit{i.e.} $k_2$).
Examples of 2$D$ Laws kernels are shown in Fig.~\ref{fig:Laws}.
Note that the above definitions for Laws kernels were normalised, and in this sense deviate from those originally defined by Laws himself~\cite{Law1980}.
\subsection[Laws Texture Energy Images]{Laws Texture Energy Images \id{PQSD}}
Laws used his kernels to generate texture energy images~\cite{Law1980}. This is done in two steps. In the first step, a response map $h$ is generated by convolving the image with a Laws kernel along each image direction, as described above. Then, a smoothed image is computed where the absolute intensities\footnote{It is worth noting that whereas "energy" often involves the computation of squared quantities, the absolute value was proposed by Laws. The goal is to regroup negative and positive filter responses.} of voxels in $h$ within Chebyshev distance $\delta$ (\textid{I176}) of a centre voxel are summed to create an energy image $h_{\text{energy}}$:
$$h_{\text{energy}}[\boldsymbol{k}]=\frac{1}{W}\sum_{k_{0,1}=-\delta}^{\delta}
\dots
\sum_{k_{0,d}=-\delta}^{\delta}\left|h[\boldsymbol{k}+\boldsymbol{k}_0]\right|,$$
where $\boldsymbol{k}_0=(k_{0,1},\dots,k_{0,d})$ and $W=\left(2\delta+1\right)^N$ the number of voxels in the $N$-dimensional neighbourhood.
In practice, \(h_{\text{energy}}\) can be computed using kernel convolutions by convolving $|h|$ with a $2\delta+1$ element long 1$D$ kernel with constant values $1/\left(2\delta+1\right)$ along each of the image directions, \textit{i.e.} a mean filter as described in Section~\ref{subsec:meanFilter}.

Note that the definition given above deviates from the one given by Laws~\cite{Law1980} by introducing the normalisation factor \(W\).
Laws moreover suggested using a sliding window of $15\times 15$ pixels, which corresponds to $\delta=7$.
We recommend that $\delta$ is chosen within the context of a given application.

To summarize, Laws filtering requires the following sequence of operations: (i) pad the input image, (ii) filter with a given Laws kernel, (iii) pad the response map and (iv) compute the energy image via the mean filter.
We suggest using the same padding (see Section~\ref{sec:constantValuePadding}) used to compute the initial response of the Laws kernel to compute the energy image.
Laws filtering is not rotation-invariant.
However, since the kernels are separable, rotational invariance can be efficiently approximated using permutations and unidimensional filter flipping, followed by orientation pooling with \textit{e.g.} the max (see Section~\ref{sec:combiningDSandRotInv} and Appendix~\ref{app:separableConvRightAngleEquivariant}).
In this case, we recommend computing the texture energy image after orientation pooling.
%%% RECOMMENDATION to compute the texture energy image after orientation pooling

An example of image filtering with the $g_{L5S5}$ kernel is shown in Fig.~\ref{fig:LawsExample}.

\begin{figure}[!t] 
\centering
   \begin{minipage}[b]{0.45\textwidth} 
     \centering
     \includegraphics[trim = 0 0 0 0, clip,scale=0.40]{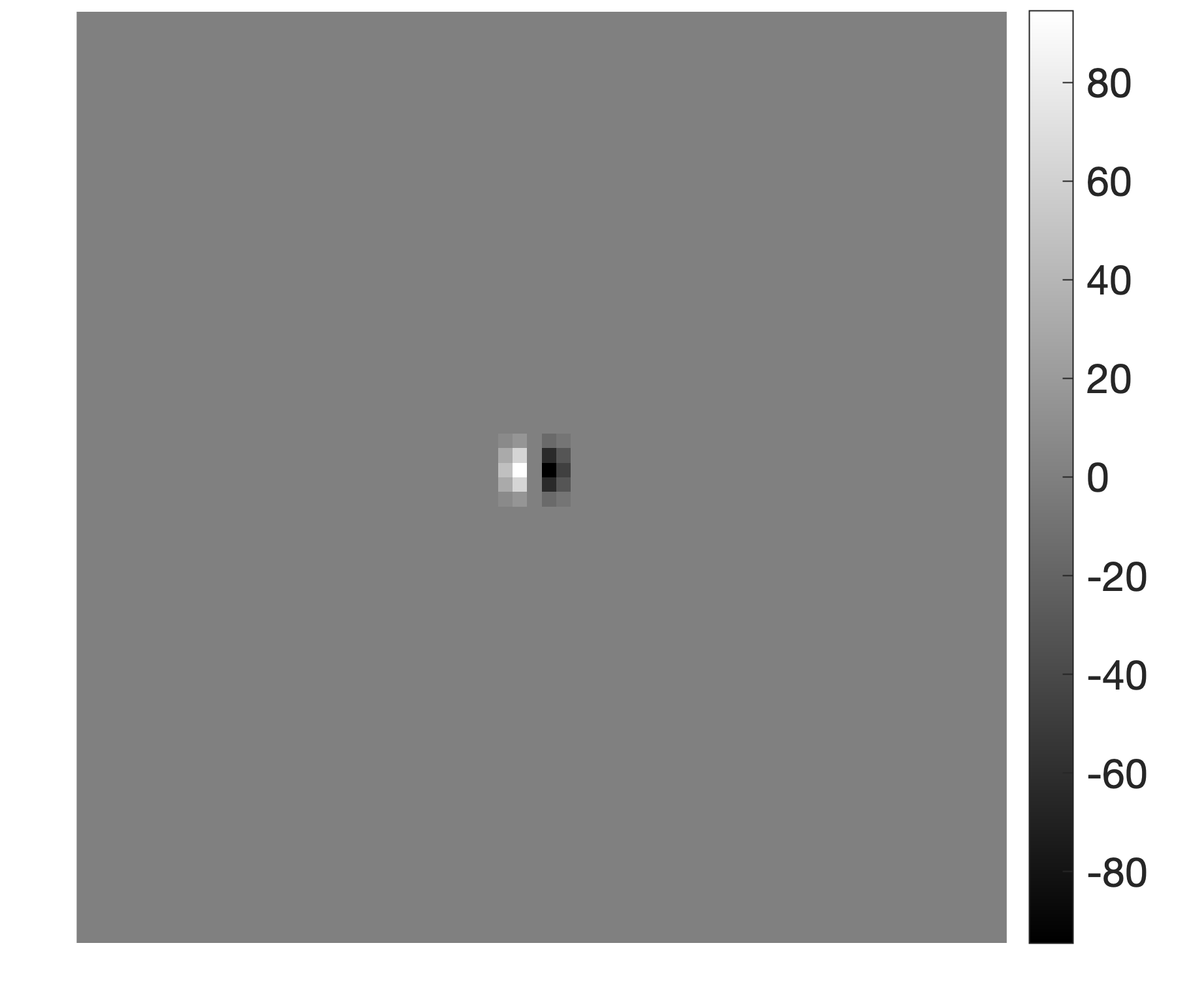}
     \subcaption{\footnotesize Outlier submission.}
     %\hspace{50pt}
   \end{minipage}
   \begin{minipage}[b]{0.45\textwidth}
     \centering
     \includegraphics[trim = 0 0 0 0, clip, scale=0.40]{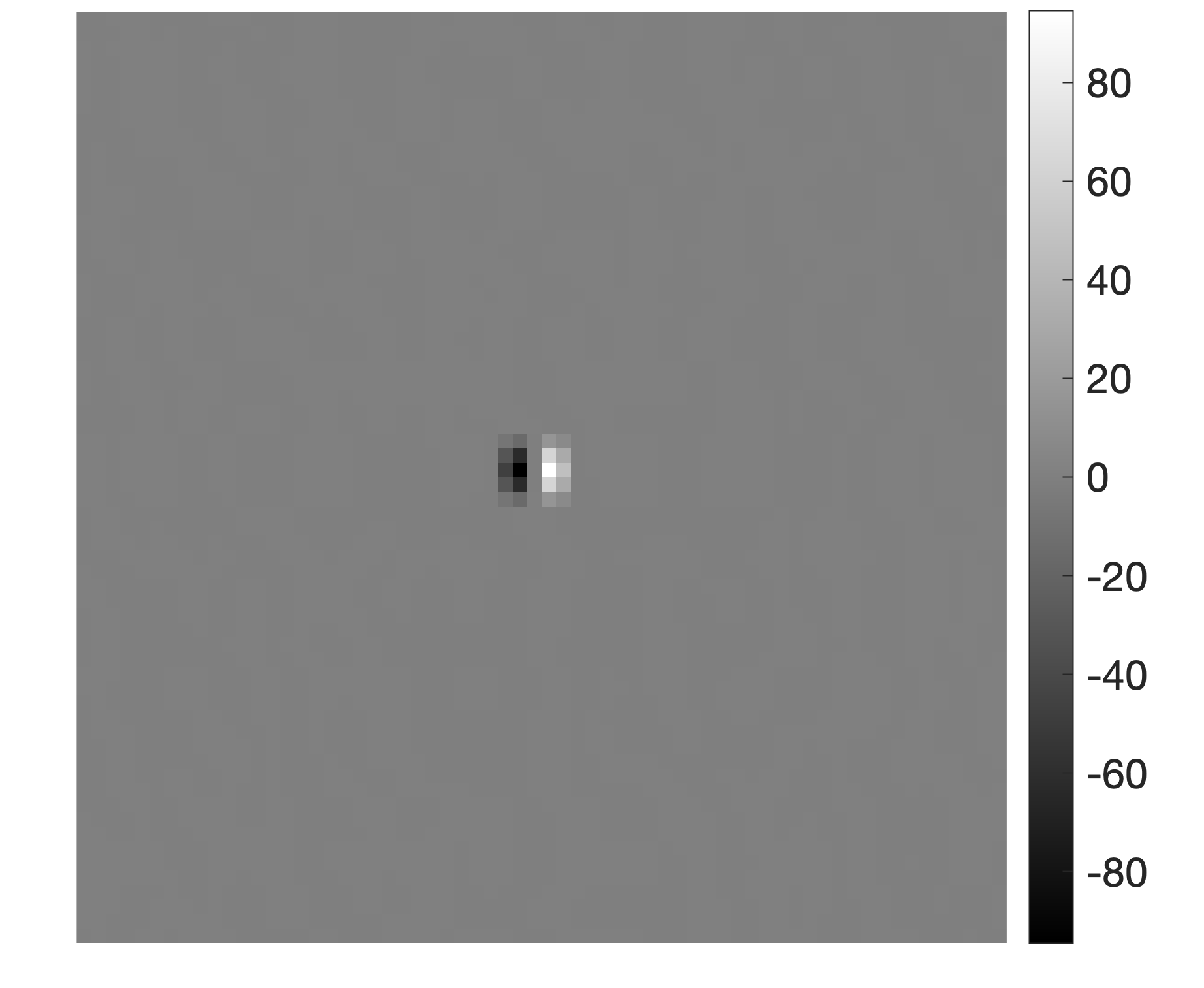}
     \subcaption{\footnotesize Valid CRM.}
     %\hspace{50pt}
   \end{minipage}
      \begin{minipage}[b]{0.45\textwidth}
     \centering
     \includegraphics[trim = 0 0 0 0, clip, scale=0.40]{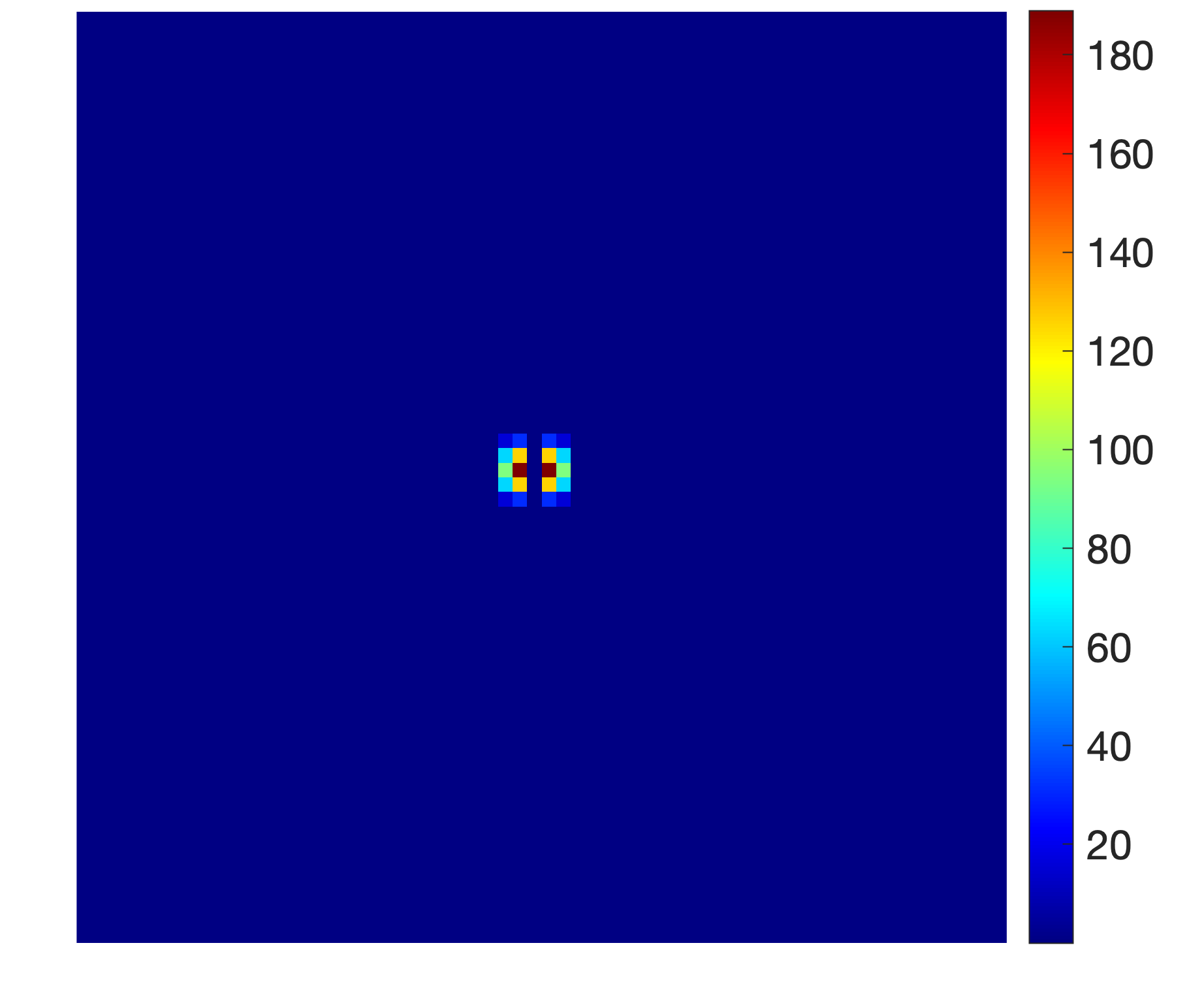}
     \subcaption{\footnotesize Absolute difference.}
     %\hspace{50pt}
   \end{minipage}
   \begin{minipage}[b]{0.45\textwidth}
     \centering
     \includegraphics[trim = 0 0 0 0, clip, scale=0.40]{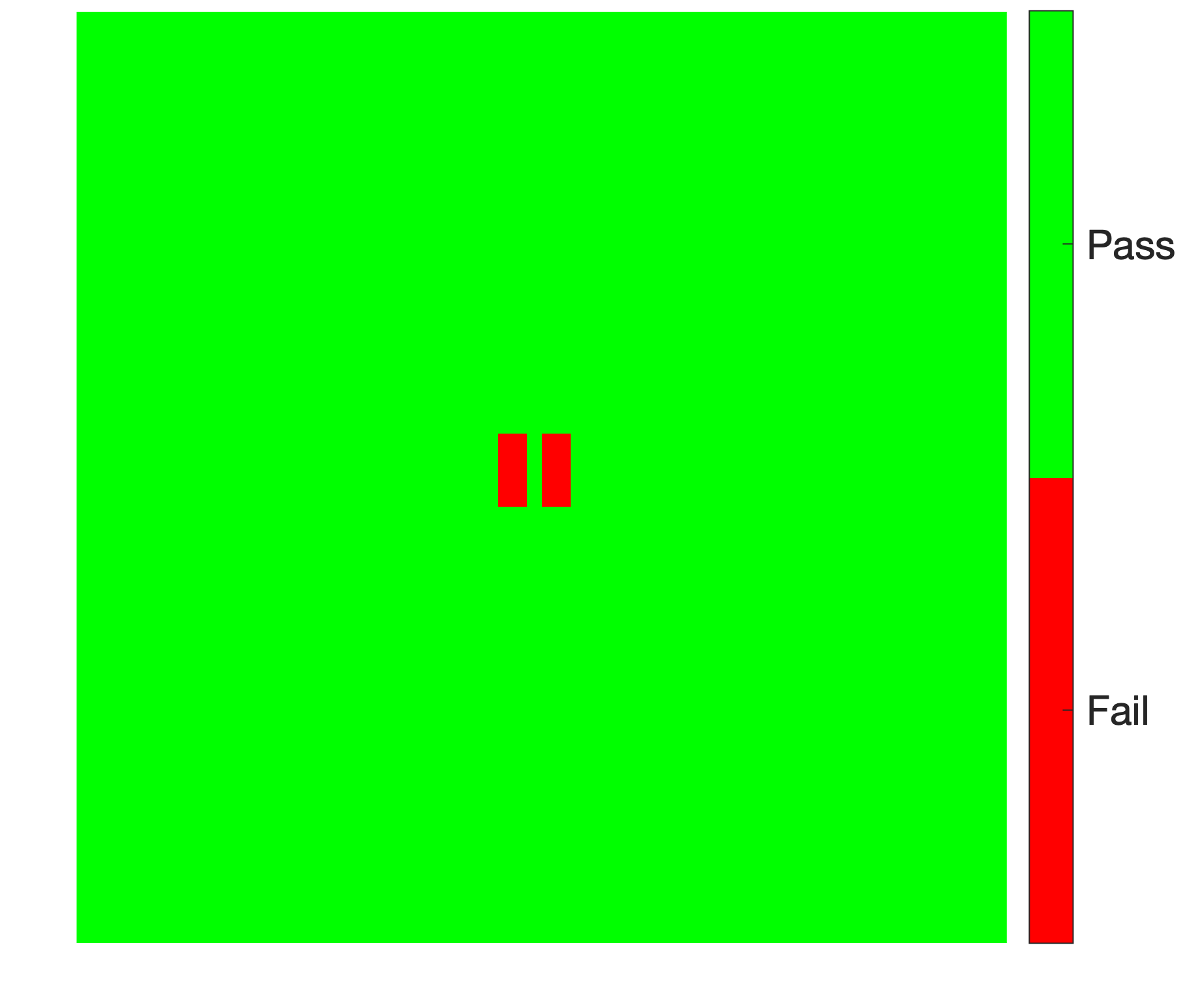}
     \subcaption{\footnotesize Voxel-wise \textit{passing map}.}
   \end{minipage} 
   \caption{Example of outlier discrepancy for Laws filter test 3.a.1. (Table \ref{tab:benchmark_filter_settings}) The central 2D slice of the 3D volumes are visualised for: \textbf{(a)} an outlier submission, \textbf{(b)} the valid \textit{consensus response map} (CRM) found, \textbf{(c)} the absolute difference between the outlier and consensus,  \textbf{(d)} and voxel-wise passing map of this difference. The two response maps (\textbf{(a)} and \textbf{(b)}) are identical apart from orientation. }
   \label{fig:3a1_Laws_example}
\end{figure}

\vspace{2mm}
\begin{tcolorbox}[width=150mm, halign=left, colframe=black, colback=white, boxsep=0mm, arc=3mm, colframe=black!50!white,
title=Implementation Troubleshooting, title filled=true, fonttitle=\bfseries]
\begin{itemize}
\item Filter orientation and convolution direction is a key cause of discrepancy, as some Laws kernels are not symmetric. %(for filter tests that are not rotationally invariant). 
An example of an outlier response map for filter test 3.a.1\ (Table \ref{tab:benchmark_filter_settings}) that appears to have a different orientation is shown in Fig.\ \ref{fig:3a1_Laws_example}.
Check that filter kernels are applied in the direction consistent with Section \ref{sec:imageDirections}. Use the orientation phantom to check if your software orients the image as aspected. 
% \item Ensure that the header information correctly orientates the saved response map.
\item Only one Energy Image should be calculated. This is performed after orientation pooling and on absolute intensity values.
\end{itemize}
\end{tcolorbox}

\begin{figure}
\centering
\includegraphics[trim = 0 0 0 0, clip, width=\linewidth]{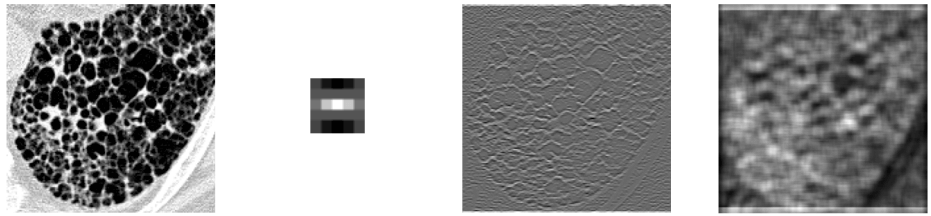}\\
$f[\boldsymbol{k}]$
\hspace{2.5cm}
$g_{L5S5}[\boldsymbol{k}]$
\hspace{2.1cm}
$h_{L5S5}[\boldsymbol{k}]$
\hspace{2.3cm}
$h_{\text{energy}}[\boldsymbol{k}]$
\caption{Example of image filtering with the $g_{L5S5}$ kernel. The zero padding boundary condition was used to calculate both $h_{L5S5}$ and $h_{\text{energy}}$.}
  \label{fig:LawsExample}
\end{figure}

\section[Gabor]{Gabor \id{Q88H}}\label{sec:Gabor}
\begin{figure}
\centering
\includegraphics[trim = 0 0 0 0, clip, scale=0.8]{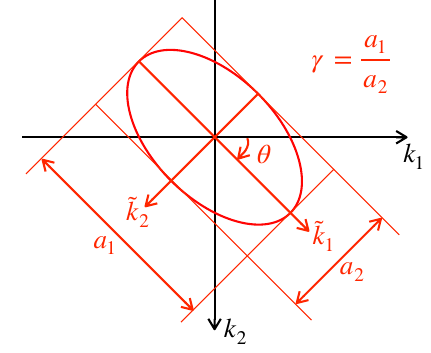}\\
\vspace{.3cm}
\includegraphics[trim = 0 0 0 0, clip, scale=0.77]{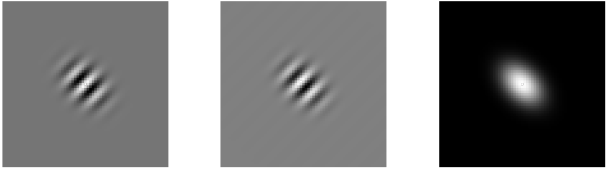}\\
$\text{Re}\left(g_{5,\frac{2}{\pi},\frac{3}{2},\frac{\pi}{4}}[\boldsymbol{k}]\right)$
\hspace{1.6cm}
$\text{Im}\left(g_{5,\frac{2}{\pi},\frac{3}{2},\frac{\pi}{4}}[\boldsymbol{k}]\right)$
\hspace{1.8cm}
$\lvert g_{5,\frac{2}{\pi},\frac{3}{2},\frac{\pi}{4}}[\boldsymbol{k}]\rvert$
\includegraphics[trim = 0 0 0 0, clip, scale=0.77]{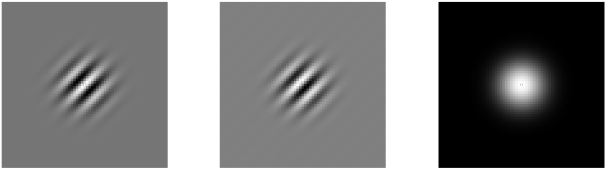}\\
$\text{Re}\left(g_{5,\frac{2}{\pi},1,\frac{\pi}{4}}[\boldsymbol{k}]\right)$
\hspace{1.6cm}
$\text{Im}\left(g_{5,\frac{2}{\pi},1,\frac{\pi}{4}}[\boldsymbol{k}]\right)$
\hspace{1.8cm}
$\lvert g_{5,\frac{2}{\pi},1,\frac{\pi}{4}}[\boldsymbol{k}]\rvert$
\includegraphics[trim = 0 0 0 0, clip, scale=0.77]{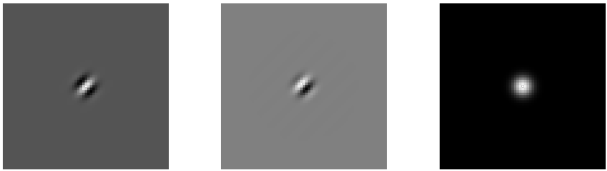}\\
$\text{Re}\left(g_{2,\frac{2}{\pi},1,\frac{\pi}{4}}[\boldsymbol{k}]\right)$
\hspace{1.6cm}
$\text{Im}\left(g_{2,\frac{2}{\pi},1,\frac{\pi}{4}}[\boldsymbol{k}]\right)$
\hspace{1.8cm}
$\lvert g_{2,\frac{2}{\pi},1,\frac{\pi}{4}}[\boldsymbol{k}]\rvert$
\includegraphics[trim = 0 0 0 0, clip, scale=0.77]{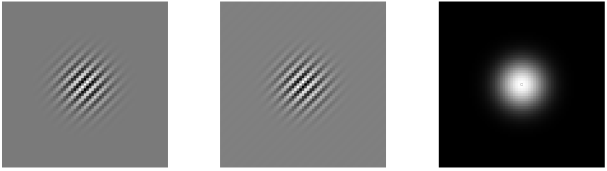}\\
$\text{Re}\left(g_{5,\frac{1}{\pi},1,\frac{\pi}{4}}[\boldsymbol{k}]\right)$
\hspace{1.6cm}
$\text{Im}\left(g_{5,\frac{1}{\pi},1,\frac{\pi}{4}}[\boldsymbol{k}]\right)$
\hspace{1.8cm}
$\lvert g_{5,\frac{1}{\pi},1,\frac{\pi}{4}}[\boldsymbol{k}]\rvert$
\caption{Coordinate system and examples of 2$D$ Gabor filters in the spatial domain 
%with $\sigma=5$mm, $\lambda=\frac{2}{\pi}$mm, $\gamma=\frac{3}{2}$, $\theta=\frac{\pi}{4}$, 
computed on $65\times 65$ grids with a pixel spacing of 0.8mm.}
  \label{fig:Gabor}
\end{figure}

Gabor filter banks allow for extracting multi-directional and multi-scale texture information via a systematic parcellation of the Fourier domain with elliptic Gaussian windows~\cite{Bianconi2007} (see Fig.~3.2 of Depeursinge \textit{et al.}~\cite{DeF2017}).
In the spatial domain, 2$D$ Gabor kernels are complex Gaussian-windowed oscillatory functions defined as
\begin{equation}\label{eq:Gabor2D}
g_{\sigma,\lambda,\gamma,\theta}[\boldsymbol{k}]=
\mathrm{e}^{
-\frac{
\tilde{k}_1^2 +
\gamma^2\tilde{k}_2^2
}
{2\sigma^2}
+\mathrm{j}\frac{2\pi\tilde{k}_1}{\lambda}}
,
\end{equation}
where $\sigma$ controls the scale of the filter (standard deviation of the Gaussian envelope; \textid{41LN}) and $\lambda$ is the wavelength (\textit{i.e.} inverse of the frequency $F=\frac{1}{\lambda}$ of the oscillations; \textid{S4N6}). Both $\sigma$ and $\lambda$ are implied to be in voxel units, i.e. $\sigma = \sigma^* / s$ and $\lambda = \lambda^* / s$, with $s$ the voxel spacing, and $\sigma^*$ and $\lambda^*$ defined in mm.
\marginnote{\footnotesize v4: Clarified units of $\sigma$ and $\lambda$ parameters.}
$\gamma$ is the spatial aspect ratio (\textit{i.e.} the ellipticity of the support of the filter as $\gamma=a_1/a_2$; \textid{GDR5}), \mbox{$(\tilde{k}_1,\tilde{k}_2) =  \mathrm{R}_{\theta} \boldsymbol{k}$} defines the radial and orthoradial elliptic Gaussian axes at the orientation $\theta$ (\textid{FQER}) via the 2$D$ rotation matrix \mbox{$\mathrm{R}_{\theta} = 
\left(\begin{matrix}
\cos\theta &  \phantom{+}\sin\theta \\
\sin\theta & - \cos\theta
\end{matrix}
\right)$}.
As a convention, we define $\theta$ to turn clockwise in the axial plane $(k_1,k_2)$.
The considered coordinate system follows the conventions introduced in Section~\ref{sec:imageDirections}. 
It is depicted in Fig.~\ref{fig:Gabor} along with an example of a 2$D$ Gabor filter seen in the spatial domain.

\begin{figure}
\centering
\includegraphics[trim = 0 0 0 0, clip, width=\linewidth]{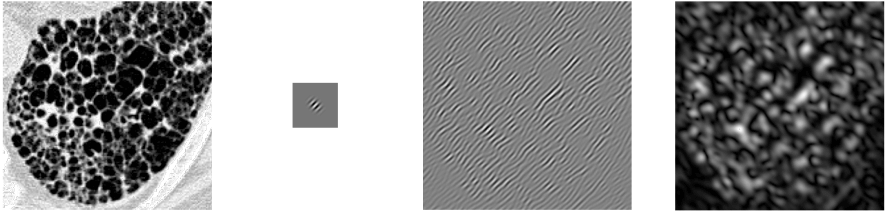}\\
\hspace{.5cm}
$f[\boldsymbol{k}]$
\hspace{1.6cm}
$\text{Re}\left(g_{5,\frac{2}{\pi},\frac{3}{2},\frac{\pi}{4}}[\boldsymbol{k}]\right)$
\hspace{.8cm}
$\text{Re}\left(h_{5,\frac{2}{\pi},\frac{3}{2},\frac{\pi}{4}}[\boldsymbol{k}]\right)$
\hspace{1.6cm}
$\lvert h_{5,\frac{2}{\pi},\frac{3}{2},\frac{\pi}{4}}[\boldsymbol{k}]\rvert$\\
\vspace{.3cm}
\includegraphics[trim = 0 0 0 0, clip, width=\linewidth]{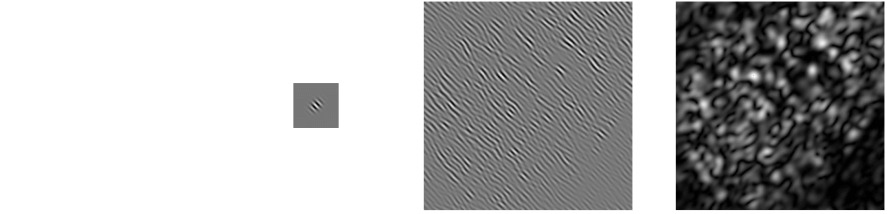}\\
\hspace{.5cm}
\phantom{$f[\boldsymbol{k}]$}
\hspace{1.6cm}
$\text{Re}\left(g_{5,\frac{2}{\pi},\frac{3}{2},-\frac{\pi}{4}}[\boldsymbol{k}]\right)$
\hspace{.7cm}
$\text{Re}\left(h_{5,\frac{2}{\pi},\frac{3}{2},-\frac{\pi}{4}}[\boldsymbol{k}]\right)$
\hspace{1.5cm}
$\lvert h_{5,\frac{2}{\pi},\frac{3}{2},-\frac{\pi}{4}}[\boldsymbol{k}]\rvert$
\caption{Example of image filtering with Gabor filters (the pixel spacing is 0.8mm, mirror boundary conditions used for the convolution).
Collagen junctions oriented at $\theta=-\frac{\pi}{4}$ (top) and $\theta=\frac{\pi}{4}$ (bottom) are highlighted.
}
  \label{fig:Gabor_example}
\end{figure}

In practice, the Gabor function is used as a filter kernel and the spatial frequency bandwidth of the filter needs to be defined. 
Petkov \textit{et al.}~\cite{Petkov1997} established the relation between the half-response spatial frequency bandwidth $F_b$ (in units of octaves) and the ratio $\sigma / \lambda$ as
\begin{equation}\label{eq:GaborSFB}
F_b = \mathrm{log}_2 \left( \frac{\frac{\sigma}{\lambda}\pi+\sqrt{\frac{\ln 2}{2}}}{\frac{\sigma}{\lambda}\pi-\sqrt{\frac{\ln 2}{2}}} \right), \text{ and inversely}\quad \frac{\sigma}{\lambda} = \frac{1}{\pi} \sqrt{\frac{\ln 2}{2}} \cdot \frac{2^{F_b}+1}{2^{F_b}-1}.
\end{equation}
Gabor filters can then be constructed at multiple scales and orientations to explore the spectrum of patterns in an image. 
Bianconi \textit{et al.}~\cite{Bianconi2007} proposed to extract response maps at multiple orientations $\{\theta_1,\dots,\theta_P\}$ and frequencies $\{F_1,\dots,F_Q\}$ along with $\gamma$ are defined to cover all directions and scales up to the maximum frequency $F_Q=\frac{1}{\lambda_Q}$.
%%% RECOMMENDATION ???
It is worth noting that when using a suitable sequence of scales, Gabor filters can also satisfy the wavelet admissibility condition (\textit{i.e.} referred to as ``Gabor wavelets") and can, in this particular case, fully cover the Fourier domain.
Finally, since $g_{\sigma,\lambda,\gamma,\theta}[\boldsymbol{k}]$ is complex, the modulus of the associated response map can be used before aggregation and feature calculation as $\lvert h[\boldsymbol{k}] \rvert = \lvert (g_{\sigma,\lambda,\gamma,\theta} \ast f) [\boldsymbol{k}]\rvert$.
An example of filtering with Gabor is depicted in Fig.~\ref{fig:Gabor_example}.

\marginnote{\footnotesize v4: Clarified spatial extent of Gabor filter support; v6: Updated description to produce the most consistent results.}
As with Laplacian-of-Gaussian filters the spatial support of Gabor filters is \((-\infty, \infty\)) (\emph{i.e.} not compact). However, unlike Laplacian-of-Gaussian filters, many standard implementations for Gabor filters do not use truncation. Instead, the most consistent results seem to be achieved by cropping the filter along each dimension $i$ as follows:
\begin{equation}\label{eq:gaborFilterSupport}
M_i= 
\begin{cases}
N_i + 1 & \text{if } N_i \text{ is even}\\
N_i & \text{if } N_i \text{ is odd}
\end{cases}
\end{equation}

Gabor filters are not rotation-invariant and are therefore best suited for applications where the absolute feature orientation is meaningful.
Rotation equivariance and invariance can be approximated by combining the response maps of several elements of the Gabor filterbank (see Section~\ref{sec:combiningDSandRotInv}). These response maps can be created by stepping the orientation by \(\Delta \theta\) (\textid{XTGK}).

The 3$D$ extension of Gabor filters can be found in Qian \textit{et al.}~\cite{Qian2006}. However, no recommendations are provided in the 3$D$ case to sample scales and orientations for further feature extraction.
One simple option to achieve (\textit{i.e.} approximate) 3$D$ image analysis is to extract 2$D$ Gabor features as recommended above in the three orthogonal planes  of the image frame of reference, followed by an averaging of the response maps over the three planes.
\section{Wavelets}\label{sec:wavelets}
Wavelets form a large category of filtering methods based on a collection of high-pass and low-pass filters that are designed to cover the entire image spectrum~\cite{Mallat1989} (see Section~\ref{sec:spectalCoverage}).
Pairwise combinations of one high- and one low-pass filter result usually in a sequence of band-pass response maps (\textit{i.e.} wavelet coefficients) with a factor of 2 between their scales and one remaining low-pass response map.
Two properties must be considered when implementing wavelet transforms, namely \emph{decimation} and \emph{separability}.
Decimation relates to the downsampling operation of the response maps and is compared in Sections~\ref{sec:decimatedWT} and~\ref{sec:undecimatedWT}.
Separable and non-separable wavelets concern the separability of high- and low-pass filters and are detailed in Sections~\ref{sec:separableWavelets} and~\ref{sec:nonseparableWavelets}.
\subsection[Decimated Transform]{Decimated Transform \id{PH3R}}\label{sec:decimatedWT}
The decimated transform is not redundant and allows coding images with a minimal number of coefficients.
However, the response maps containing the coefficients are iteratively decimated, which means that their size decreases throughout the levels of the decomposition, leading to a lack of translation invariance.

\begin{figure}
\centering
\includegraphics[trim = 0 0 0 0, clip, width=0.4\linewidth]{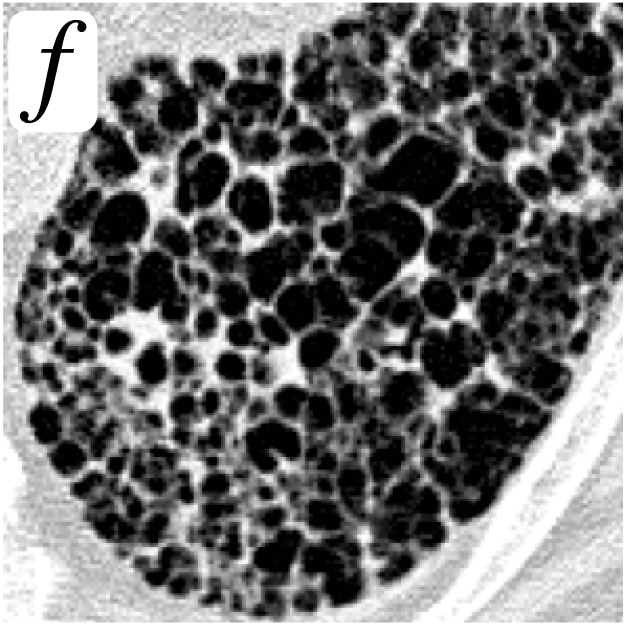}
\includegraphics[trim = 0 0 0 0, clip, width=0.4\linewidth]{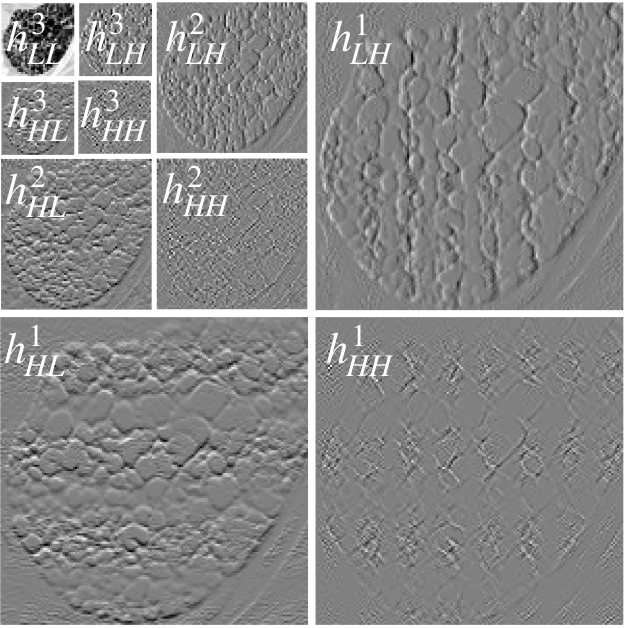}\\
\caption{Response maps from the 2$D$ decimated separable wavelet transform of the input image $f$.
The first three iterations are shown when using the Haar wavelet.}
  \label{fig:decimatedWT}
\end{figure}

For instance, with separable wavelets (see Section~\ref{sec:separableWavelets}), the image $f[\boldsymbol{k}]$ is first convolved with a high-pass filter $g_H$ and a matching low-pass filter $g_L$ along each image direction.
In 2$D$, this yields four response maps: $h_{LL}$, $h_{LH}$, $h_{HL}$ and $h_{HH}$.
All response maps are then downsampled by a factor of two in all directions to become $h_{LL}^1$, $h_{LH}^1$, $h_{HL}^1$ and $h_{HH}^1$.
This concludes the first iteration of the discrete wavelet transform.
For the next iterations, the low-pass coefficients $h_{LL}^j$ are subsequently convolved with $g_H$ and $g_L$ along each image direction and downsampled. 
This means that the $h_{LL}^j$ response map of each decomposition level is used as input image for the next level $j+1$ (\textid{GCEK}).
It is worth noting that $h_{LL}^j$ is traditionally discarded when the wavelet transform is used for image compression and reconstruction.
Because the response maps $h_{LL}^j$ are downsampled $j$ times, this has the same effect as dilating (\textit{i.e.} upsampling) the filters by $2^j$ in Eq.~\eqref{eq:wavelets}.
The downsampling of the response maps yields the wavelet coefficients of iteration $j+1$.
The response maps resulting from the decimated separable wavelet decomposition is illustrated in Fig.~\ref{fig:decimatedWT} for the 2$D$ case.
Although illustrated in the context of separable wavelets, decimated transforms can also be used with non-separable wavelets.

It is worth noting that the convolution is, in this case, modified because the shifts are restricted to the resolution of the $j$ times downsampled response maps~\cite{Dau1992}. In addition, this must be taken into account when aggregating the response maps (see Section~\ref{sec:overview}), where the ROI mask $\boldsymbol{R}$ must also be downsampled to match the dimensions of the corresponding response maps $h^j[\boldsymbol{k}]$.

\subsection[Undecimated Transform]{Undecimated Transform \id{CVCQ}}\label{sec:undecimatedWT}
The undecimated transform, also called \emph{stationary} transform, yields a translation-invariant decomposition by obviating the downsampling steps required by the decimated transform. Although the transform becomes redundant (\textit{i.e.} it yields more coefficients than strictly required for a perfect reconstruction), it is better suited to our case where we are not interested in image coding but rather image analysis.

The separable undecimated transform involves upsampling of the wavelet filters $g_H^j$ and $g_L^j$ to achieve the multiscale decomposition via the matched-size filters in Eq.~\eqref{eq:wavelets}.
While others exist, a common upsampling approach is to use the \emph{\`{a} trous} algorithm~\cite{Dut1989}.
\marginnote{\footnotesize v4: Clarified implementation of the \emph{\`{a} trous} algorithm.}
The \emph{\`{a} trous} algorithm involves inserting zeros, or holes, into the filter kernel. For example, the high-pass kernel of the Haar wavelet is \(\left[ -1 / \sqrt{2}, 1 / \sqrt{2}\right]\). The first level decomposition of this high-pass kernel is \(\left[ -1 / \sqrt{2}, 0, 1 / \sqrt{2}\right]\), or alternatively \(\left[ -1 / \sqrt{2}, 0, 1 / \sqrt{2}, 0\right]\). The second level decomposition of the same kernel is formed by inserting zeros between the values of the first level kernel, forming \(\left[ -1 / \sqrt{2}, 0, 0, 0, 1 / \sqrt{2}\right]\), or alternatively \(\left[ -1 / \sqrt{2}, 0, 0, 0, 1 / \sqrt{2}, 0, 0, 0\right]\). Both alternatives are valid, but result in different response maps. Popular standard implementations in \texttt{MATLAB} and \texttt{pywavelets} use the second alternative. We therefore recommend using the second alternative for reproducibility.

The response maps $h^j[\boldsymbol{k}]$ produced through undecimated transform have the same dimensionality as the input image $f$ and are simply obtained via the convolution of $f[\boldsymbol{k}]$ with the filters $g_H^j$ and $g_L^j$ along each image direction.
In 2$D$, this yields four response maps for every iteration: $h_{LL}^j$, $h_{LH}^j$, $h_{HL}^j$ and $h_{HH}^j$.

\begin{figure}
\centering
\includegraphics[trim = 0 0 0 0, clip, width=0.2\linewidth]{f.png}\\
\includegraphics[trim = 0 0 0 0, clip, width=0.8\linewidth]{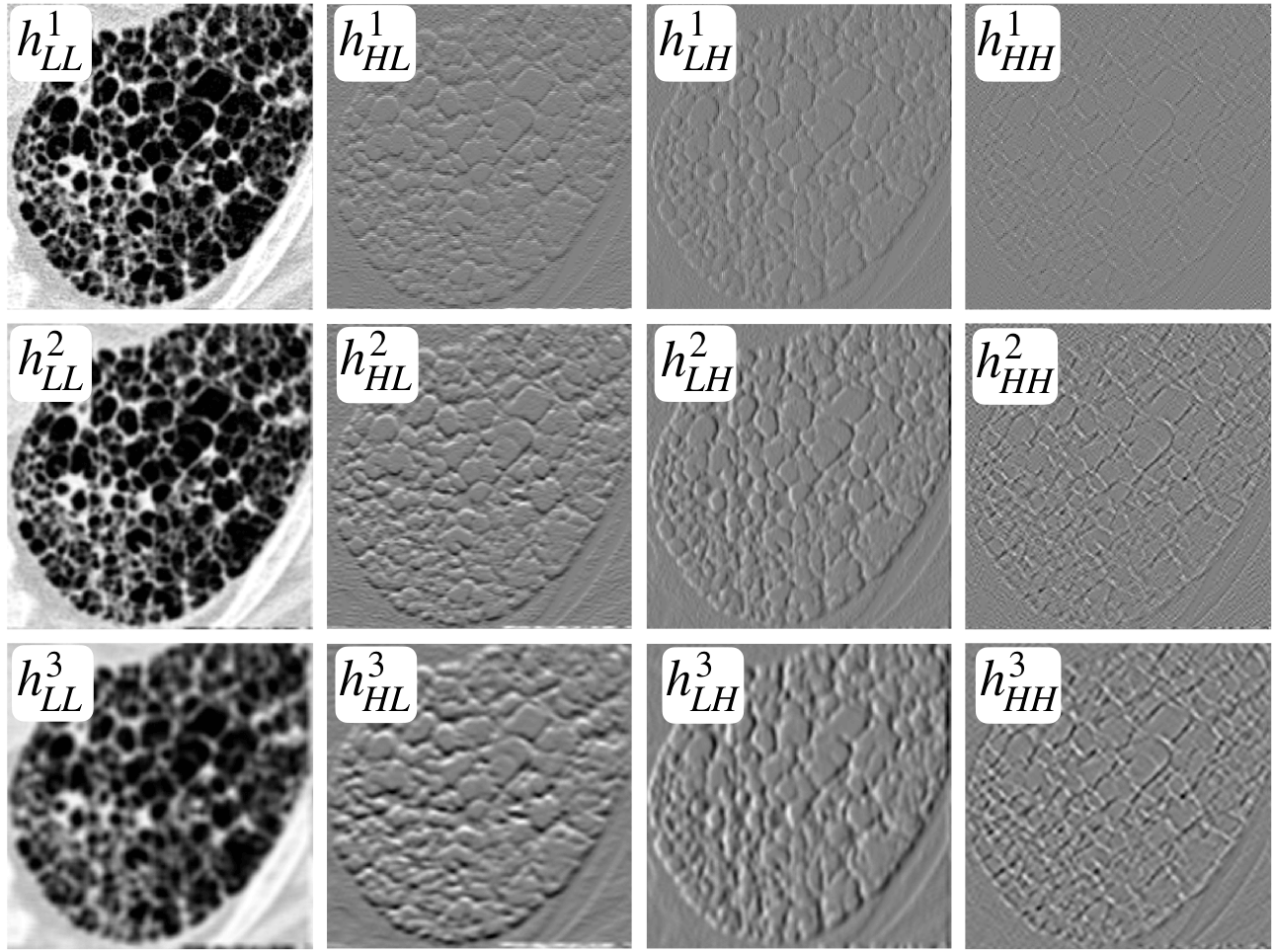}\\
\caption{Response maps from the 2$D$ undecimated separable wavelet transform of the input image $f$.
The first three iterations are shown when using the Haar wavelet.}
  \label{fig:undecimatedWT}
\end{figure}

The response maps resulting from the undecimated separable wavelet decomposition is illustrated in Fig.~\ref{fig:undecimatedWT} for the 2$D$ case.
Again, although illustrated here for separable wavelets, non-decimated transforms can also be used with non-separable wavelets.
\subsection[Separable Wavelets]{Separable Wavelets \id{25BO}}\label{sec:separableWavelets}
The discrete separable wavelet transform yields a collection of 1$D$ wavelet kernels obtained from $J$ dilations of one unique mother wavelet function, which is a high-pass filter $g_H[k]$~\cite{Dau1992}.
The remaining low frequencies are covered by a low-pass filter $g_L$ called scaling function.
When considering dyadic dilations, the scale of the kernels $g_H$ and $g_L$ are indexed by $j$ as
\begin{equation}\label{eq:wavelets}
g^j[k] = 2^{j/2}g[2^j k].
\end{equation}
The distinctive property of the separable wavelet transform is to use the separable convolution. 
The latter is computationally efficient but for radiomics has two distinct disadvantages: it is not rotationally invariant, and only strictly separable wavelets can be used (see Section~\ref{sec:separableConv}). 

\begin{figure}
\centering
\includegraphics[trim = 0 0 0 0, clip, width=0.2\linewidth]{f.png}\\
\includegraphics[trim = 0 0 0 0, clip, width=0.6\linewidth]{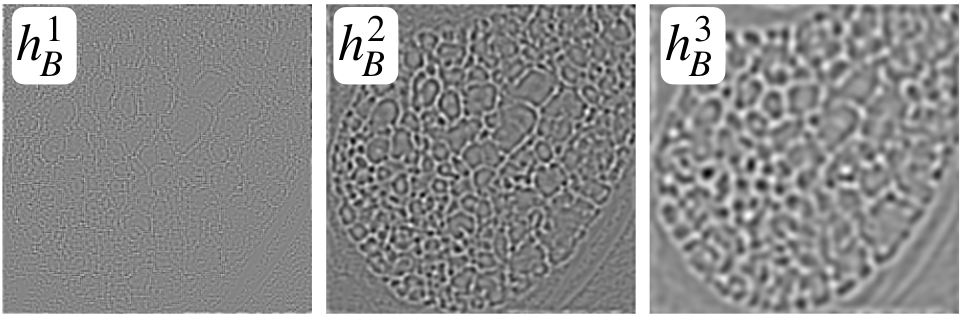}\\
\caption{Response maps from the 2$D$ undecimated non-separable wavelet transform of the input image $f$.
The first three iterations are shown when using the Simoncelli wavelet.}
  \label{fig:nonseparableWT}
\end{figure}
\subsubsection{Directionality Considerations}\label{sec:directionalityWT}
In 3$D$, the convolution of the low-pass and high-pass filters in the three directions of space yields eight different wavelet response maps:  $h_{LLL}$, $h_{LLH}$, $h_{LHL}$, $h_{LHH}$, $h_{HLL}$, $h_{HLH}$, $h_{HHL}$ and $h_{HHH}$. Within the image frame of reference\footnote{The image frame of reference of every image set (\textit{e.g.} a scan) of a study should have the same common orientation relative to the patient reference frame, see Section~\ref{sec:imageDirections}.}, these three directions are $k_1$, $k_2$ and $k_3$. For example, let us consider the $h_{LHL}$ response map: a low-pass filter is applied in the $k_1$ direction, a high-pass filter is applied in the $k_2$ direction and a low-pass filter is applied in the $k_3$ direction (see Section~\ref{sec:separableConv}).
%(\textit{i.e.} "$h_{XYZ}^0$"). 
%
\subsubsection[Haar]{Haar\id{UOUE}}\label{sec:haar}
The simplest separable wavelet is the Haar wavelet.
The Haar wavelet and scaling function form the simplest admissible function pair and are defined as~\cite{Lee2018}
$$g_{H,\text{Haar}} = \left[ \frac{-1}{\sqrt{2}}, \frac{1}{\sqrt{2}}\right], \quad g_{L,\text{Haar}} = \left[ \frac{1}{\sqrt{2}}, \frac{1}{\sqrt{2}}\right].$$

\subsubsection{Daubechies}\label{sec:daubechies}
Daubechies wavelets and scaling functions are characterised by their number of vanishing moments $p$, \textit{i.e.} their ability to approximate polynomials of order up to $p-1$~\cite{Dau1992}.
For $p=1$ Daubechies kernels are equivalent to Haar.
Kernel values of wavelet and scaling functions for $p>1$ can be obtained from the PyWavelets website\footnote{http://wavelets.pybytes.com/family/db/, as of July 2018.}.
\subsubsection{Other}\label{sec:otherWavelets}
There exists a large number of other separable wavelet/scaling function pairs (\textit{e.g.} Meyer, Coiflets), each targeting different objectives in terms of signal analysis. The kernel values of most common wavelets can be obtained from the Wavelet Browser of PyWavelets\footnote{http://wavelets.pybytes.com, as of July 2020.}.
\marginnote{\footnotesize v6: Added implementation troubleshooting to clarify how undecimated wavelet decomposition should be implemented.}
\vspace{2mm}
\begin{tcolorbox}[width=150mm, halign=left, colframe=black, colback=white, boxsep=0mm, arc=3mm, colframe=black!50!white,
title=Implementation Troubleshooting, title filled=true, fonttitle=\bfseries]
\begin{itemize}
\item Undecimated wavelet decomposition of separable wavelets require a sequence of steps. Here we illustrate these steps using the Haar wavelet for two iterations, where we want to compute the pseudo-rotation invariant response map of the high-pass ($HHH$) filter in the final iteration:
    \begin{enumerate}
        \item iteration $j=1$: 1\textsuperscript{st} decomposition level
            \begin{itemize}
                \item Define the basic low-pass $g_L$ kernel: $\left[1/\sqrt{2}, 1/\sqrt{2}\right]$
                \item If necessary, append a zero to obtain an odd filter dimension: $g^1_{L_{odd}}=\left[1/\sqrt{2}, 1/\sqrt{2}, \mathbf{0}\right]$
                \item Create a filter bank of 4 (2$D$) or 24 (3$D$) low-pass filter sets based on $g^1_{L_{odd}}$ for rotation invariance, see appendix \ref{app:separableConvRightAngleEquivariant}.
                \item Apply each filter bank to obtain 4 (2$D$) or 24 (3$D$) response maps $h^1_j$.
            \end{itemize}
        \item iteration $j=2$: 2\textsuperscript{nd} decomposition level
            \begin{itemize}
                \item Define the basic high-pass $g_H$ kernel: $\left[-1/\sqrt{2}, 1/\sqrt{2}\right]$
                \item Apply the à trous algorithm and insert zeros: $g^2_H=\left[-1/\sqrt{2}, \mathbf{0}, 1/\sqrt{2}, \mathbf{0} \right]$
                \item If necessary, append a zero to obtain an odd filter dimension: $g^2_{H_{odd}}=\left[-1/\sqrt{2}, 0, 1/\sqrt{2}, 0, \mathbf{0} \right]$
                \item Create a filter bank of 4 (2$D$) or 24 (3$D$) low-pass filter sets based on $g^2_{H_{odd}}$ for rotation invariance, see appendix \ref{app:separableConvRightAngleEquivariant}.
                \item Apply each filter bank $j$ to the corresponding low-pass response maps of the previous iteration ($h^1_j$) obtain 4 (2$D$) or 24 (3$D$) response maps $h^2_j$.
                \item Pool the response maps, for example through averaging, to obtain response map $h^2_{HHH}$.
            \end{itemize}
    \end{enumerate}
\end{itemize}
\end{tcolorbox}

\subsection[Non-separable Wavelets]{Non-separable Wavelets \id{LODD}}\label{sec:nonseparableWavelets}
\marginnote{\footnotesize v4: Extended description of non-separable wavelets.}
As motivated in  Section~\ref{sec:geomInvariances}, filters (and more generally texture operators) should be invariant or equivariant to local rotations in most cases. 
However, with the exception of the Gaussian filter, all separable filters (including separable wavelets) are not invariant/equivariant to rotations.
Therefore, it is interesting to consider non-separable wavelets to achieve isotropic image analysis.
The starting point for the definition of non-separable wavelets is a single (unidimensional) radial profile in the Fourier domain as $\hat{g}[\boldsymbol{\nu}]$, where $\hat{g}$ is a function of $||\boldsymbol{\nu}||$ defining the radial coordinate~\cite{UCV2011}.
As a consequence, these functions are circularly symmetric and can be further combined with directional analysis such as the Riesz transform (Section~\ref{sec:Riesz}).

Non-separable wavelets are easily implemented directly in the Fourier domain using their unidimensional radial profile.
The support of the filter in the Fourier domain is the same as the considered $N_1\times\cdots\times N_D$ image $f$ to allow for efficiently computing the convolution as a simple Hadamard product via Eq.~\eqref{eq:convolutionFFT}.
Along dimension $i$, each coordinate $\nu_i$ is contained in $[-\frac{N_i}{2},\frac{N_i}{2}]$ which is centered around the null frequency $\boldsymbol{\nu}=\boldsymbol{0}$.
The maximum coordinate value $\frac{N_i}{2}$ will be assigned to the value $\nu_{B}=\pi$ radians/sample, which is called the normalized Nyquist frequency.
The unidimensional radial coordinate is then obtained as $||\boldsymbol{\nu}||=\sqrt{\nu_1^2+\nu_2^2+\dots+\nu_D^2}$.
It is worth noting that the value of $||\boldsymbol{\nu}||$ will exceed $\nu_B$ in the corners of the image. These ``corner" values should be ignored as the corresponding frequencies can only be measured along diagonal directions, yielding anisotropic image analysis. They are ignored by construction for all proposed radial wavelets, e.g.~\eqref{eq:shannonWavelet} and~\eqref{eq:simoncelliWavelet}.

\begin{figure}
\centering
\includegraphics[trim = 0 0 0 0, clip, width=.8\linewidth]{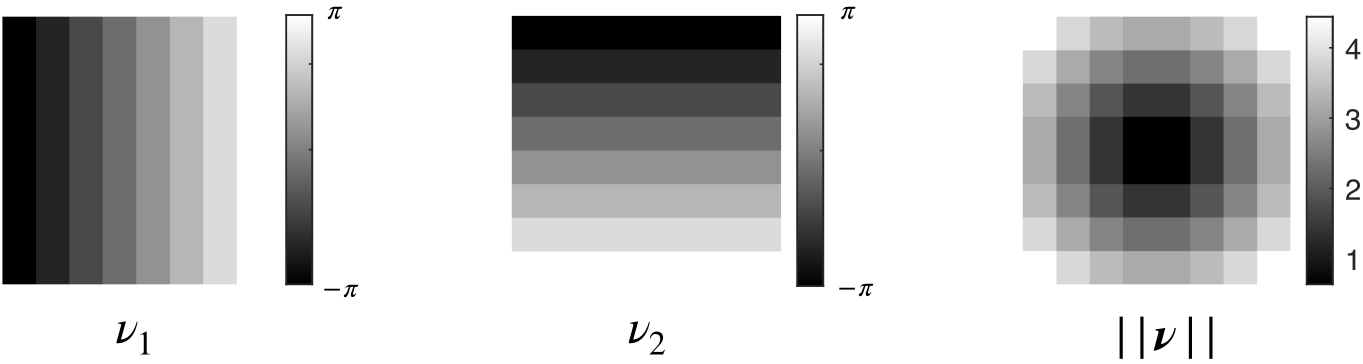}\\
\caption{Coordinate grids in voxel dimensions for for $N=8$.}
  \label{fig:coordinateGridsFourier}
\end{figure}

In order to implement non-separable wavelets in practice, we recommend defining a coordinate grid in the Fourier domain, where each dimension $\nu_i$ is sampled in the interval $[-\pi,\pi]$ with a step of $\frac{2\pi}{N_i}$.
Coordinate grids in voxel dimensions are illustrated in Fig.~\ref{fig:coordinateGridsFourier} for $N=8$.
%%% RECOMMENDATION on how to define coordinate grids in Fourier for implementing nonseparable wavelets

%

Whereas wavelet functions are usually defined as a cascade of high- and low-pass filters, here we focus on the band-pass filters corresponding to the consecutive subbands (i.e. response maps) that would results from combined high- and low- pass filtering (see introductory paragraph of Section~\ref{sec:wavelets}). In other words, these response maps can be directly obtained using a corresponding band-pass filter $h_B[\boldsymbol{k}]$, without the need of re-convolving with previous iterations of the wavelet transform. The resulting response maps are referred to as ``B maps''.

The simplest non-separable and circularly symmetric wavelet is the Shannon wavelet~\cite{UCV2011} (\textid{GWM2}) characterised by the following mother function
\begin{equation}\label{eq:shannonWavelet}
\hat{g}_{\text{sha}}[\boldsymbol{\nu}]=
\begin{cases}
1\quad\text{if}\quad\frac{\nu_B}{2}< ||\boldsymbol{\nu}|| \leq \nu_B,\\
0\quad\text{otherwise}.
\end{cases}
\end{equation}
Eq.~\eqref{eq:shannonWavelet} corresponds to the sinc wavelet in the spatial domain, which has the disadvantage of having large spatial supports. Therefore, an interesting alternative is the smoother and more compactly supported (in space) Simoncelli wavelet~\cite{PoS2000} (\textid{PRT7}). Its band-pass function is defined as
\begin{equation}\label{eq:simoncelliWavelet}
\hat{g}_{\text{sim}}[\boldsymbol{\nu}]=
\begin{cases}
\cos\left(\frac{\pi}{2}\log_2\left(\frac{2||\boldsymbol{\nu}||}{\nu_B}\right)\right)\quad\text{if}\quad\frac{\nu_B}{4}\leq ||\boldsymbol{\nu}|| \leq \nu_B,\\
0\quad\text{otherwise}.
\end{cases}
\end{equation}

\begin{figure}
\centering
\includegraphics[trim = 0 0 0 0, clip, width=1\linewidth]{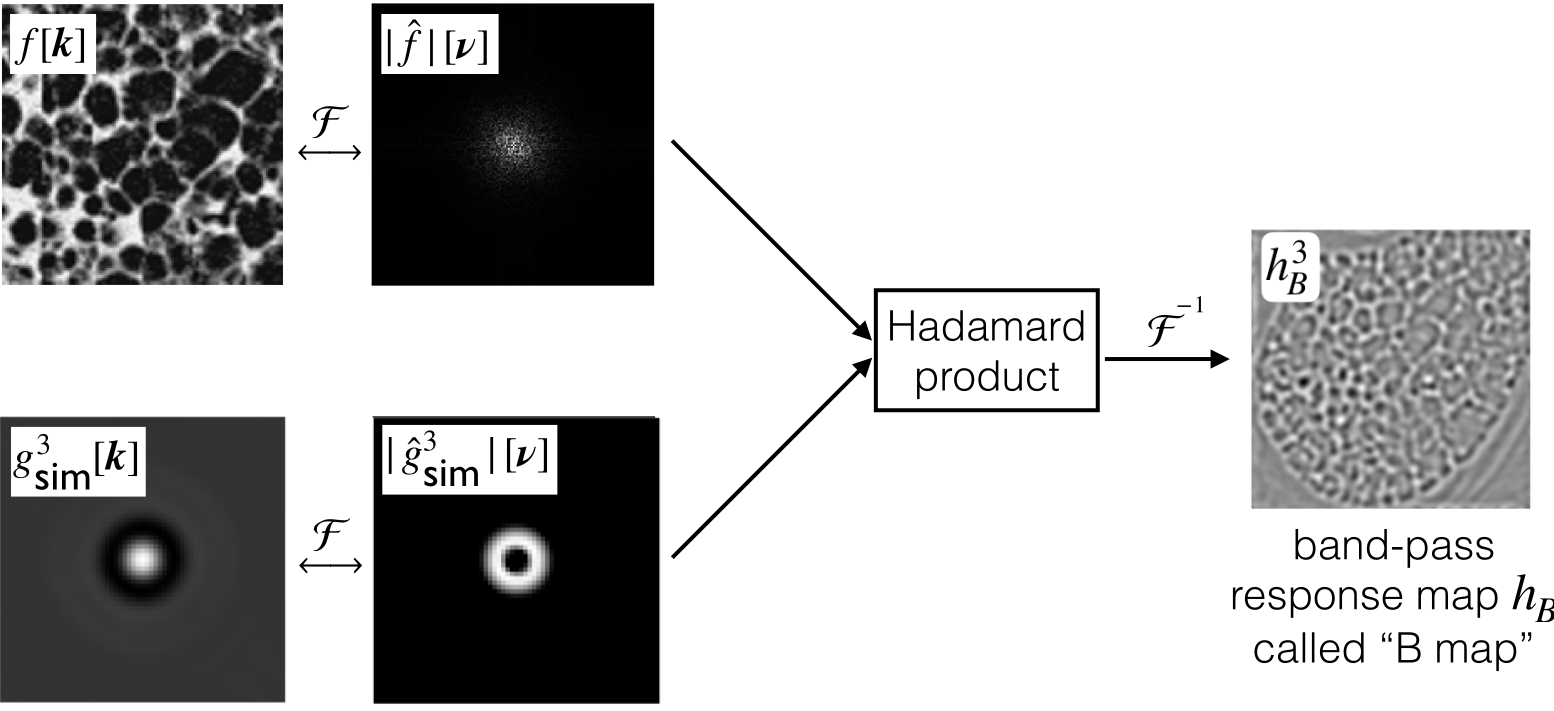}\\
\caption{Convolution in the Fourier domain using a Hadamard product. Complex modulus of the Fourier images are shown only for display purposes: the Hadamard product must be done with complex values.}
  \label{fig:hadamardConv}
\end{figure}

An easy way to implement several consecutive iterations of such a non-separable wavelet transform is to start from the initial definition of the band-pass filter (the one with highest frequency band, \emph{e.g.} Eq.~\eqref{eq:simoncelliWavelet}), and constructing the filter with half $\nu_B$ at every consecutive step. Concretely, the bandpass filter at a given scale level $j$ is obtained by replacing $\nu_B$ by $\frac{\nu_B}{2^j}$ in~\eqref{eq:shannonWavelet} or~\eqref{eq:simoncelliWavelet}.
%%% RECOMMENDATION to compute B maps at any level of the transform
The response map can then be computed by taking the Hadamard product of the bandpass filter and the image in Fourier domain, and then performing an inverse Fourier transform.
This process is illustrated in Fig.~\ref{fig:hadamardConv}.

\begin{figure}
\centering
\includegraphics[trim = 0 0 0 0, clip, width=0.6\linewidth]{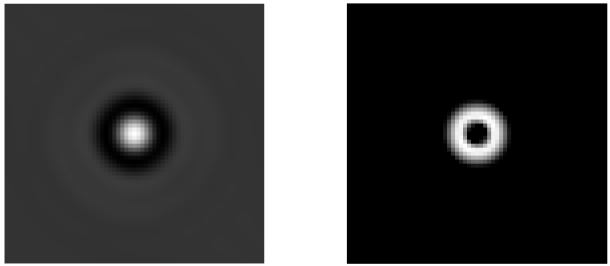}\\
$g_{\text{sim}}[\boldsymbol{k}]$
\hspace{3.05cm}
$|\hat{g}_{\text{sim}}[\boldsymbol{\nu}]|$
\caption{Example of a 2$D$ band-pass Simoncelli wavelet $g_{\text{sim}}$ after three decomposition levels, computed on a $66\times 66$ grid.
The wavelet is shown in the spatial (left) and in the Fourier (right, complex modulus shown) domains.}
  \label{fig:simoncelli}
\end{figure}

The response maps resulting from the undecimated non-separable wavelet decomposition is illustrated in Fig.~\ref{fig:nonseparableWT} for the 2$D$ case.
An example of a Simoncelli wavelet is shown in Fig.~\ref{fig:simoncelli}.
Other alternatives can be found in Table 1 of Unser \textit{et al.}~\cite{UCV2011}.

\subsection{Wavelets: Considerations for Radiomics}\label{sec:waveletConsiderations}
To summarise, undecimated non-separable wavelets have the advantage of yielding isotropic (\textit{i.e.} rotation invariant/equivariant, see Section~\ref{sec:combiningDSandRotInv}) and translation equivariant image analysis. Moreover, they yield only one response per decomposition level (\textit{i.e.} one per scale), which significantly reduces the number of radiomics features when compared to their separable counterpart.
\marginnote{\footnotesize v4: Added recommendation to not perform decomposition using separable wavelets beyond the first level.}
Separable wavelets were mostly designed for image coding, which has very different design constraints.
Such wavelets yield a large collection of response maps that are biased towards image axes and lack rotational invariance.
While it is possible to make them approximately rotation invariant using orientation pooling over equivariant right angle representations as suggested in Section~\ref{sec:combiningDSandRotInv} and Appendix~\ref{app:separableConvRightAngleEquivariant}, we do not recommend to use this orientation pooling procedure for decomposition levels larger than 1.
Applying orientation pooling after every convolution operation of a multi-level separable wavelet decomposition would result in an overly complicated algorithm, yielding only approximate rotational invariance where the granularity of the approximation depends on the number of group elements (\emph{e.g.} right angles only or using more angular samples) in the used equivariant representation.
Therefore, we recommend not to use separable wavelet for rotation invariant image analysis, or to limit the wavelet decomposition to only one level.
Isotropic non-separable wavelets such as the Simoncelli wavelet allow achieving multi-level, truly isotropic and rotation invariant image analysis by design.
When the directionality of image patterns of interest (\textit{e.g.}, tumour margin, vessels) is expected to be important, aligned directional wavelet filters such as Riesz (Section~\ref{sec:Riesz}) that allow combining directional sensitivity with local invariance to rotations can be considered (see Section~\ref{sec:combiningDSandRotInv}).

\section[Riesz]{Riesz \id{AYRS}}\label{sec:Riesz}
\marginnote{\footnotesize v4: Extended description of the Riesz transform.}
Whereas non-separable wavelets like Simoncelli constitute an excellent first solution for using wavelets in radiomics studies, they are also circularly symmetric and therefore cannot characterise directional patterns.
To address directionality of patterns, these non-separable circularly symmetric wavelets can be advantageously combined with directional analysis methods such as directional derivatives~\cite{Prasanna2014} or spherical harmonics.

An elegant approach to obtain features measuring directional transitions between pixel values is to use image derivatives. 
Besides being simple to compute, these have the advantage of being interpretable (at least for the first- and second-order), which makes them attractive for understanding the meaning of the texture measures in a particular medical or biological application context. 
For instance, the first-order derivative is called the gradient and informs about the slope of the transitions in the image, \emph{e.g.} sharp versus smooth transitions.
Second-order derivatives are called Hessian and quantify the curvature of the transitions.
It is worth noting that the spatial scale of these transitions will be controlled by a preliminary isotropic filtering step based on \emph{e.g.} a Gaussian smoother (low-pass) or a Simoncelli wavelet (band-pass). 

\begin{figure}
\centering
\includegraphics[trim = 0 0 0 0, clip, width=1\linewidth]{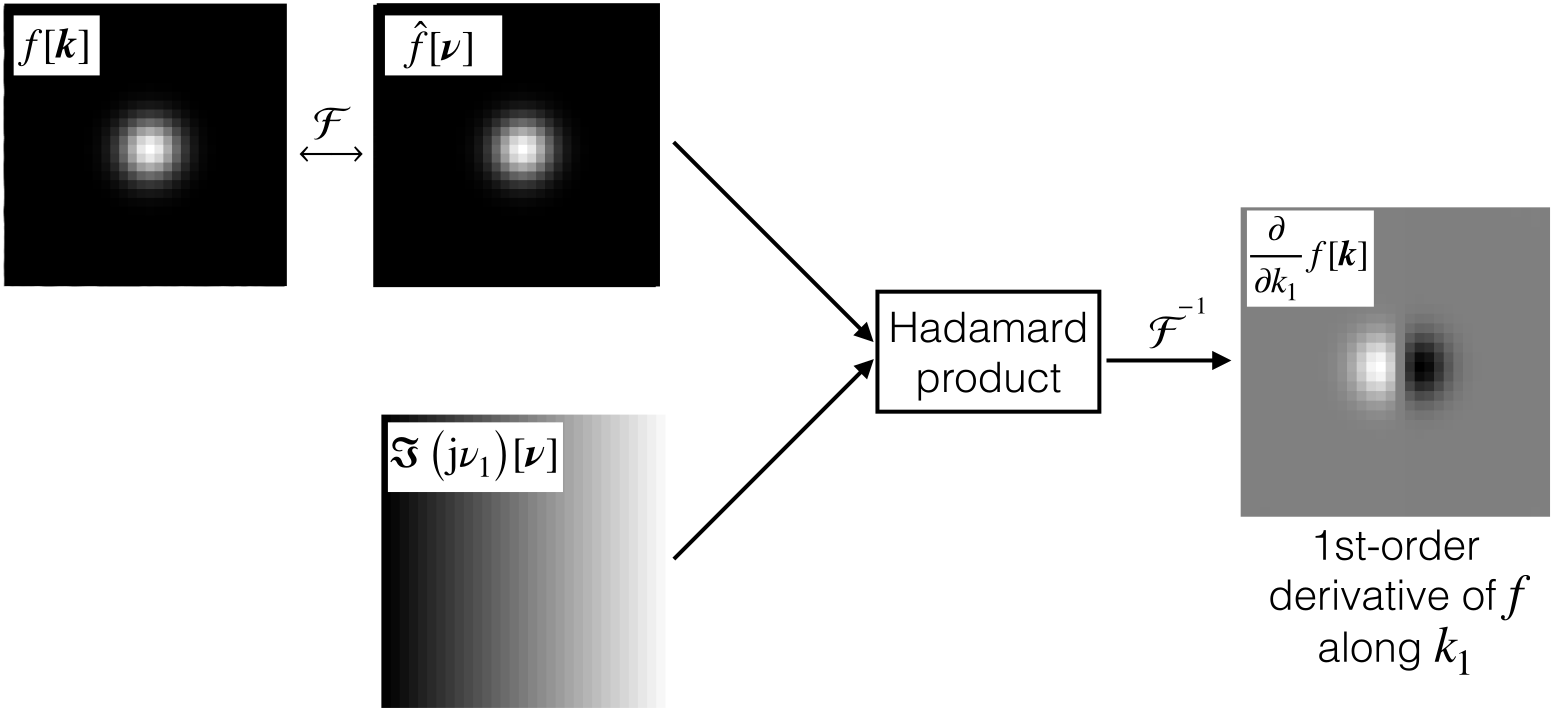}\\
\caption{Deriving an image in the Fourier domain (illustrated with a Gaussian function). The notation $\Im(\cdot)$ denotes the imaginary part.}
  \label{fig:GaussianDerivative}
\end{figure}

An interesting option to compute the derivatives is to do so in the Fourier domain~\cite{DeF2017}.
This also provides the opportunity to easily compute higher-order image derivatives of order $l$ (\textid{OC48}) as
\begin{equation}\label{eq:FourierDerivative}
\frac{\partial^l}{\partial k_i^l} f[\boldsymbol{k}] \overset{\mathcal{F}}{\longleftrightarrow} (\,\mathrm{j}\nu_i\,)^l \hat{f}[\boldsymbol{\nu}],
\end{equation}
where $1\leq i \leq D$.
It can be noticed that differentiating an image along the direction $k_i$ only requires multiplying its Fourier transform by $\mathrm{j}\nu_i$.
Computing $l^{\text{th}}$-order derivatives has an intuitive interpretation (\emph{e.g.}, gradient for $l=1$, curvature for $l=2$).
Let us illustrate this by differentiating a simple 2$D$ Gaussian function of dimension $32\times 32$ (see Fig.~\ref{fig:GaussianDerivative}).
A Gaussian function $f[\boldsymbol{k}]$ is first transformed in the Fourier domain resulting in $\hat{f}[\boldsymbol{\nu}]$. It is worth noting that the Fourier transform of a Gaussian is real-valued and also a Gaussian.
In parallel, the derivative kernel $\mathrm{j}\nu_1$ can be simply defined by multiplying the Fourier coordinate grid $\nu_1$ (see Fig.~\ref{fig:coordinateGridsFourier}) by the imaginary unit $\mathrm{j}$.
Whe using the Hadamard product in Fourier and then back to the space domain (\emph{i.e.} convolution operation as depicted in Fig.~\ref{fig:hadamardConv}), the result corresponds to the derivative of $f$ along $k_1$, $\frac{\partial}{\partial k_1} f[\boldsymbol{k}]$.

Unfortunately, a pure image derivative filter as computed in Eq.~(\ref{eq:FourierDerivative}) is high-pass (because multiplied by $\nu_i$) and accentuates high frequencies along $k_i$.
Therefore, it is desirable to implement image derivatives as all-pass filters instead of high-pass, by simply normalising the derivative kernel $\mathrm{j}\nu_i$ by the coordinate norm $\boldsymbol{\nu}$.
This is exactly what the first-order Riesz transform yields $\boldsymbol{\mathcal{R}}\{f\}[\boldsymbol{k}]$ as~\cite{UnV2010}
\begin{equation}\label{eq:RieszTranform}
\boldsymbol{\mathcal{R}}\{f\}[\boldsymbol{k}]=
\left(
\begin{array}{c}
{\mathcal R}_1 \{f\}[\boldsymbol{k}]   \\
\vdots \\
{\mathcal R}_D \{f\}[\boldsymbol{k}] 
\end{array}
\right)
\overset{\mathcal{F}}{\longleftrightarrow} -\mathrm{j}\frac{\boldsymbol{\nu}}{||\boldsymbol{\nu}||}\hat{f}[\boldsymbol{\nu}].
\end{equation}
It can be noticed that dividing the Fourier representation by the norm of $\boldsymbol{\nu}$ transforms Eq.~\eqref{eq:FourierDerivative} in $D$ all-pass operators $\mathcal{R}_i$.
Eq.~\eqref{eq:RieszTranform} can be used to compute first-order directional derivatives (\textit{i.e.} gradient-like for characterising image edges).
However, higher-order derivatives can be relevant for radiomics studies (\textit{e.g.} second-order or Hessian-like for characterising image ridges like tumour margin or vessels).
% add example difference between high-pass derivative and "Riesz-based" derivative. REMARK: zero average along x, hence the vertical black stripe in the spectrums

\begin{figure}
\centering
\includegraphics[trim = 0 0 0 0, clip, width=1\linewidth]{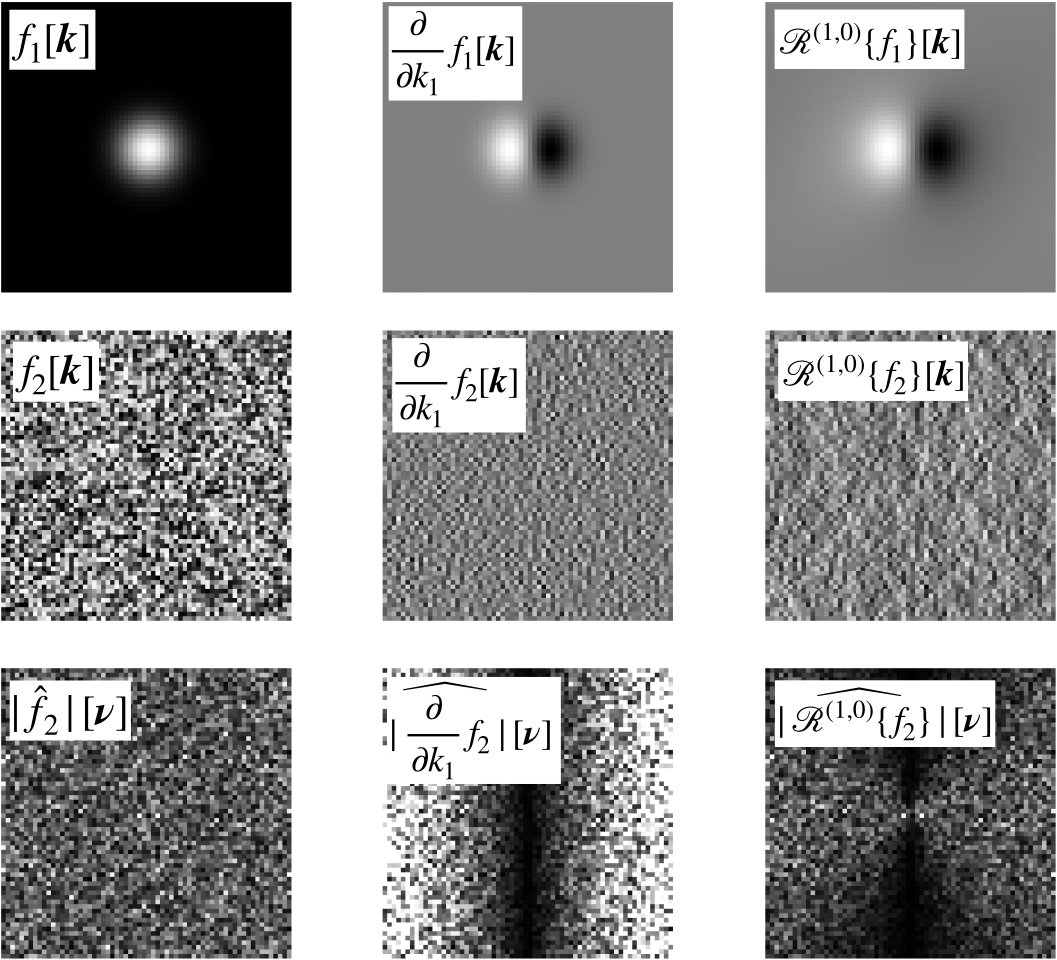}\\
\caption{Qualitative comparison between classical image derivatives (high-pass) as defined in~\eqref{eq:FourierDerivative} and Riesz-based image derivatives (all-pass) as defined in~\eqref{eq:RieszTranform}.\\
\underline{First row:} Gaussian image $f_1$ and its derivatives in the spatial domain. The classical derivative $\frac{\partial}{\partial k_1} f_1[\boldsymbol{k}]$ and the Riesz transform $\mathcal{R}^{(1,0)}\{f_1\}[\boldsymbol{k}]$ look very similar because the Gaussian image is essentially composed of low frequencies.\\
\underline{Second row:} White noise image $f_2$ and its derivatives in the spatial domain. The classical derivative $\frac{\partial}{\partial k_1} f_2[\boldsymbol{k}]$ contains mostly high frequencies, whereas the Riesz derivative $\mathcal{R}^{(1,0)}\{f_2\}[\boldsymbol{k}]$  retains all frequencies.\\
\underline{Third row:} White noise image $\hat{f}_2$ and its derivatives in the Fourier domain (moduluses are shown). The difference between the two derivatives approaches is striking, where the boost of high frequencies along $\nu_1$ for the classical derivative approach can be clearly seen in $|\widehat{\frac{\partial}{\partial k_1} f_2}|[\boldsymbol{\nu}]$. The vertical and centered black line results from the fact that first order derivatives are zero mean (\emph{i.e.} they contain no null frequencies).}
  \label{fig:rieszVSderivative}
\end{figure}

A qualitative comparison between classical image derivatives as defined in~\eqref{eq:FourierDerivative} and Riesz-based image derivatives~\eqref{eq:RieszTranform} is proposed in Fig.~\ref{fig:rieszVSderivative} for both a Gaussian image ($f_1$) and a white noise image ($f_2$).

For a fixed (maximal) order $L$, the collection of higher-order all-pass image derivatives are defined in the Fourier domain as
\begin{equation}\label{eq:RieszTransformHigherOrder}
\widehat{\boldsymbol{\mathcal{R}}}^{\boldsymbol{l}}\{\hat{f}\}[\boldsymbol{\nu}]=
(-\mathrm{j})^L
\sqrt{\frac{L!}{l_1!\cdots l_D!}}
\frac{\nu_1^{l_1}\cdots\,\nu_D^{l_D}}
{\left( \nu_1^2+\cdots+\nu_D^2 \right)^{L/2}}
\,\hat{f}[\boldsymbol{\nu}],
\end{equation}
which yields a total of ${{L+D-1}\choose{D-1}}=\frac{(L+D-1)!}{L!(D-1)!}$ all-pass filters for all combinations of the elements $l_i$ of the vector $\boldsymbol{l}$ as $|\boldsymbol{l}|=l_1+\dots+l_D=L$.
The collection of Riesz operators $\boldsymbol{\mathcal{R}}^{(l_1,l_2,\dots,l_D)}$ of order $L$ is denoted by $\boldsymbol{\mathcal{R}}^L$.
For instance, in 3$D$ and with $L=2$, the element $\widehat{\mathcal{R}}^{(0,2,0)}\{f\}[\boldsymbol{\nu}]$ corresponds qualitatively to a second-order derivative of $f$ along the direction $k_2$ and we have $$\widehat{\mathcal{R}}^{(0,2,0)}\{f\}[\boldsymbol{\nu}]=
\frac{-\nu_2^{2}}
{\nu_1^2+\nu_2^2+\nu_3^2}
\,\hat{f}[\boldsymbol{\nu}].$$

A set of band-pass, multi-scale and multi-orientation filters $\boldsymbol{g}_{\sigma,\boldsymbol{l}}[\boldsymbol{k}]$ can be obtained by simply applying the Riesz transform to circularly symmetric non-separable wavelets \textit{e.g.} Eq.~\eqref{eq:simoncelliWavelet} or multi-scale filters \textit{e.g.} the LoG filter $g_{\sigma}$ Eq.~\eqref{eq:LoG}, as
\begin{equation}\label{eq:rieszFilter}
\boldsymbol{g}_{\sigma,\boldsymbol{l}}[\boldsymbol{k}] = 
\boldsymbol{\mathcal{R}}^{\boldsymbol{l}}
\{g_{\sigma}\}[\boldsymbol{k}].
\end{equation}

\begin{figure}
\centering
\includegraphics[trim = 0 0 0 0, clip, width=0.7\linewidth]{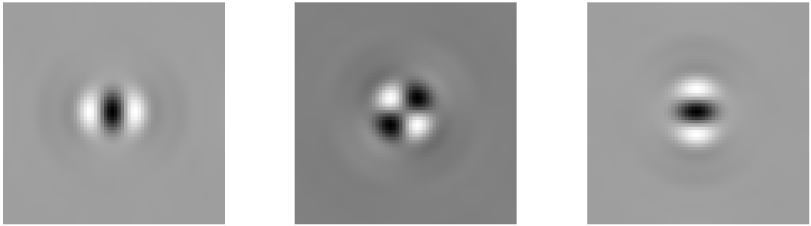}\\
$\mathcal{R}^{(2,0)}
\{g_{\text{sim}}\}[\boldsymbol{k}]$
\hspace{1.3cm}
$\mathcal{R}^{(1,1)}
\{g_{\text{sim}}\}[\boldsymbol{k}]$
\hspace{1.3cm}
$\mathcal{R}^{(0,2)}
\{g_{\text{sim}}\}[\boldsymbol{k}]$
\caption{A 2$D$ Riesz filterbank for $L=2$ and combined with a Simoncelli wavelet.}
  \label{fig:Riesz}
\end{figure}

An example of a 2$D$ Riesz filterbank for $L=2$ and combined with a Simoncelli wavelet is shown in Fig~\ref{fig:Riesz}.
Eq.~\eqref{eq:rieszFilter} yields a collection of filters measuring $L$-th order (scaled) derivatives. However, these derivatives are computed along image axes $k_1,\dots,k_D$, which entails similar challenges as separable wavelets: (i) analyzing images along their axes does not have a particular signification for the problem at hand (\textit{e.g.} characterising local tissue structures within a tumour) and (ii) they are not rotation invariant/equivariant.

\subsection[Aligning Riesz kernels]{Aligning Riesz kernels \id{1SD3}}\label{sec:RieszAlignment}
\marginnote{\footnotesize v6: Additional clarifications on how to align Riesz kernels were provided.}
To address (i) and (ii), one can locally align all Riesz filters using a given alignment criterion to combine directional sensitivity with local invariance to rotations, as motivated in Section~\ref{sec:combiningDSandRotInv} and Fig.~\ref{fig:LRI_overview}.
Whereas the max orientation pooling operation can be seen as an alignment criterion, the latter only focuses on the maximum response of one single filter.
Since Riesz filter banks include several filters with distinct profiles (see \textit{e.g.} Fig.~\ref{fig:Riesz} with order $L=2$), other alignment criteria based on meaningful quantities can be used.
We will seek for alignment criteria that are optimizing the response of the entire filterbank and not only one single kernel.

Two suitable alignment criteria are those maximizing the gradients or the Hessian~\cite{DMP2017}, which can be efficiently computed directly from the Riesz coefficients using the structure (or Hessian) tensor $\mathrm{J}[\boldsymbol{k}_0]$~\cite{ChU2012}.
It is worth noting that the while classical definition of the structure tensor is based on simple (high-pass) image derivatives, we can use in this context the all-pass image derivatives provided by the Riesz transform.
In practice, it is beneficial to regularise this structure tensor using a Gaussian window $g_{\sigma_{\text{tensor}}}[\boldsymbol{k}]$ (noted as $\nu(\boldsymbol{x})$ in Section IV-B of~\cite{ChU2012}).
Regularizing the structure tensor is important to determine the scale of the structures from which the orientation is important at a given position $\boldsymbol{k}_0$.
This must be optimized for the problem at hand via the variance $\sigma_{\text{tensor}}$ of a Gaussian window $g_{\sigma_{\text{tensor}}}[\boldsymbol{k}]$.
It is worth noting that $g_{\sigma_{\text{tensor}}}$ is independent from the multi-scale framework used to compute the Riesz transform (e.g. Simoncelli or the LoG filter, see Eq.~\eqref{eq:rieszFilter}). 

In 3$D$, the nine elements of the regularised structure tensor computed from Riesz coefficients at the position $\boldsymbol{k}_0$ are defined using (see Section IV-B of Chenouard \textit{et al.}~\cite{ChU2012})
% \begin{equation}
% \mathrm{J}[\boldsymbol{k}_0]=
% \begin{pmatrix} 
% \mathcal{R}_1^2\{g_{\sigma_{\text{tensor}}}\ast f\}[\boldsymbol{k}_0] & 
% \mathcal{R}_1\mathcal{R}_2\{g_{\sigma_{\text{tensor}}}\ast
% f\}[\boldsymbol{k}_0] 
% & 
% \mathcal{R}_1\mathcal{R}_3\{g_{\sigma_{\text{tensor}}}\ast
% f\}[\boldsymbol{k}_0]\\ 
% \mathcal{R}_2\mathcal{R}_1\{g_{\sigma_{\text{tensor}}}\ast f\}[\boldsymbol{k}_0] & 
% \mathcal{R}_2^2\{g_{\sigma_{\text{tensor}}}\ast f\}[\boldsymbol{k}_0] 
% & 
% \mathcal{R}_2\mathcal{R}_3\{g_{\sigma_{\text{tensor}}}\ast
% f\}[\boldsymbol{k}_0]
% \\ 
% \mathcal{R}_3\mathcal{R}_1\{g_{\sigma_{\text{tensor}}}\ast f\}[\boldsymbol{k}_0] & 
% \mathcal{R}_3\mathcal{R}_2\{g_{\sigma_{\text{tensor}}}\ast f\}[\boldsymbol{k}_0] 
% & 
% \mathcal{R}_3^2\{g_{\sigma_{\text{tensor}}}\ast
% f\}[\boldsymbol{k}_0]
% \end{pmatrix},
% \end{equation}
\begin{equation}\label{eq:structureTensor}
\Big[\mathrm{J}[\boldsymbol{k}_0]\Big]_{mn}=
\sum_{\boldsymbol{k}\in\mathbb{Z}^D} g_{\sigma_{\text{tensor}}}[\boldsymbol{k}-\boldsymbol{k}_0]\cdot
\mathcal{R}_m\{f\}[\boldsymbol{k}]\cdot\mathcal{R}_n\{f\}[\boldsymbol{k}],
\end{equation}
where the indices $m,n\in\{1,2,3\}$ identify the combinations of all first-order 3$D$ Riesz components\footnote{To simplify the notations, the element e.g. $\mathcal{R}^{(1,0,0)}$ is noted $\mathcal{R}_1$, the element $\mathcal{R}^{(0,1,0)}$ is noted $\mathcal{R}_2$, etc.}.
%applied on the regularized image \mbox{$g_{\sigma_{\text{tensor}}}\ast f$}.
It is worth noting that while the sum is defined over the entire image domain $N_1\times\cdots\times N_D\subset\mathbb{Z}^D$, a smaller window can be considered in practice based on the decay of the Gaussian window $g_{\sigma_{\text{tensor}}}[\boldsymbol{k}]$.
The collection of eigenvectors $(\boldsymbol{u}_1,\boldsymbol{u}_2,\boldsymbol{u}_3)$ of $\mathrm{J}$, sorted by eigenvalues, defines a rotation matrix $\mathrm{U}$ for every position $\boldsymbol{k}_0$.
In 3$D$ for instance, $\mathrm{U}$ is a $3\times 3$ rotation matrix that maximizes the energy of $\mathcal{R}_1$, then maximizes the residual energy of $\mathcal{R}_2$, and then $\mathcal{R}_3$.
Therefore, $\mathrm{U}$ constitutes an interesting alignment criteria for the position $\boldsymbol{k}_0$.
Computing $\mathrm{J}$ and then $\mathrm{U}$ for every position in the image defines alignment maps, which however come with a high computational cost.

\begin{figure}
\centering
\includegraphics[trim = 0 0 0 0, clip, width=0.5\linewidth]{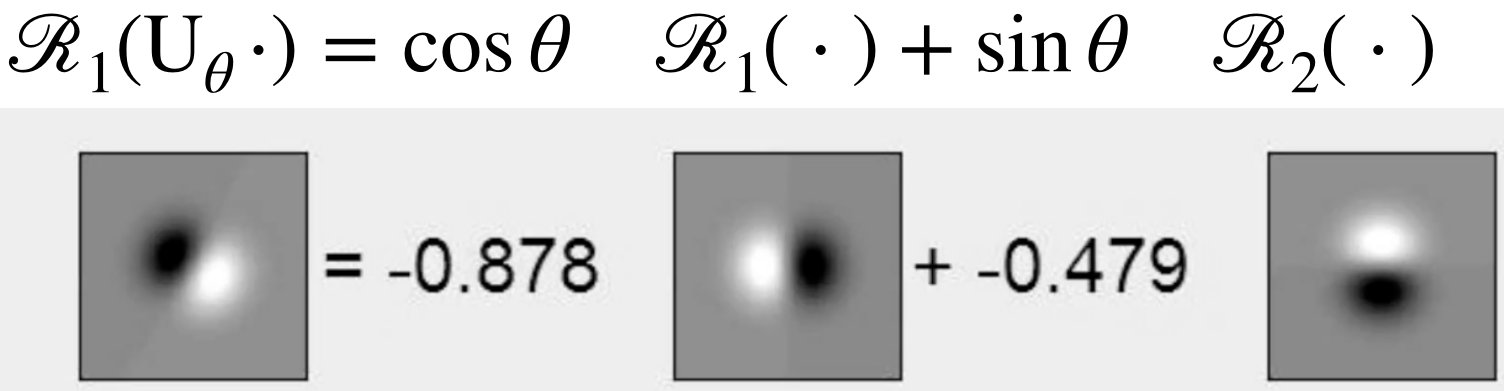}
\caption{Illustration of the steerability property of the Riesz transform in 2$D$. Rotating the Riesz kernel $\mathcal{R}_1$ using a rotation matrix $\mathrm{U}_\theta$ only requires computing a linear combination~\cite{UnV2010}.}
  \label{fig:2dsteerability}
\end{figure}

Computing these alignment maps allows determining how to further orient all elements of the Riesz filter bank to achieve locally rotation invariant image analysis.
%More specifically, the collection of eigenvectors (sorted by corresponding eigenvalues) of the structure tensor at the position $\boldsymbol{k}_0$ yields a rotation matrix that can be used to align the Riesz filters. 
Thanks to the steerability property of the Riesz transform~\cite{UnV2010}, no additional convolution operations are required to compute the response of rotated filters.
A steerable filter~\cite{FrA1991} is a filter that can be obtained as a linear combination of steerable kernels, and all rotations of the former can also be obtained via another linear combination of the filter responses parameterised by the desired rotation (\textit{e.g.} one rotation angle in 2$D$), which obviates the need to re-convolve the oriented filter with the input image.
In 2$D$, steering the kernel $\mathcal{R}_1$ using a rotation matrix 
$\mathrm{U}_\theta
=\begin{pmatrix} 
\cos{\theta} & -\sin{\theta} \\
\sin{\theta} & \cos{\theta} \\
\end{pmatrix}
$ can simply be obtained using
$\mathcal{R}_1(\mathrm{U}_{\theta}\,\cdot)=\cos\theta\,\mathcal{R}_1(\cdot)+\sin\theta\,\mathcal{R}_2(\cdot)$, which only requires computing a linear combination parameterised by the rotation.
This is illustrated in Fig.~\ref{fig:2dsteerability}.

More generally, rotating all elements of the Riesz transform can be obtained with the linear combination
\begin{equation}\label{eq:steerability}
    \boldsymbol{\mathcal{R}}\{f_{\mathrm{U}}\}=
    \mathrm{S}_{\mathrm{U}}\boldsymbol{\mathcal{R}}\{f\},
\end{equation}
where $\mathrm{S}_{\mathrm{U}}$ is called the steering matrix corresponding to the rotation matrix $\mathrm{U}$.
$\mathrm{S}_{\mathrm{U}}$ is a square matrix with a number of row/columns equal to the number of elements in the Riesz filterbank, see Eq.\eqref{eq:RieszTransformHigherOrder}.
Let us detail the form of $\mathrm{S}_{\mathrm{U}}$ in the 3$D$ case, for a rotation matrix $\mathrm{U} = (\boldsymbol{u}_1,\boldsymbol{u}_2,\boldsymbol{u}_3)^T$ with $\boldsymbol{u}_i\in\mathbb{R}^3$.
For the 2$D$ case, we refer to the Section II-E of Unser et al.~\cite{UnV2010}.
To specifically define the elements of this steering matrix in the 3$D$ case, let us start with the introduction of the multi-index notation as used by Chenouard et al.~\cite{ChU2012}.
We consider index vectors of the form $\boldsymbol{n} = (n_1, n_2, n_3) \in \mathbb{N}^3$. The following multi--index notations and operators are used:
\begin{itemize}
\item Sum of components: $|\boldsymbol{n}| = n_1+n_2+n_3$,
\item Max of components: $\max(\boldsymbol{n}) = \max(n_1,n_2,n_3)$,
\item Factorial: $\boldsymbol{n}! = n_1!n_2!n_3!$,
\item Exponentiation of a vector $\boldsymbol{x} = (x_1, x_2, x_3) \in \mathbb{R}^3$: $\boldsymbol{x}^{\boldsymbol{n}}~=~x_1^{n_1}x_2^{n_2}x_3^{n_3}$.
\end{itemize}
Using this multi-index notation, we can define the steering coefficients as elements $s_{\boldsymbol{n},\boldsymbol{m}}$ of the steering matrix $\mathrm{S}_{\mathrm{U}}$ in the following compact form (Section III-E of~\cite{ChU2012}): 
\begin{equation}\label{eq:steeringMatrix3D}
s_{\boldsymbol{n},\boldsymbol{m}} = \sqrt{\frac{\boldsymbol{m}!}{\boldsymbol{n}!}}
\sum_{\mid \boldsymbol{v}_1 \mid = n_1}{\sum_{\mid \boldsymbol{v}_2 \mid = n_2}{\sum_{\mid \boldsymbol{v}_3 \mid = n_3}{
\delta_{\boldsymbol{v}_1 + \boldsymbol{v}_2 + \boldsymbol{v}_3,\boldsymbol{m}}}}}
\cdot \frac{\boldsymbol{n}!}{\boldsymbol{v}_1!\boldsymbol{v}_2!\boldsymbol{v}_3!}
\boldsymbol{u}_1^{\boldsymbol{v}_1}\boldsymbol{u}_2^{\boldsymbol{v}_2}\boldsymbol{u}_3^{\boldsymbol{v}_3}, \nonumber
\end{equation}
where $\boldsymbol{v}_i \in \mathbb{N}^3$ and $\delta_{\boldsymbol{v}_1 + \boldsymbol{v}_2 + \boldsymbol{v}_3,\boldsymbol{m}}$ is the Kronecker symbol used to exclude the summation terms with $\boldsymbol{v}_1 + \boldsymbol{v}_2 + \boldsymbol{v}_3 \neq \boldsymbol{m}$.

\begin{figure}
\centering
\includegraphics[trim = 0 0 0 0, clip, width=\linewidth]{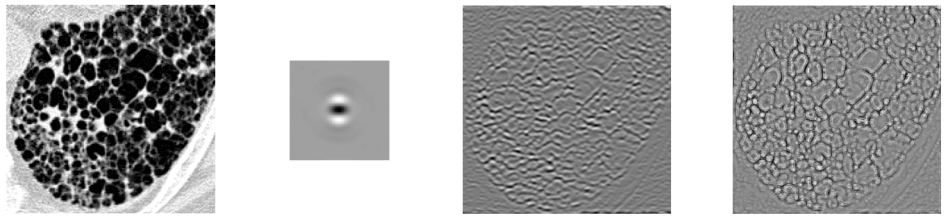}\\
$f[\boldsymbol{k}]$
\hspace{2cm}
$\mathcal{R}^{(0,2)}
\{g_{\text{sim}}\}[\boldsymbol{k}]$
\hspace{1.8cm}
$h[\boldsymbol{k}]$
\hspace{2.8cm}
$h_{\text{aligned}}[\boldsymbol{k}]$
\caption{Example of image filtering in 2$D$ with the Riesz transform.
$h_{\text{aligned}}[\boldsymbol{k}]$ was obtained by aligning the responses of $\mathcal{R}^{(0,2)}
\{g_{\text{sim}}\}[\boldsymbol{k}]$ based on the structure tensor and allows highlighting of collagen fibers at any orientation.}
  \label{fig:RieszExample}
\end{figure}

Applying~\eqref{eq:steerability} at every image position will align the entire filterbank and yield aligned response maps $h_{\text{aligned}}[\boldsymbol{k}]$ allowing both directional and rotation-invariant image analysis.
An example of image filtering in 2$D$ with the Riesz kernel $\mathcal{R}^{(0,2)}
\{g_{\text{sim}}\}[\boldsymbol{k}]$ is shown in Fig.~\ref{fig:RieszExample}.
It is worth noting that although being slightly sub-optimal from a computational aspect, bypassing steerability is possible. In that case, one need to rotate the Riesz kernels using $\mathrm{U}$ and interpolation, and then to (re-)convolve the image with the rotated kernels to obtain the aligned response maps $h_{\text{aligned}}[\boldsymbol{k}]$.

To summarise, aligned Riesz filters allow directional and rotation-invariant image analysis that can characterise interpretable transitions in medical images.
Rotationally-invariant steerable representations can also be learned to be specific to the problem at hand~\cite{Depeursinge2017,AFO2019}, also with CNNs~\cite{AFO2019,AFO2019b,WHS2018,Winkels2019,Weiler2018,bekkers2018roto}.
Implementations of the Riesz transform are available, e.g. in~\cite{Her2016}$^($\footnote{\texttt{\url{https://pypi.org/project/itk-isotropicwavelets/}}, as of July 2019.}$^)$,~\cite{ChU2012}$^($\footnote{\texttt{\url{http://bigwww.epfl.ch/demo/steerable-wavelets-3d/}}, as of July 2019.}$^)$ and its adaptation for texture feature extraction, including filter alignment, in~\cite{DMP2017}$^($\footnote{\texttt{\url{http://publications.hevs.ch/index.php/publications/show/2035/}}, as of July 2019.}$^)$.

\section{Qualitative Comparison of Linear Filter Properties}\label{sec:qualitativeComparison}
Based on the quantitative comparison criteria introduced in Section~\ref{sec:linearFilterProperties}, it appears that not all approaches can fulfill all properties that are desirable for local medical image analysis.
One typically observes a trade off between implementation simplicity and efficiency versus filter specificity and invariance/equivariance to geometric transforms~\cite{Depeursinge2018}.
Table~\ref{tab:quantitativeComparison} provides a qualitative comparison of the filtering methods detailed in Section~\ref{sec:filtersDescription}.

While approximated rotation invariance can be artificially added to all methods by computing the average of directional response maps, this often annihilates directional sensitivity (see Section~\ref{sec:combiningDSandRotInv}).
Other designs allow combining rotation invariance with directional sensitivity (see Section~\ref{sec:combiningDSandRotInv}).
%For instance, max pooling over uniformly sampled filter orientations or rotational invariants via spherical harmonics~\cite{Depeursinge2018} can be used to successfully combine the two properties.

\begin{table}\scriptsize
\caption{Qualitative comparison of common linear filtering approaches. ``yes/no" indicates filters that can be but are not necessarily designed as wavelets.}
\label{tab:quantitativeComparison}
\begin{center}
\begin{tabular}{c|cccc|}
 & \begin{minipage}{50pt}\centering\vspace{3pt} directional sensitivity \vspace{3pt}\end{minipage} & \begin{minipage}{60pt}\centering\vspace{3pt} local rotation invariance \vspace{3pt}\end{minipage} & \begin{minipage}{90pt}\centering\vspace{3pt} separability of the convolution\vspace{3pt}\end{minipage} & wavelet \\
\hline
LoG & no & yes & no (yes with DoG) & yes/no \\
\begin{minipage}{50pt}\centering\vspace{3pt} Laws kernels \vspace{3pt}\end{minipage} & yes & no & yes & no \\
\begin{minipage}{50pt}\centering\vspace{3pt} Gabor \vspace{3pt}\end{minipage} & yes & no & no & yes/no \\
\begin{minipage}{50pt}\centering\vspace{3pt} separable wavelets \vspace{3pt}\end{minipage} & yes & no & yes & yes\\
\begin{minipage}{50pt}\centering\vspace{3pt} isotropic non-separable wavelets \vspace{3pt}\end{minipage} & no & yes & no & yes \\
\begin{minipage}{50pt}\centering\vspace{3pt} aligned wavelets \hbox{(\textit{e.g.} Riesz)} \vspace{3pt}\end{minipage} & yes & yes & no & yes \\
\hline
\end{tabular}
\end{center}
\end{table}
\section{Interpolation and convolutional filtering}\label{sec:interpolationAndFiltering}

When applied to an image, displacements of one voxel may not correspond to a displacement of the same physical distance depending on the direction. 
Physical distance is defined by voxel spacing which, in 3$D$ imaging, commonly differs between directions and images. This case is referred to as ``anisotropic'' image resolution. For example, in-plane voxel spacings in computed tomography (CT) images are usually smaller than the slice thickness. Also, voxel spacing can differ between acquisition protocols. Thus, simply applying filters to all images generates response maps that cannot be directly compared, because the frequency response of filters is different. To avoid this problem, we recommend performing filtering after resampling the image to uniform voxel spacing using interpolation (see Fig.~\ref{fig:imageProcessing}).
%%% RECOMMENDATION

\begin{figure}
\centering
   \begin{minipage}[b]{140pt}
     \centering
     \includegraphics[trim = 0 0 0 0, clip, scale=0.38]{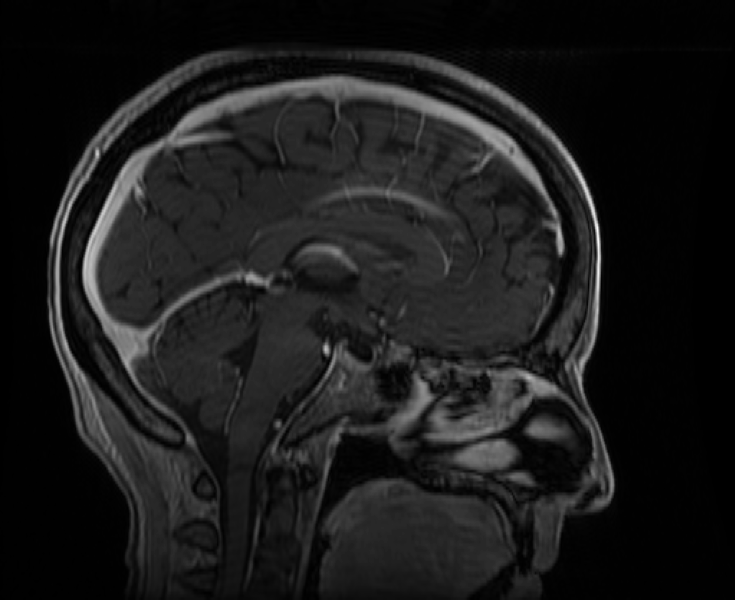}
     \subcaption{original, 1 mm}\label{fig:radiomics_filter_1mm_original}
     \hspace{100pt}
   \end{minipage}
   \begin{minipage}[b]{140pt}
     \centering
     \includegraphics[trim = 0 0 0 0, clip, scale=0.38]{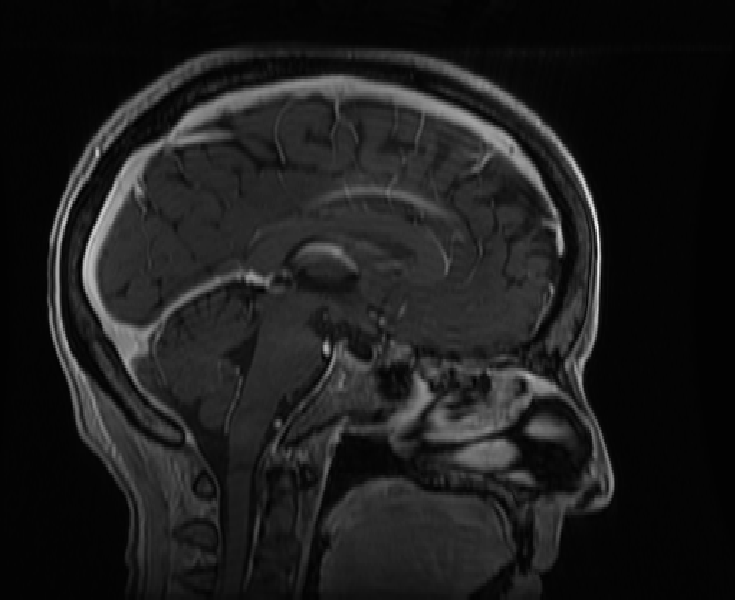}
     \subcaption{low-pass, 1 mm}\label{fig:radiomics_filter_1mm_low_pass}
     \hspace{100pt}
   \end{minipage}
   \begin{minipage}[b]{140pt}
     \centering
     \includegraphics[trim = 0 0 0 0, clip, scale=0.38]{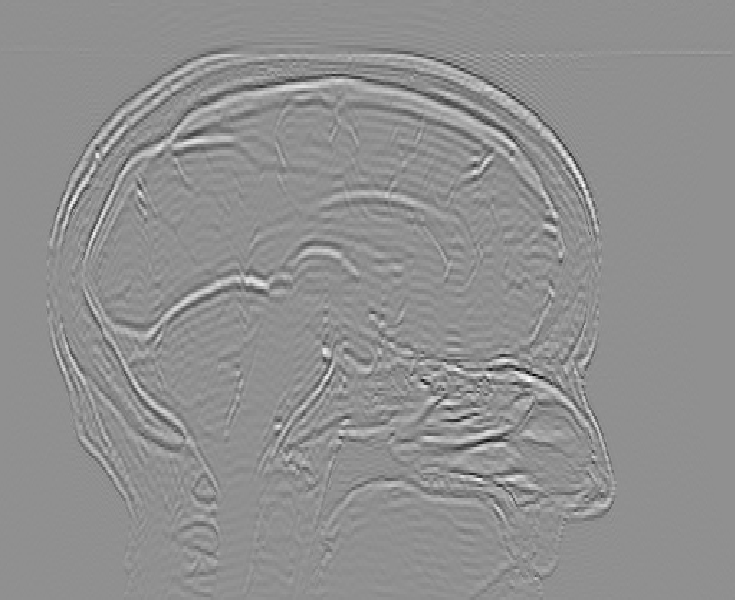}
     \subcaption{high-pass, 1 mm}\label{fig:radiomics_filter_1mm_high_pass}
     \hspace{100pt}
   \end{minipage}
   \begin{minipage}[b]{140pt}
     \centering
     \includegraphics[trim = 0 0 0 0, clip, scale=0.38]{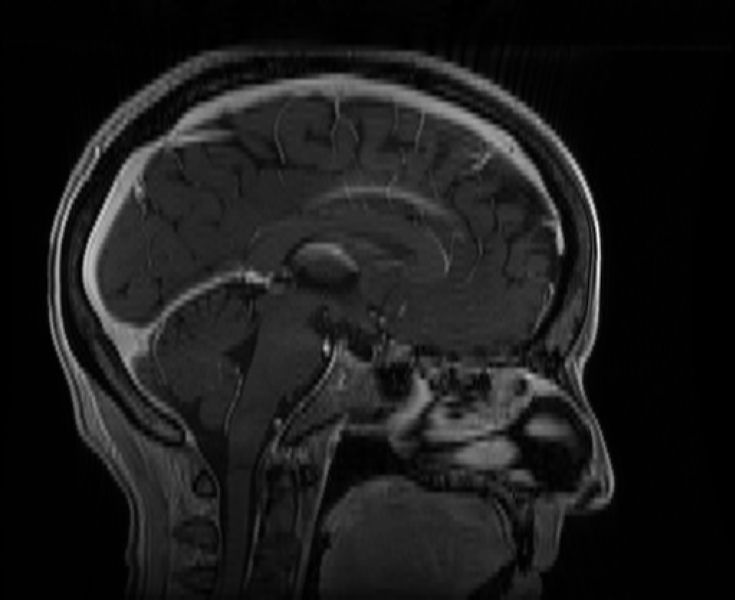}
     \subcaption{original, 2 mm}\label{fig:radiomics_filter_2mm_original}
     \hspace{100pt}
   \end{minipage}
   \begin{minipage}[b]{140pt}
     \centering
     \includegraphics[trim = 0 0 0 0, clip, scale=0.38]{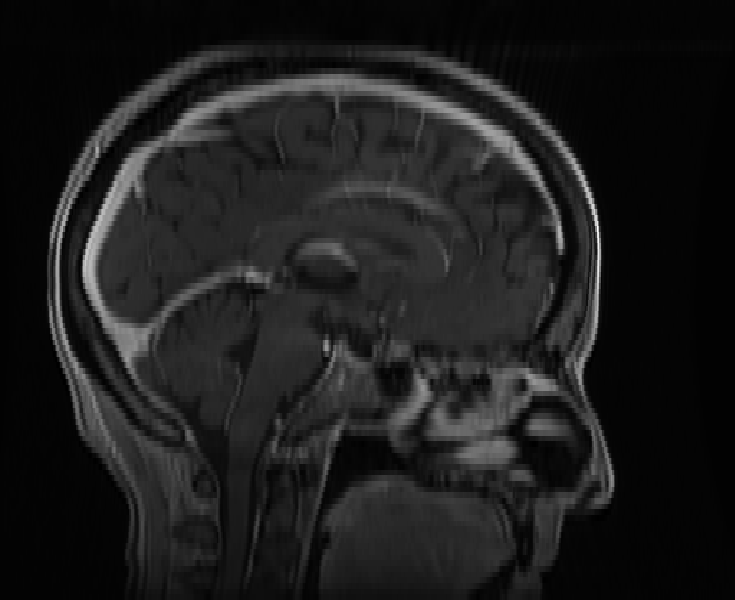}
     \subcaption{low-pass, 2 mm}\label{fig:radiomics_filter_2mm_low_pass}
     \hspace{100pt}
   \end{minipage}
   \begin{minipage}[b]{140pt}
     \centering
     \includegraphics[trim = 0 0 0 0, clip, scale=0.38]{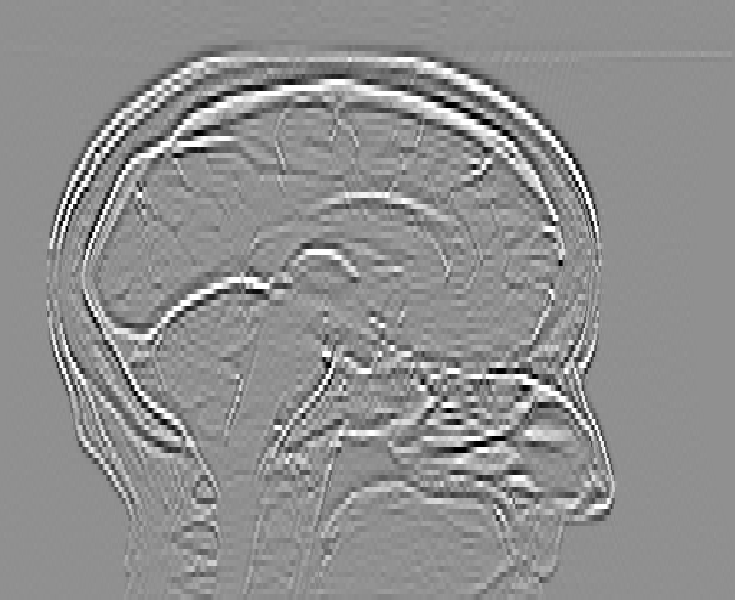}
     \subcaption{high-pass, 2 mm}\label{fig:radiomics_filter_2mm_high_pass}
     \hspace{100pt}
   \end{minipage}
   \begin{minipage}[b]{140pt}
     \centering
     \includegraphics[trim = 0 0 0 0, clip, scale=0.38]{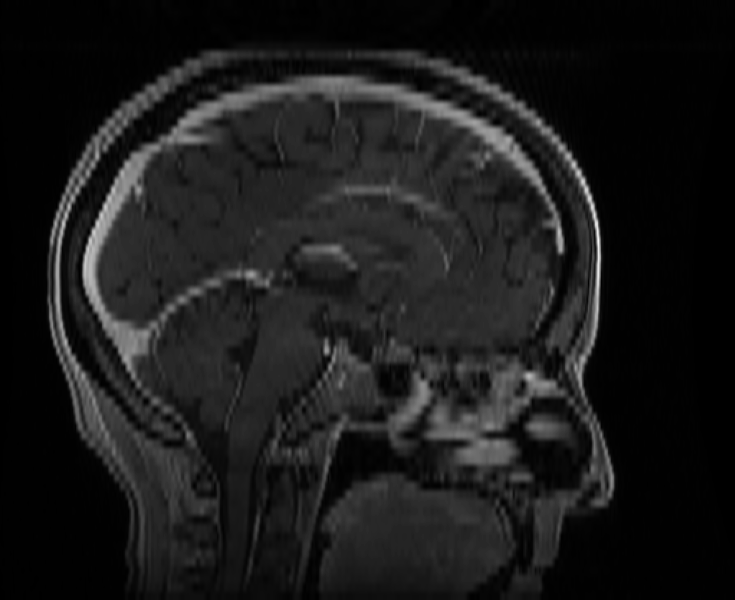}
     \subcaption{original, 3 mm}\label{fig:radiomics_filter_3mm_original}
     \hspace{100pt}
   \end{minipage}
   \begin{minipage}[b]{140pt}
     \centering
     \includegraphics[trim = 0 0 0 0, clip, scale=0.38]{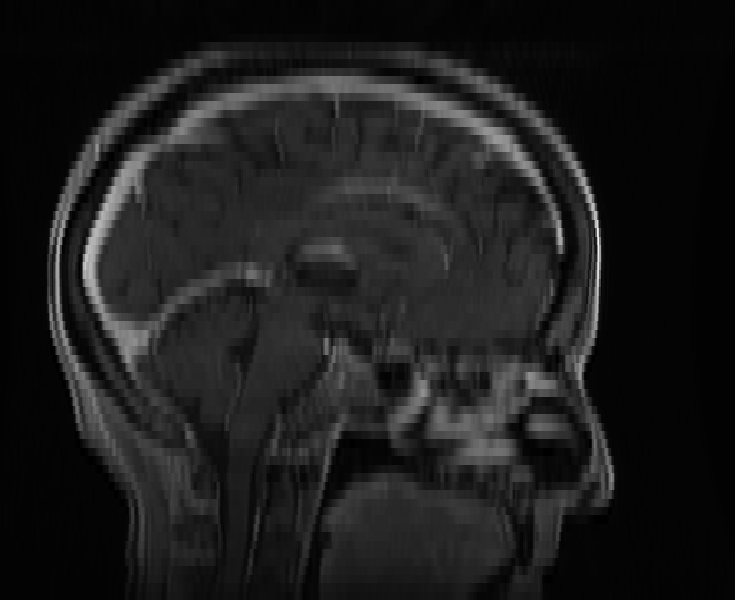}
     \subcaption{low-pass, 3 mm}\label{fig:radiomics_filter_3mm_low_pass}
     \hspace{100pt}
   \end{minipage}
   \begin{minipage}[b]{140pt}
     \centering
     \includegraphics[trim = 0 0 0 0, clip, scale=0.38]{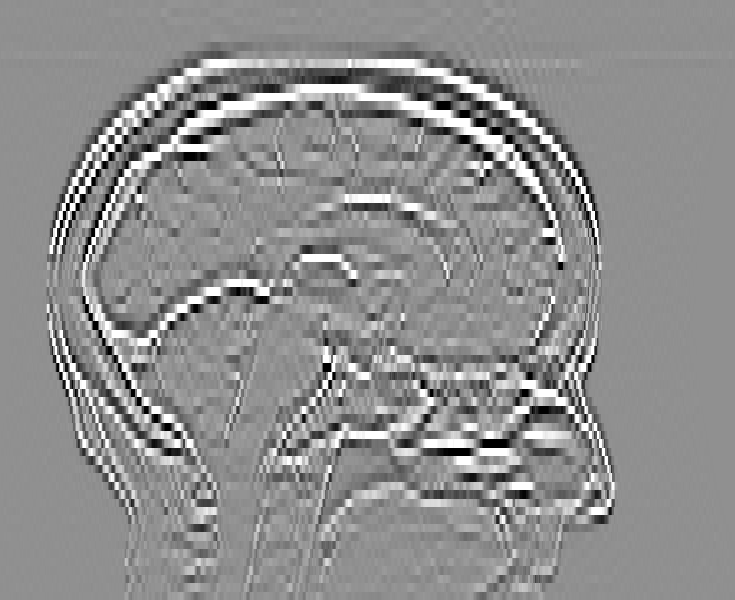}
     \subcaption{high-pass, 3 mm}\label{fig:radiomics_filter_3mm_high_pass}
     \hspace{100pt}
   \end{minipage}
   \begin{minipage}[b]{140pt}
     \centering
     \includegraphics[trim = 0 0 0 0, clip, scale=0.38]{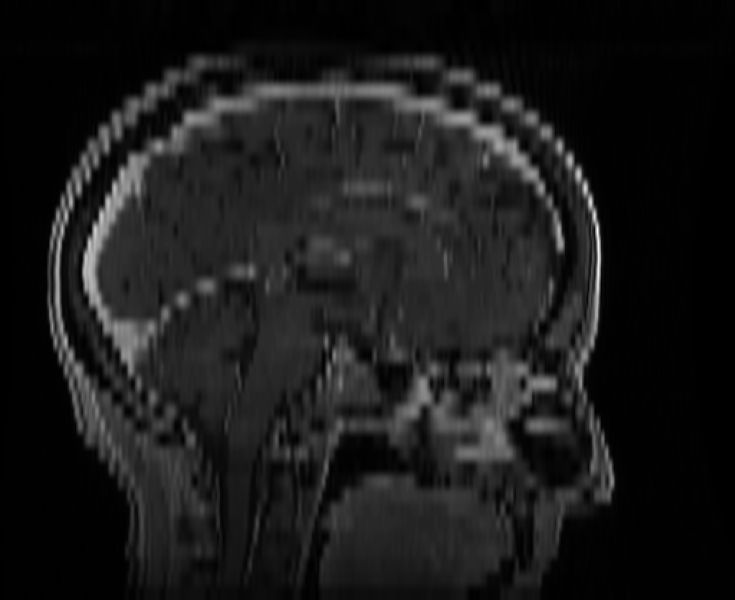}
     \subcaption{original, 5 mm}\label{fig:radiomics_filter_5mm_original}
     \hspace{100pt}
   \end{minipage}
   \begin{minipage}[b]{140pt}
     \centering
     \includegraphics[trim = 0 0 0 0, clip, scale=0.38]{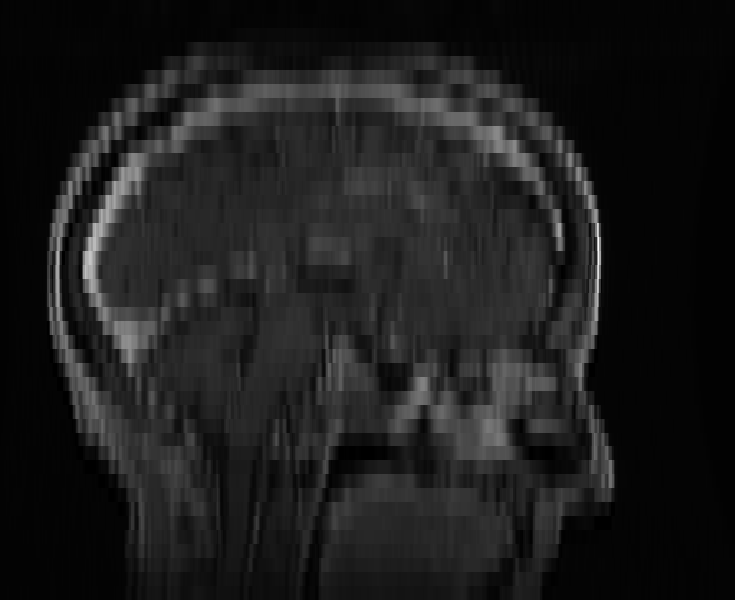}
     \subcaption{low-pass, 5 mm}\label{fig:radiomics_filter_5mm_low_pass}
     \hspace{100pt}
   \end{minipage}
   \begin{minipage}[b]{140pt}
     \centering
     \includegraphics[trim = 0 0 0 0, clip, scale=0.38]{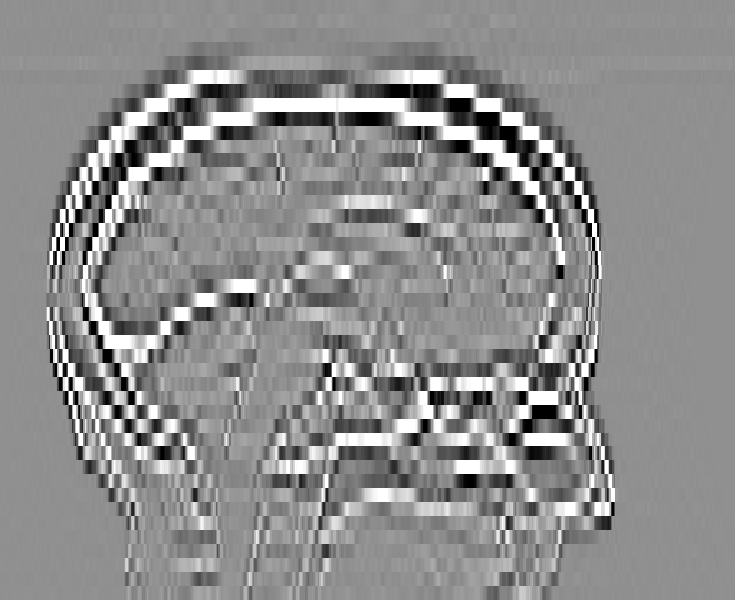}
     \subcaption{high-pass, 5 mm}\label{fig:radiomics_filter_5mm_high_pass}
     \hspace{100pt}
   \end{minipage}
  \caption{Comparison of the low- and high-pass filters (LLL and HHH, respectively) of the Daubechies 2 wavelet for different slice spacing, without interpolation prior to filtering. All original images, all low-pass and all high-pass response maps share the same intensity scale. Notice that the high-pass filter response visibly depends on the slice spacing. This effect is also present in the low-pass image, but less so. The T1-weighted spoiled gradient echo image dataset was generated by the National Cancer Institute Clinical Proteomic Tumor Analysis Consortium (CPTAC)~\cite{CPTAC2018-ot,Clark2013-sv}.
  }
  \label{fig:radiomics_filter_before_interpolation}
\end{figure}

\begin{figure}
\centering
   \begin{minipage}[b]{150pt}
     \centering
     \includegraphics[trim = 0 0 0 0, clip, scale=0.4]{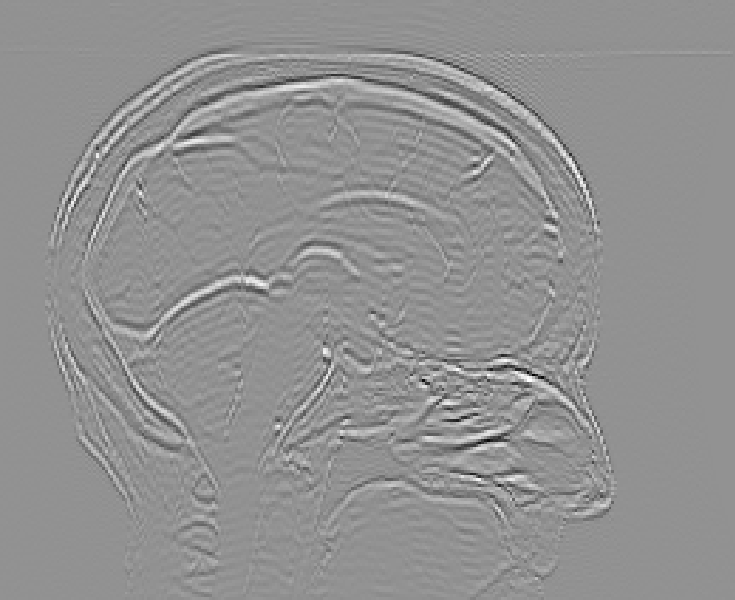}
     \subcaption{linear, 1 mm spacing}\label{fig:radiomics_linear_1mm_high_pass}
     \hspace{100pt}
   \end{minipage}
   \begin{minipage}[b]{150pt}
     \centering
     \includegraphics[trim = 0 0 0 0, clip, scale=0.4]{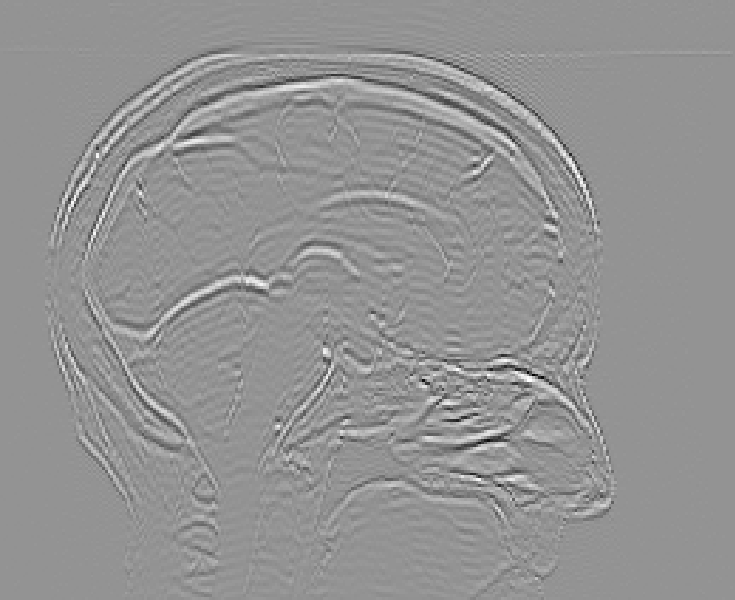}
     \subcaption{cubic spline, 1 mm spacing}\label{fig:radiomics_spline_1mm_high_pass}
     \hspace{100pt}
   \end{minipage}
   \begin{minipage}[b]{150pt}
     \centering
     \includegraphics[trim = 0 0 0 0, clip, scale=0.4]{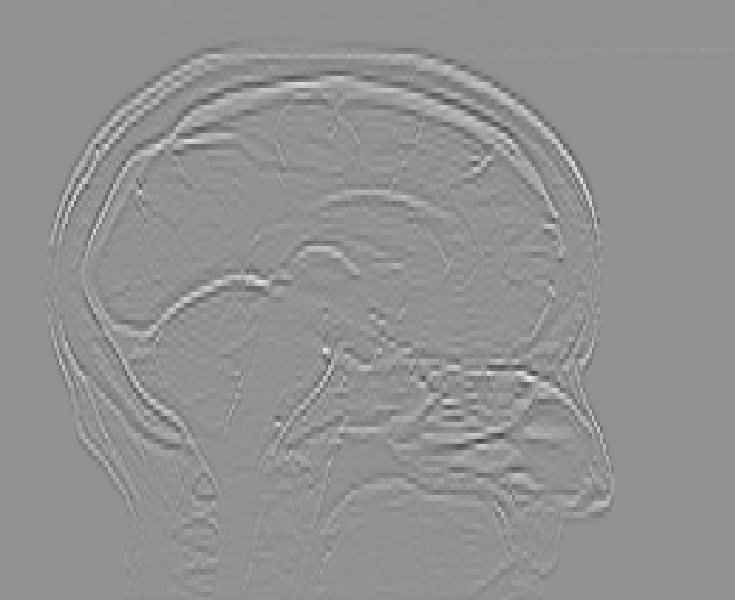}
     \subcaption{linear, 2 mm spacing}\label{fig:radiomics_linear_2mm_high_pass}
     \hspace{100pt}
   \end{minipage}
   \begin{minipage}[b]{150pt}
     \centering
     \includegraphics[trim = 0 0 0 0, clip, scale=0.4]{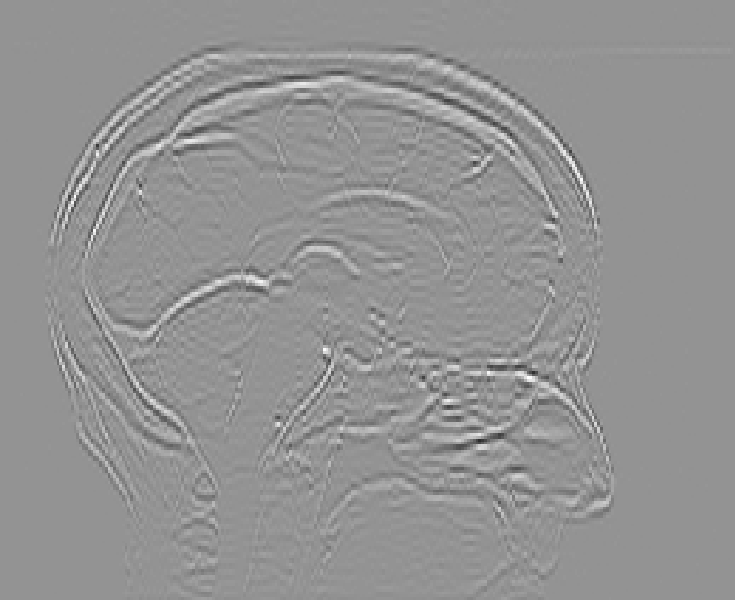}
     \subcaption{cubic spline, 2 mm spacing}\label{fig:radiomics_spline_2mm_high_pass}
     \hspace{100pt}
   \end{minipage}
   \begin{minipage}[b]{150pt}
     \centering
     \includegraphics[trim = 0 0 0 0, clip, scale=0.4]{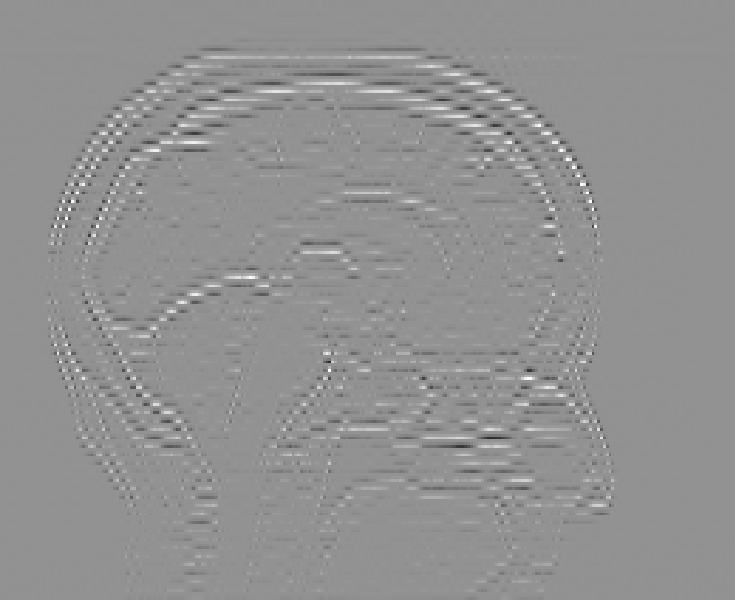}
     \subcaption{linear, 3 mm spacing}\label{fig:radiomics_linear_3mm_high_pass}
     \hspace{100pt}
   \end{minipage}
   \begin{minipage}[b]{150pt}
     \centering
     \includegraphics[trim = 0 0 0 0, clip, scale=0.4]{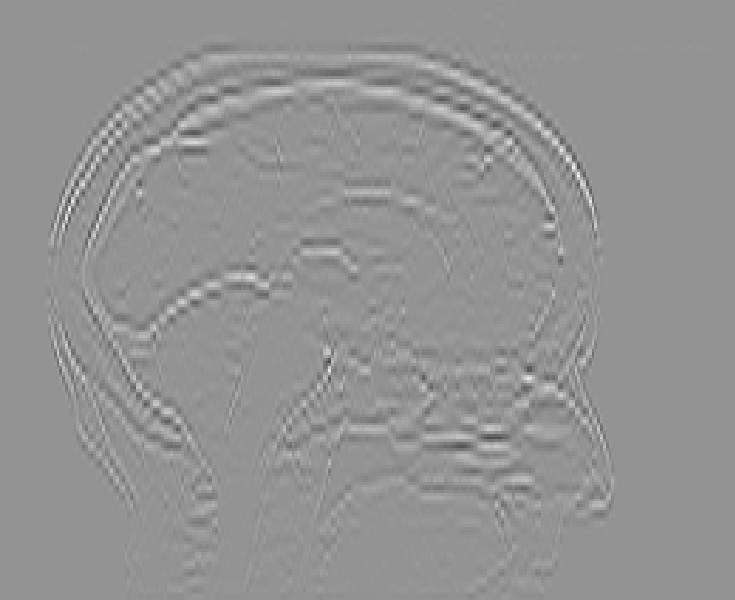}
     \subcaption{cubic spline, 3 mm spacing}\label{fig:radiomics_spline_3mm_high_pass}
     \hspace{100pt}
   \end{minipage}
      \begin{minipage}[b]{150pt}
     \centering
     \includegraphics[trim = 0 0 0 0, clip, scale=0.4]{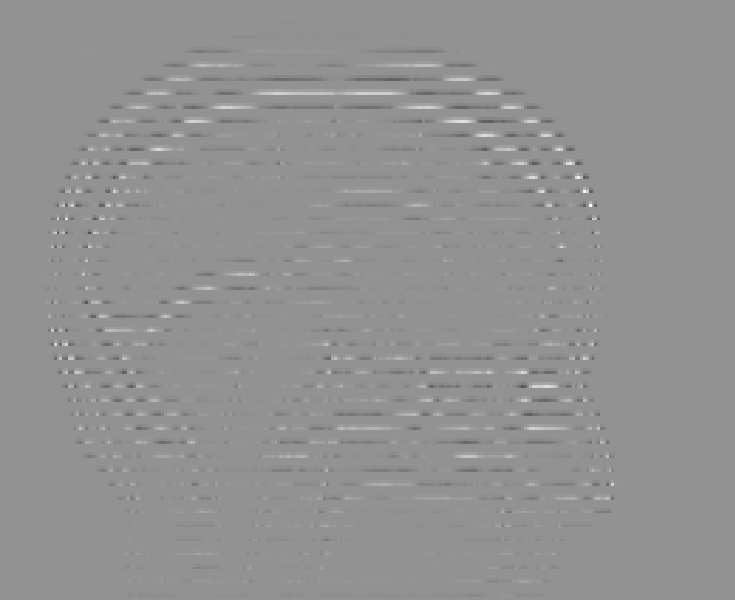}
     \subcaption{linear, 5 mm spacing}\label{fig:radiomics_linear_5mm_high_pass}
     \hspace{100pt}
   \end{minipage}
   \begin{minipage}[b]{150pt}
     \centering
     \includegraphics[trim = 0 0 0 0, clip, scale=0.4]{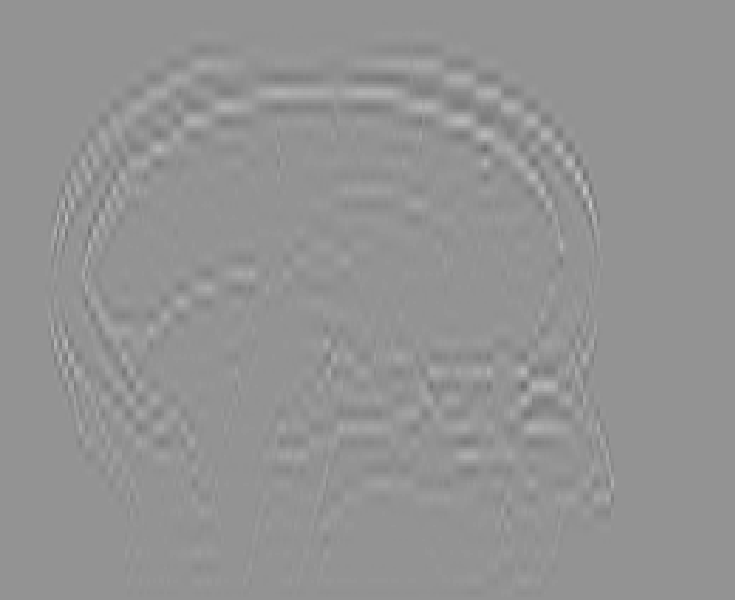}
     \subcaption{cubic spline, 5 mm spacing}\label{fig:radiomics_spline_5mm_high_pass}
     \hspace{100pt}
   \end{minipage}
  \caption{Comparison of response maps of the Daubechies 2 wavelet high-pass filter (HHH) for images with a different slice spacing that are resampled to 1 mm isotropic voxels using trilinear or tricubic spline interpolation prior to filtering. Images with the same interpolation method have the same intensity scale. Unlike in Figure \ref{fig:radiomics_filter_before_interpolation}, image intensities remain similar, though the intensity range degrades with greater slice spacing. The response maps of 3 mm and 5 mm images obtained after trilinear interpolation show clear intermittent gaps between locations of the original slice. Within these slices, linear interpolation acts as a low-pass filter that suppresses high-frequency content. The cubic spline interpolation does not suffer from this problem. The T1-weighted spoiled gradient echo image dataset (\(0.5 \times 0.5 \times 1.0\) mm) was generated by the National Cancer Institute Clinical Proteomic Tumor Analysis Consortium (CPTAC)~\cite{CPTAC2018-ot,Clark2013-sv}.
  }
  \label{fig:radiomics_filter_after_interpolation}
\end{figure}

However, performing image interpolation prior to filtering yields its own issues (see Fig.~\ref{fig:radiomics_filter_before_interpolation}). Interpolation itself, by definition, alters the frequency content of the image, which may be particularly visible in response maps created by  high-pass filters (see Fig.~\ref{fig:radiomics_filter_after_interpolation}). As can be observed from the figure, trilinear interpolation acts as a low-pass filter when upsampling, whereas the tricubic spline method keeps preserves more high-frequency content. Ideal interpolation methods for upsampling do not exist as it is impossible to completely infer missing data. While we expect that superresolution based on deep learning could be used to improve interpolation results in the future~\cite{Dong2014}, in the meantime features derived from response maps created by high-pass filters may lack robustness.

% "interpolating" the filter itself
Alternatively, when an analytical expression of the filter is available, the latter is defined in the continuous domain, for example for a Gaussian filter. This provides the opportunity to (re)compute filter values for anisotropic voxel grids as well as inter-image/patient differences in image resolution and obviates interpolation. However, for clarity, we will not consider this approach in the benchmarking process (Section \ref{sec:benchmarking}).

% Boundary conditions
For most applications, boundary conditions are not critical as ROIs do not tend to be located at an image boundary. Thus, for many medical images, the nearest value boundary conditions may be used. If the ROI is close to a boundary, mirror boundary conditions may be preferable (see Section~\ref{sec:boundaryConditions}).

% Feature computation
In general, image features, or a subset of features, may be calculated from response maps in the same way as for the original image. Image filtering also affects the discretisation method that can be used prior to computing, for example texture features. Intensities in most response maps no longer have a direct physical meaning. Hence, \textit{fixed bin size} methods can no longer be used for response maps. \textit{Fixed bin number} or similar discretisation methods should be used instead~\cite{ZLV2017}.

\chapter{Reporting guidelines and detailed reporting}\label{sec:guidelinesDetails}

The section describes detailed reporting guidelines for radiomics analyses, with an added focus on the application of convolutional filters.

\section{Reporting guidelines}\label{sec_reporting_guidelines}

The IBSI previously produced and released a reporting guideline \cite{Zwanenburg2020-jt}. Below are the updated and revised reporting guidelines. These guidelines are technical in nature, to allow for complete and detailed reporting. For more general reporting guidelines, see e.g. the CLAIM \cite{Mongan2020-rn} checklist and the TRIPOD \cite{Collins2015-lx} statement. Topic-specific guidelines may also be helpful, such as the joint EANM/SNMMI guideline on radiomics \cite{Hatt2022-uo} or the RELAINCE guideline \cite{Jha2022-jw}.

\newcounter{itemcounter}
\newcounter{subitemcounter}[itemcounter]

\newcommand\stepitemcounter{\stepcounter{itemcounter}\theitemcounter}
\newcommand\startsubitemcounter{\stepcounter{itemcounter}\stepcounter{subitemcounter}\theitemcounter\alph{subitemcounter}}
\newcommand\stepsubitemcounter{\stepcounter{subitemcounter}\theitemcounter\alph{subitemcounter}}

\footnotesize
\begin{longtable}{p{3.5cm}ccp{7cm}}

\toprule
\textbf{topic} & & \textbf{item} & \textbf{description}\\
\midrule
\endhead

\bottomrule
\multicolumn{4}{r}{\textit{continued on next page}}
\endfoot

\bottomrule
\caption[Reporting guidelines]{Guidelines for reporting on radiomics studies. Not all items may be applicable.} \label{table_guidelines}
\endlastfoot

\multicolumn{4}{l}{\textbf{Patient}}\\
\midrule
Region of interest\footnote{Also referred to as volume of interest.} & &
\stepitemcounter & Describe the region of interest that is being imaged.\\
Patient preparation & & \startsubitemcounter & Describe specific instructions given to patients prior to image acquisition, e.g. fasting prior to imaging.\\
& & \stepsubitemcounter & Describe administration of drugs to the patient prior to image acquisition, e.g. muscle relaxants. \\
& & \stepsubitemcounter & Describe the use of specific equipment for patient comfort during scanning, e.g. ear plugs. \\
Radioactive tracer & PET, SPECT & \startsubitemcounter & Describe which radioactive tracer was administered to the patient, e.g. 18F-FDG. \\
& PET, SPECT & \stepsubitemcounter & Describe the administration method. \\
& PET, SPECT & \stepsubitemcounter & Describe the injected activity of the radioactive tracer at administration. \\
& PET, SPECT & \stepsubitemcounter & Describe the uptake time prior to image acquisition.\\
& PET, SPECT & \stepsubitemcounter & Describe how competing substance levels were controlled.\footnote{An example is glucose present in the blood which competes with the uptake of 18F-FDG tracer in tumour tissue. To reduce competition with the tracer, patients are usually asked to fast for several hours and a blood glucose measurement may be conducted prior to tracer administration.} \\
Contrast agent & & \startsubitemcounter & Describe which contrast agent was administered to the patient. \\
& & \stepsubitemcounter & Describe the administration method. \\
& & \stepsubitemcounter & Describe the injected quantity of contrast agent.\\
& & \stepsubitemcounter & Describe the uptake time prior to image acquisition.\\
& & \stepsubitemcounter & Describe how competing substance levels were controlled.\\
Comorbidities & & \stepitemcounter & Describe if the patients have comorbidities that affect imaging.\footnote{An example of a comorbidity that may affect image quality in 18F-FDG PET scans are type I and type II diabetes melitus, as well as kidney failure.} \\
\multicolumn{4}{l}{\textbf{Acquisition}\footnote{Many acquisition parameters may be extracted from DICOM header meta-data, or calculated from them.}}\\
\midrule
Acquisition protocol & & \stepitemcounter & Describe whether a standard imaging protocol was used, and where its description may be found.\\
Scanner type & & \stepitemcounter & Describe the scanner type(s) and vendor(s) used in the study. \\
Imaging modality & & \stepitemcounter & Clearly state the imaging modality that was used in the study, e.g. CT, MRI. \\
Scanner calibration & & \stepitemcounter & Describe how and when the scanner was calibrated. \\
Acquisition type & & \startsubitemcounter & State if the scans were static, dynamic or gated, if this could be unclear.\\
& dynamic & \stepsubitemcounter & Describe the acquisition time per time frame.\\
& dynamic & \stepsubitemcounter & Describe any temporal modelling technique that was used.\\
& gated & \stepsubitemcounter & Describe what signal is used for gated acquisition.\\
Scan duration & & \stepitemcounter & Describe the duration of the complete scan or the time per bed position. \\
Patient instructions & & \stepitemcounter & Describe specific instructions given to the patient during acquisition, e.g. breath holding.\\
Anatomical motion correction & & \stepitemcounter & Describe the method used to minimise the effect of anatomical motion. \\
Tube voltage & CT & \stepitemcounter & Describe the peak kilo voltage output of the X-ray source.\\
Tube current & CT & \startsubitemcounter & Describe the tube current in mA.\\
& CT & \stepsubitemcounter & Describe the technique used to beam modulate intensity, if any.\\
& CT & \stepsubitemcounter & Describe the average exposure in the region of interest.\\
Spectral CT technique & spectral CT & \stepitemcounter & Describe the technique used to acquire spectral CT imaging.\\
Time-of-flight & PET & \startsubitemcounter & State if scanner time-of-flight capabilities are used during acquisition. \\
& PET & \stepsubitemcounter & Describe the temporal resolution of the scanner.\\
Collimator & SPECT & \stepitemcounter & Describe the type of collimator used on the detector\\
Magnetic field strength & MRI & \stepitemcounter & Describe the nominal strength of the main magnetic field, e.g 1.5T.\\
RF coil & MRI & \stepitemcounter & Describe what kind RF coil used for acquisition, e.g. body coil, incl. vendor. \\
Acquisition type & MRI & \stepitemcounter & Describe the acquisition type of the MRI scan, e.g. 2D, 3D, parallel.\\
Scanning sequence & MRI & \startsubitemcounter & Describe which scanning sequence was acquired.\\
& MRI & \stepsubitemcounter & Describe which sequence variant was acquired.\\
& MRI & \stepsubitemcounter & Describe which scan options apply to the current sequence, e.g. flow compensation, cardiac gating, etc.\\
& MRI & \stepsubitemcounter & Describe the repetition time in ms between subsequent pulse sequences.\\
& MRI & \stepsubitemcounter & Describe the echo time in ms.\\
& MRI & \stepsubitemcounter & Describe the inversion time in ms between the middle of the inverting RF pulse to the middle of the excitation pulse, if applicable.\\
& MRI & \stepsubitemcounter & Describe the flip angle produced by the RF pulses, if applicable.\\
& MRI & \stepsubitemcounter & Describe the acquisition trajectory of the k-space, if other than linear.\\
& MRI & \stepsubitemcounter & Describe the number of times each point in k-space is sampled, if more than once.\\
& MRI & \stepsubitemcounter & Describe the number of lines or traversals in k-space that are acquired per RF excitation pulse, if more than 1.\\
\multicolumn{4}{l}{\textbf{Reconstruction}\footnote{Many reconstruction parameters may be extracted from DICOM header meta-data.}}\\
\midrule
Resolution & & \startsubitemcounter & Describe the distance between pixels/voxels in the axial plane, or alternatively the field of view and matrix size.\\
& & \stepsubitemcounter & Describe the distance between pixels/voxels along the axial direction, e.g. the slice spacing.\\
CT reconstruction & CT & \startsubitemcounter & Describe the convolution kernel used to reconstruct the image.\\
& CT & \stepsubitemcounter & Describe settings pertaining to iterative reconstruction algorithms.\\
& CT & \stepsubitemcounter & Describe artefact-reduction techniques used for e.g. beam hardening artefacts, metal artefacts, partial volume effects.\\
PET/SPECT reconstruction & PET, SPECT & \startsubitemcounter & Describe which reconstruction method was used, e.g. 3D OSEM.\\
& PET, SPECT & \stepsubitemcounter & Describe the number of iterations and subsets for iterative reconstruction.\\
& PET, SPECT & \stepsubitemcounter & Describe if and how point-spread function modelling was performed.\\
PET/SPECT image corrections & PET, SPECT & \startsubitemcounter & Describe if and how attenuation correction was performed.\\
& PET, SPECT & \stepsubitemcounter & Describe if and how scatter correction was performed.\\
& PET, SPECT & \stepsubitemcounter & Describe if and how other forms of correction were performed, e.g. randoms correction, dead time correction etc.\\
MRI reconstruction method & MRI & \startsubitemcounter & Describe the reconstruction method used to reconstruct the image from the k-space information, if not obvious.\\
& MRI & \stepsubitemcounter & Describe any artifact suppression methods used during reconstruction to suppress artifacts due to undersampling of k-space.\\
Diffusion-weighted imaging & DWI-MRI & \stepitemcounter & Describe the b-values used for diffusion-weighting. \\
\multicolumn{4}{l}{\textbf{Image registration}}\\
\midrule
Registration method & & \stepitemcounter & Describe the method used to register multi-modality imaging. \\
\multicolumn{4}{l}{\textbf{Image processing - data conversion}} \\
\midrule
SUV normalisation & PET & \stepitemcounter & Describe which standardised uptake value (SUV) normalisation method is used.\\
ADC computation & DWI-MRI & \stepitemcounter & Describe how apparent diffusion coefficient (ADC) values were calculated.\\
Other data conversions & & \stepitemcounter & Describe any other conversions that are performed to generate e.g. perfusion maps.\\
\multicolumn{4}{l}{\textbf{Image processing - post-acquisition processing}} \\
\midrule
Anti-aliasing & & \stepitemcounter & Describe the method used to deal with anti-aliasing when down-sampling during interpolation.\\
Noise suppression & & \stepitemcounter & Describe methods used to reduce image noise.\\
Post-reconstruction smoothing filter & PET & \stepitemcounter & Describe the width of the Gaussian filter (FWHM) to spatially smooth intensities.\\
Skull stripping & MRI (brain) & \stepitemcounter & Describe method used to perform skull stripping.\\
Non-uniformity correction\footnote{Also known as bias-field correction.} & MRI & \stepitemcounter & Describe the method and settings used to perform non-uniformity correction.\\
Intensity normalisation & & \stepitemcounter & Describe the method and respective parameters used to normalise intensity distributions within a patient or patient cohort, if applicable.\\
Augmentation/perturbation methods & & \stepitemcounter & Describe method and respective parameters used to augment and/or perturb images and ROI masks.\\
Other post-acquisition processing methods & & \stepitemcounter & Describe any other methods that were used to process the image and are not mentioned separately in this list.\\
\multicolumn{4}{l}{\textbf{Segmentation}} \\
\midrule
Segmentation method & & \startsubitemcounter & Describe according to which guidelines regions of interest were segmented.\\
& & \stepsubitemcounter & Describe how regions of interest were segmented, e.g. manually.\\
& & \stepsubitemcounter & Describe any details concerning the region of interest that may not be obvious, e.g. the segmented gross tumour volume did not include grossly involved lymph nodes.\\
& & \stepsubitemcounter & Describe the number of experts, their expertise and consensus strategies for manual delineation.\\
& & \stepsubitemcounter & Describe methods and settings used for semi-automatic and fully automatic segmentation.\\
& & \stepsubitemcounter & Describe which image was used to define segmentation in case of multi-modality imaging.\\
Conversion to mask & & \stepitemcounter & Describe the method used to convert polygonal or mesh-based segmentations to a voxel-based mask.\\
\multicolumn{4}{l}{\textbf{Image processing - image interpolation}} \\
\midrule
Interpolation method & & \startsubitemcounter & Describe which interpolation algorithm was used to interpolate the image.\\
& & \stepsubitemcounter & Describe how the position of the interpolation grid was defined, e.g. align by center.\\
& & \stepsubitemcounter & Describe how the dimensions of the interpolation grid were defined, e.g. rounded to nearest integer.\\
& & \stepsubitemcounter & Describe how extrapolation beyond the original image was handled.\\
Voxel dimensions & & \stepitemcounter & Describe the size of the interpolated voxels.\\
Intensity rounding & CT & \stepitemcounter & Describe how fractional Hounsfield Units are rounded to integer values after interpolation.\\
\multicolumn{4}{l}{\textbf{Image processing - ROI interpolation}} \\
\midrule
Interpolation method & & \stepitemcounter & Describe which interpolation algorithm was used to interpolate the region of interest mask.\\
Partially masked voxels & & \stepitemcounter & Describe how partially masked voxels after interpolation are handled.\\
\multicolumn{4}{l}{\textbf{Image processing - re-segmentation}} \\
\midrule
Re-segmentation methods & & \stepitemcounter & Describe which methods and settings are used to re-segment the ROI intensity mask.\\
\multicolumn{4}{l}{\textbf{Image processing - discretisation}} \\
\midrule
Discretisation method\footnote{Discretisation may be performed separately to create intensity-volume histograms. If this is indeed the case, this should be described as well.} & & \startsubitemcounter & Describe the method used to discretise image intensities.\\
& & \stepsubitemcounter & Describe the number of bins (FBN) or the bin size (FBS) used for discretisation, if applicable.\\
& & \stepsubitemcounter & Describe the lowest intensity in the first bin for FBS discretisation.\footnote{This is typically set by range re-segmentation.}\\
\multicolumn{4}{l}{\textbf{Image processing - image transformation}} \\
\midrule
Image filters & & \startsubitemcounter & Describe which transformations are performed, if any.\\
& & \stepsubitemcounter & Describe if the image transformation was applied after interpolation or before. \footnote{Note that, generally, interpolation should be performed prior to transformation to ensure that spatial frequency response is the same for all relevant image directions. Only some filters allow for defining spatial response directly, e.g. Gaussian. In such cases, interpolation is not strictly required.}\\
& & \stepsubitemcounter & Describe transformation-specific parameters.\\
& & \stepsubitemcounter & Describe which features were computed from the response map.\\
IBSI compliance & & \stepitemcounter & State if the software used to perform the image transformation is able to reproduce the IBSI reference response maps and reference feature values for the relevant image filters.\\
\multicolumn{4}{l}{\textbf{Feature computation}} \\
\midrule
Feature set & & \stepitemcounter & Describe which hand-crafted features are computed, and refer to their definitions or provide these.\\
Texture parameters & & \startsubitemcounter & Define how texture-matrix based features were aggregated from underlying texture matrices.\\
& & \stepsubitemcounter & Define how CM, RLM, NGTDM and NGLDM weight distances, e.g. no weighting.\\
& & \stepsubitemcounter & Define whether symmetric or asymmetric co-occurrence matrices were computed.\\
& & \stepsubitemcounter & Define the (Chebyshev) distance at which co-occurrence of intensities is determined for co-occurrence matrices, e.g. 1.\\
& & \stepsubitemcounter & Define the distance and distance norm for which voxels with the same intensity are considered to belong to the same zone for the purpose of constructing an SZM and/or DZM, e.g. Chebyshev distance of 1.\\
& & \stepsubitemcounter & Define the distance norm for determining the distance of zones to the border of the ROI, e.g. Manhattan distance.\\
& & \stepsubitemcounter & Define the neighbourhood distance and distance norm for the NGTDM and/or NGLDM, e.g. Chebyshev distance of 1.\\
& & \stepsubitemcounter & Define the coarseness parameter for the NGLDM, e.g. 0.\\
IBSI compliance & & \stepitemcounter & State if the software used to extract the set of image biomarkers is able to reproduce the IBSI feature reference values.\\
\multicolumn{4}{l}{\textbf{Machine learning and radiomics analysis}} \\
\midrule
Diagnostic and prognostic modelling & & \stepitemcounter & See the TRIPOD guidelines for reporting on diagnostic and prognostic modelling.\\
Robustness & & \stepitemcounter & Describe how robustness of the features was assessed, e.g. test-retest analysis.\\
Comparison with known factors & & \stepitemcounter & Describe performance of radiomics models when compared with known (clinical) factors.\\
Multicollinearity & & \stepitemcounter & Describe multicollinearity between image features in the model signature.\\
Model availability & & \stepitemcounter & Describe where radiomics models with the necessary pre-processing information may be found. \\
Data availability & & \stepitemcounter & Describe where imaging data and relevant meta-data used in the study may be found.\\
Software & & \stepitemcounter & Describe which software and version was used to perform image processing, image filters, computation of hand-crafted features, and modelling.\\
\end{longtable}

\normalsize
\FloatBarrier

\section{Detailed reporting on convolutional filters}
This reference manual describes parameters for a number of convolutional filters.

\subsection{General parameters}
The parameters listed below apply to every filter.

\begin{table}[ht]
\centering
\small
\begin{tabular}{llc}
\toprule
\textbf{parameter} & \textbf{value} & \textbf{id}\\
\midrule
filter application & & \textid{4AGS}\\
& within the axial plane (2D) & \textid{7QSE}\\
& volumetric (3D) & \textid{0ODV}\\
boundary condition & & \textid{GBYQ}\\
& constant value (\#) & \textid{Z3VE}\\
& nearest value & \textid{SIJG}\\
& periodic & \textid{Z7YO}\\
& mirror & \textid{ZDTV}\\
\bottomrule
\end{tabular}
\caption{General filter parameters. Constant value padding requires the value to be specified.}
\end{table}

\FloatBarrier

\subsection[Mean filter]{Mean filter \id{S60F}}
The parameters listed below apply to the mean filter.

\begin{table}[ht]
\centering
\small
\begin{tabular}{llc}
\toprule
\textbf{parameter} & \textbf{value} & \textbf{id}\\
\midrule
support & \(M=\#\) & \textid{YNOF}\\
\bottomrule
\end{tabular}
\caption{Filter parameter for the mean filter.}
\end{table}

\FloatBarrier

\subsection[Gaussian filter]{Gaussian filter \id{8BC3}}
The parameters listed below apply to the Gaussian filter.

\begin{table}[ht]
\centering
\small
\begin{tabular}{llc}
\toprule
\textbf{parameter} & \textbf{value} & \textbf{id}\\
\midrule
scale & \(\sigma^*=\#\) & \textid{41LN}\\
filter size cutoff & \(\#\) & \textid{WGPM}\\
\bottomrule
\end{tabular}
\caption{Parameters for Gaussian filters. Scale is expressed in physical dimensions, e.g. mm. The filter size cutoff is usually expressed in terms of the scale, e.g. \(4 \sigma^*\).}
\end{table}

\FloatBarrier

\subsection[Laplacian-of-Gaussian filter ]{Laplacian-of-Gaussian filter \id{L6PA}}
The parameters listed below apply to the Laplacian-of-Gaussian filter.

\begin{table}[ht]
\centering
\small
\begin{tabular}{llc}
\toprule
\textbf{parameter} & \textbf{value} & \textbf{id}\\
\midrule
scale & \(\sigma^*=\#\) & \textid{41LN}\\
filter size cutoff & \(\#\) & \textid{WGPM}\\
\bottomrule
\end{tabular}
\caption{Parameters for Laplacian-of-Gaussian filters. Scale is expressed in physical dimensions, e.g. mm. The filter size cutoff is usually expressed in terms of the scale, e.g. \(4 \sigma^*\).}
\end{table}

\FloatBarrier

\subsection[Laws kernels]{Laws kernels \id{JTXT}}
The parameters listed below apply to the Laws kernels.

\begin{table}[ht]
\centering
\small
\begin{tabular}{llc}
\toprule
\textbf{parameter} & \textbf{value} & \textbf{id}\\
\midrule
Laws kernel & & \textid{41LN}\\
& Level (L3) kernel & \textid{B5BZ}\\
& Level (L5) kernel & \textid{6HRH}\\
& Edge (E3) kernel & \textid{LJ4T}\\
& Edge (E5) kernel & \textid{2WPV}\\
& Spots (S3) kernel & \textid{MK5Z}\\
& Spots (S5) kernel & \textid{RXA1}\\
& Wave (W5) kernel & \textid{4ENO}\\
& Ripples (R3) kernel & \textid{3A1W}\\
pseudo-rotational invariance & no/yes & \textid{O1AQ}\\
response map pooling & & \textid{SVKW}\\
& average & \textid{M1CY}\\
& max & \textid{0TAW}\\
& min & \textid{DUX4}\\
energy map & no/yes & \textid{PQSD}\\
energy map distance & \(\delta=\#\) & \textid{I176}\\
\bottomrule
\end{tabular}
\caption{Parameters for Laws kernel filters. Unlike the scale parameter of Gaussian and Laplacian-of-Gaussian, the energy map distance \(\delta\) is defined in voxel units.}
\end{table}

\FloatBarrier

\subsection[Gabor filter]{Gabor filter \id{Q88H}}
The parameters listed below apply to Gabor filters.

\begin{table}[ht]
\centering
\small
\begin{tabular}{llc}
\toprule
\textbf{parameter} & \textbf{value} & \textbf{id}\\
\midrule
scale & \(\sigma^*=\#\) & \textid{41LN}\\
wavelength & \(\lambda^*=\#\) & \textid{S4N6}\\
ellipticity & \(\gamma=\#\) & \textid{GDR5}\\
filter orientation & \(\theta=\#\) & \textid{FQER}\\
response map & & \textid{5P3T}\\
& modulus, magnitude or absolute & \textid{0J1M}\\
& angle, phase or argument & \textid{V530}\\
& real & \textid{F7DI}\\
& imaginary & \textid{LI3Y}\\
pseudo-rotational invariance & no/yes & \textid{O1AQ}\\
response map pooling & & \textid{SVKW}\\
& average & \textid{M1CY}\\
& max & \textid{0TAW}\\
& min & \textid{DUX4}\\
filter orientation stepping & \(\Delta\theta=\#\) & \textid{XTGK}\\
\bottomrule
\end{tabular}
\caption{Parameters for Gabor filters. The scale parameter of the Gaussian envelope and the wavelength parameters are expressed in physical dimensions, e.g. mm.}
\end{table}

\FloatBarrier

\subsection[Separable wavelets]{Separable wavelets \id{25BO}}
The parameters listed below apply to separable wavelet filters.

\begin{table}[ht]
\centering
\small
\begin{tabular}{llc}
\toprule
\textbf{parameter} & \textbf{value} & \textbf{id}\\
\midrule
wavelet family & & \textid{BPXS}\\
& Haar wavelet & \textid{UOUE}\\
& other wavelets & \\
wavelet filter combination & & \textid{UK1F}\\
& LL & \textid{PPF6}\\
& HL & \textid{P60O}\\
& LH & \textid{M0S1}\\
& HH & \textid{WQ2Z}\\
& LLL & \textid{KZAS}\\
& LLH & \textid{TY49}\\
& LHL & \textid{4KRX}\\
& HLL & \textid{UUV9}\\
& LHH & \textid{UIT8}\\
& HHL & \textid{QTIT}\\
& HLH & \textid{8LD2}\\
& HHH & \textid{OYUC}\\
wavelet decomposition level & \# & \textid{GCEK}\\
wavelet decimation & &\\
& decimated transform & \textid{PH3R}\\
& undecimated transform & \textid{CVCQ}\\
pseudo-rotational invariance & no/yes & \textid{O1AQ}\\
response map pooling & & \textid{SVKW}\\
& average & \textid{M1CY}\\
& max & \textid{0TAW}\\
& min & \textid{DUX4}\\
\bottomrule
\end{tabular}
\caption{Parameters for separable wavelet filters. Only the Haar wavelet is explicitly described in this reference manual. Other wavelets should be mentioned by name.}
\end{table}

\FloatBarrier

\subsection[Non-separable wavelets]{Non-separable wavelets \id{LODD}}

The parameters listed below apply to non-separable wavelet filters.

\begin{table}[ht]
\centering
\small
\begin{tabular}{llc}
\toprule
\textbf{parameter} & \textbf{value} & \textbf{id}\\
\midrule
wavelet family & & \textid{389V}\\
& Shannon wavelet & \textid{GWM2}\\
& Simoncelli wavelet & \textid{PRT7}\\
& other wavelets & \\
wavelet decomposition level & \# & \textid{GCEK}\\
\bottomrule
\end{tabular}
\caption{Parameters for non-separable wavelet filters. The Shannon and Simoncelli wavelets are explicitly described in this reference manual. Other wavelets should be mentioned by name.}
\end{table}

\FloatBarrier

\subsection[Riesz transformation]{Riesz transformation \id{AYRS}}

Riesz transformations have the parameters listed below.

\begin{table}[ht]
\centering
\small
\begin{tabular}{llc}
\toprule
\textbf{parameter} & \textbf{value} & \textbf{id}\\
\midrule
Riesz transformation order \(l\) & \((\#, \#)\) or \((\#, \#, \#)\) & \textid{OC48}\\
Riesz filter steering & no/yes & \textid{1SD3}\\
Riesz structure window scale & \(\sigma^*_{\text{tensor}}=\#\) & \textid{41LN}\\
\bottomrule
\end{tabular}
\caption{Parameters for Riesz transformations. The scale parameter of the Gaussian window used to regularise the structure tensor is expressed in physical dimensions, e.g. mm.}
\end{table}

\chapter{Benchmarking}\label{sec:benchmarking}
This section details the two benchmarking phases proposed to assess differences between software implementations, to define standard filter implementations, and to achieve consensus on reference values for filter-based radiomics features.
It includes the description of digital phantom images for controlled filter input and the description of a CT image obtained in a patient with lung cancer.
The methodology used to find a consensus on standard filter implementations is subsequently detailed in Section~\ref{sec:benchmarkingMethodo}. This includes parameter configurations for image filters, how response maps are compared, as well as the submission procedure for response maps for comparison between software implementations.
The methodology used to find reference values for filter-based radiomics features is afterwards described in Section~\ref{sec:featureBenchmarkLungCT}.
\section{Phase 1: Benchmarking filters using digital phantoms}\label{sec:benchmarkingMethodo}
\begin{figure}
\centering
   \begin{minipage}[b]{145pt}
     \centering
     \includegraphics[trim = 0 0 0 0, clip, width=\linewidth] {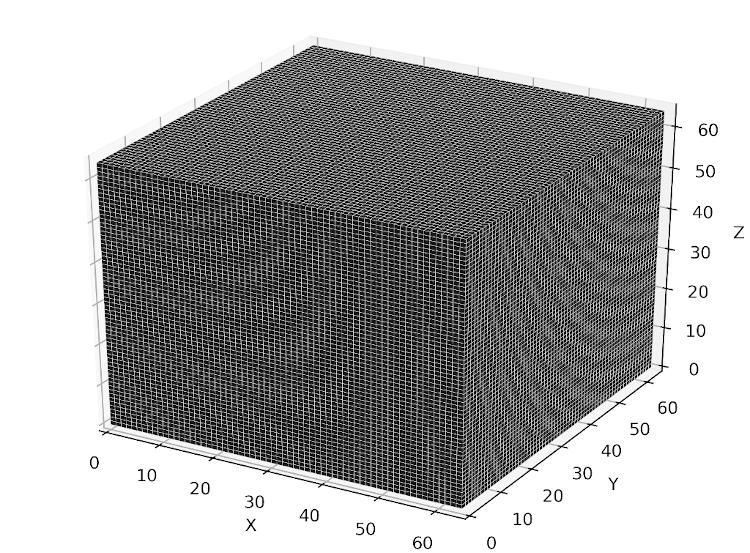}
     \subcaption{empty}\label{fig:phantom_empty}
     \hspace{100pt}
   \end{minipage}
   \begin{minipage}[b]{145pt}
     \centering
     \includegraphics[trim = 0 0 0 0, clip, width=\linewidth] {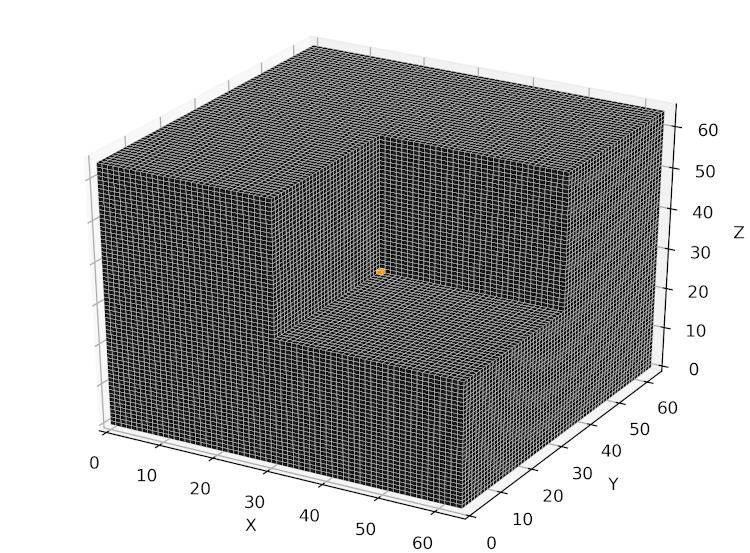}
     \subcaption{impulse response}\label{fig:phantom_impulse_response}
     \hspace{100pt}
   \end{minipage}
   \begin{minipage}[b]{145pt}
     \centering
     \includegraphics[trim = 0 0 0 0, clip, width=\linewidth] {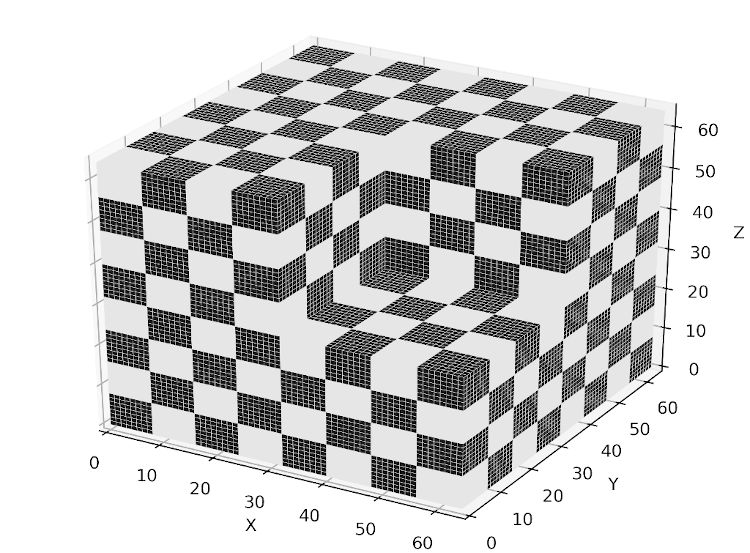}
     \subcaption{checkerboard}\label{fig:phantom_checkerboard}
     \hspace{100pt}
   \end{minipage}
   \begin{minipage}[b]{145pt}
     \centering
     \includegraphics[trim = 0 0 0 0, clip, width=\linewidth] {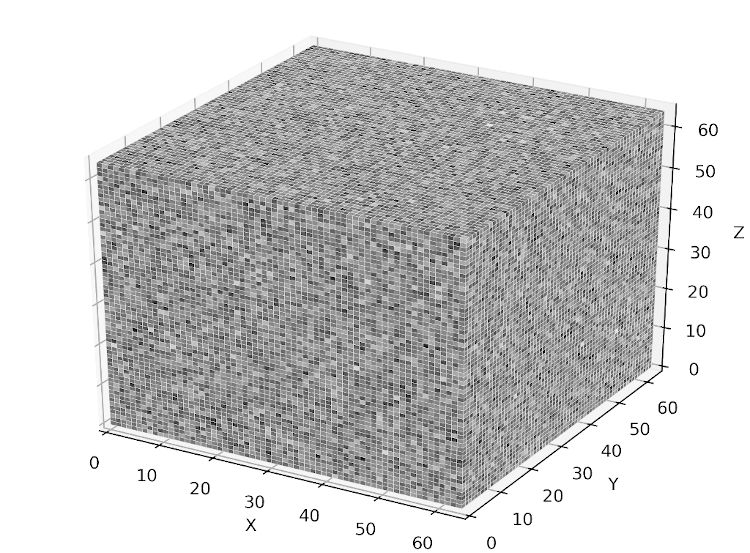}
     \subcaption{noise}\label{fig:phantom_noise}
     \hspace{100pt}
   \end{minipage}
   \begin{minipage}[b]{145pt}
     \centering
     \includegraphics[trim = 0 0 0 0, clip, width=\linewidth] {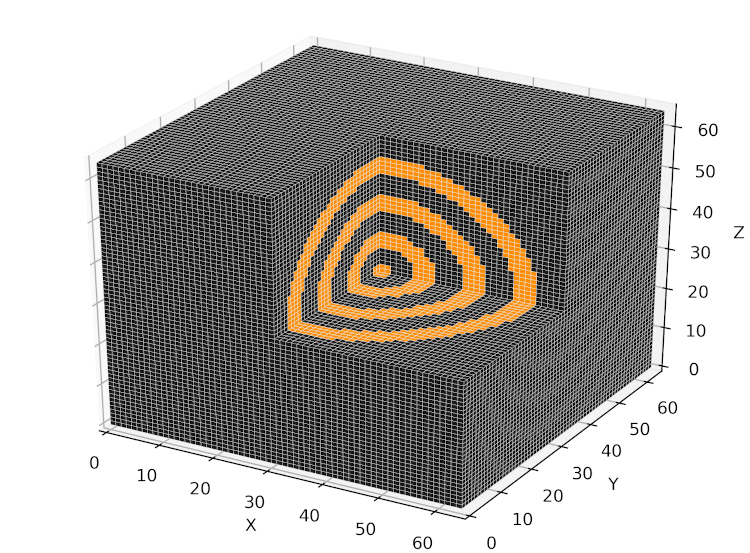}
     \subcaption{sphere}\label{fig:phantom_sphere}
     \hspace{100pt}
   \end{minipage}
   \begin{minipage}[b]{145pt}
     \centering
     \includegraphics[trim = 0 0 0 0, clip, width=\linewidth] {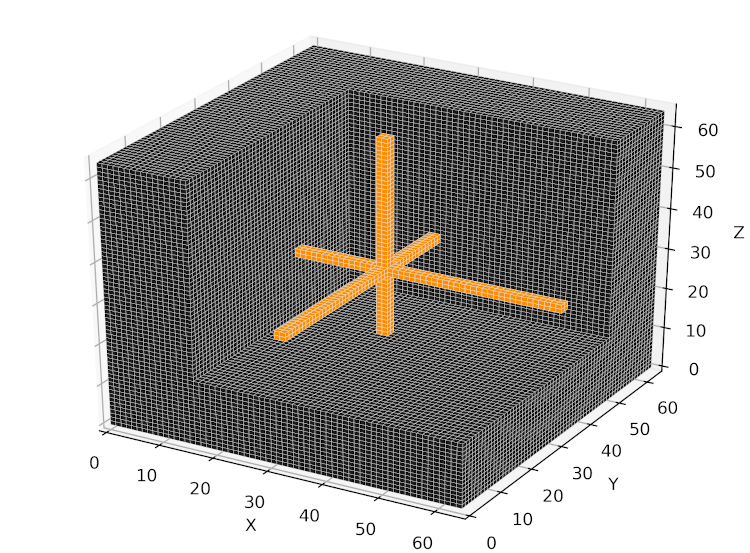}
     \subcaption{pattern \#1}\label{fig:phantom_pattern_1}
     \hspace{100pt}
   \end{minipage}
      \begin{minipage}[b]{145pt}
     \centering
     \includegraphics[trim = 0 0 0 0, clip, width=\linewidth] {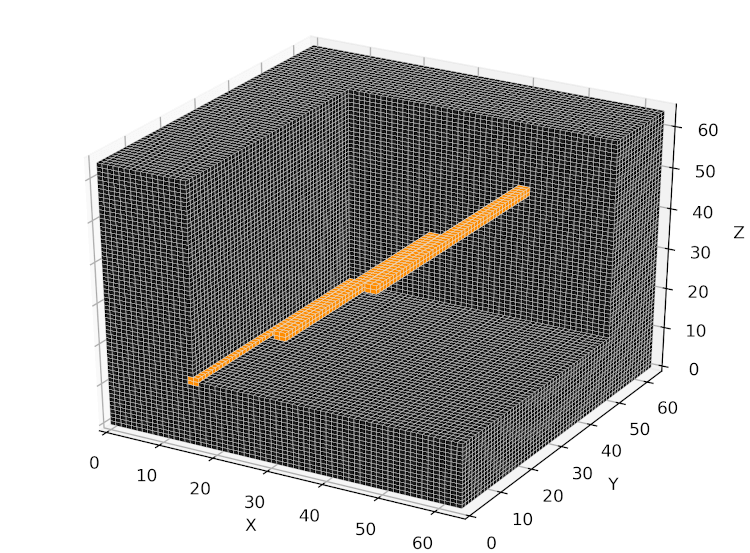}
     \subcaption{pattern \#2}\label{fig:phantom_pattern_2}
     \hspace{100pt}
   \end{minipage}
   \begin{minipage}[b]{145pt}
     \centering
     \includegraphics[trim = 0 0 0 0, clip, width=\linewidth] {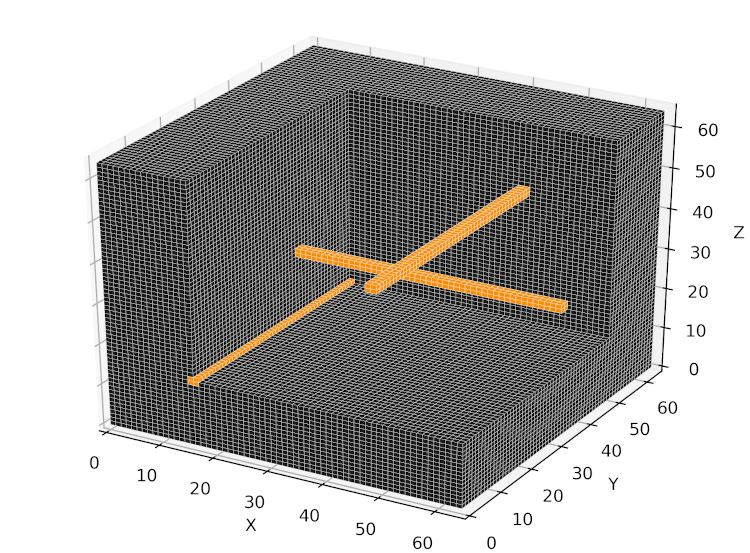}
     \subcaption{phantom \#3}\label{fig:phantom_pattern_3}
     \hspace{100pt}
   \end{minipage}
   \begin{minipage}[b]{145pt}
     \centering
     \includegraphics[trim = 0 0 0 0, clip, width=\linewidth] {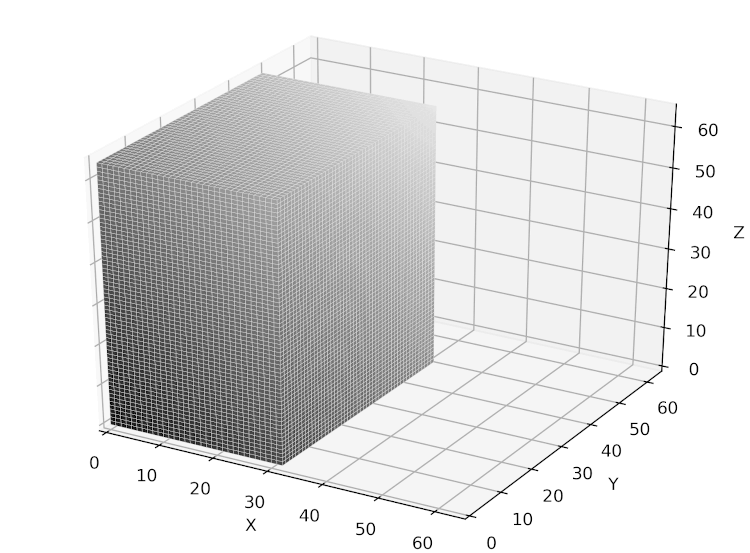}
     \subcaption{orientation}\label{fig:phantom_orientation}
     \hspace{100pt}
   \end{minipage}
  \caption{Synthetic phantoms for verifying compliance with reference filter implementation.
  }
  \label{fig:synthetic_phantom}
\end{figure}

The IBSI developed several 3$D$ phantoms to test image filter implementations. All phantoms (but the orientation phantom) have the same dimension and consist of $64\times 64\times 64$ isotropic voxels with 2.0 by 2.0 by 2.0 mm spacing. 8-bit voxel intensities in all phantoms fall in the range $[0,255]$. The phantoms are stored in the NIfTI format.
The phantoms are shown in Fig.~\ref{fig:synthetic_phantom}.
They include:
\begin{itemize}
    \item \textbf{Empty phantom} (\texttt{empty.nii.gz}): all voxels have $0$ intensity. Intended to investigate the convolution process.
    \item \textbf{Impulse response phantom} (\texttt{impulse\_response.nii.gz}): all but one voxel have intensity $0$. The single remaining centre voxel has an intensity of $255$. This allows visualisation of the filter $g[\boldsymbol{k}]$ itself.
    \item \textbf{Checkerboard phantom} (\texttt{checkerboard.nii.gz}): alternates between cubic regions with intensity $0$ and with intensity $255$.
    \item \textbf{Noise phantom} (\texttt{noise.nii.gz}): contains Gaussian noise with mean intensity of $127$ and a standard deviation of $48$. As such it has no inherent structure.
    \item \textbf{Sphere phantom} (\texttt{sphere.nii.gz}): consists of four concentric spherical hulls with intensity $255$ that are centred on the phantom centre. Thus, the phantom lacks directionality. 
    \item \textbf{Pattern \#1 phantom} (\texttt{pattern\_1.nii.gz}): this is the first of three phantoms that involve directionality. Three perpendicular lines (intensity $255$) intersect at the centre of the phantom. Along with Pattern \#2 and Pattern \#3 phantoms, it is intended to investigate filtering methods able to characterise the local organisation of image directions~\cite{DCS2017}.
    \item \textbf{Pattern \#2 phantom} (\texttt{pattern\_2.nii.gz}): this is the second of the directional phantoms. The phantom contains three parallel lines (intensity $255$).
    \item \textbf{Pattern \#3 phantom} (\texttt{pattern\_3.nii.gz}): this the last of directional phantoms. The phantom contains three lines (intensity $255$), of which two are parallel.
    \item \textbf{Orientation phantom} (\texttt{orientation.nii.gz}): Not all filters are rotationally invariant, and therefore the direction along which filters are applied affects the response map. Using the orientation phantom you can check whether the orientation of the image coordinate system in your software matches the orientation expected by the IBSI. The orientation phantom has a dimension of $(32, 48, 64)$ voxels along $k_1$ ($x$), $k_2$ ($y$) and $k_3$ ($z$) axes, respectively. The pixel intensity increases with the distance from the origin, which has an intensity of $0$. The most distal voxel has an intensity of $141$.
\end{itemize}

\FloatBarrier

The aim of benchmarking the filters using the phantom images is to arrive at a standard implementation. Therefore, filters are applied to the phantoms to create a response maps $h[\boldsymbol{k}]$, which have the same dimensions as the phantoms ($64\times 64\times 64$ voxels). Instead of calculating one or more features from each response map (see Section~\ref{sec:overview}), the entire response map is to be submitted online to ease debugging and find differences in software implementations.
Therefore it is important to note the following:
\begin{itemize}
\item The phantom data need to be converted from an integer data type to at least 32 bit floating point precision, prior to filtering.
\item Filters need to be applied using the settings provided in Table~\ref{tab:benchmark_filter_settings}. Some settings refer to real-world spacing (in millimetres), instead of voxel spacing. 
%Also make sure that the correct boundary condition is used for each phantom (see Table \ref{tab:phantom_summary}).
\item The response map needs to be exported in a (compressed) NIfTI format, with at least 32 bit floating point precision. Please make sure that the response map has the correct dimensions.
\item The response map should be uploaded to \href{https://radiomics.hevs.ch/ibsi}{\texttt{https://radiomics.hevs.ch/ibsi}}. This requires you to log in using a Github account.
\end{itemize}

%The voxel-wise distance between response maps from the different teams will then be compared to identify a consensus response map. In order to visualise and quantify the discrepancy between response maps $h[\boldsymbol{k}]$, a variational approach based on Principal Component Analysis (PCA) in the $\mathbb{R}^{64\times64\times64}$ space of voxel space is used. Each considered response map is an observation in this space.
%In addition, a consensus is defined as the average cluster (centroid in $\mathbb{R}^{64\times64\times64}$) of all submissions.
%A visualisation of the first two PCA components allows assessing the distribution and dispersion of the submissions and the consensus. 
%In addition, a boxplot is proposed to visualise the distribution of the  distances to the consensus in $\mathbb{R}^{64\times64\times64}$ and to suggest outliers.
%An example of comparison of 120 simulated response maps is shown in Fig.~\ref{fig:responseMapComparison}.

\marginnote{\footnotesize v6: Added boundary condition for Simoncelli filters (configurations 8.a.1 - 3).}
\small
\begin{longtable}{cp{3cm}cp{7.7cm}}

\toprule
\textbf{ID} & \textbf{Filter} & \textbf{Phantom} & \textbf{Filter parameters} \\
\midrule
\endfirsthead

\toprule
\textbf{ID} & \textbf{Filter} & \textbf{Phantom} & \textbf{Filter parameters} \\
\midrule
\endhead

\bottomrule
\multicolumn{4}{r}{\textit{Continued on next page}}
\endfoot

\\
\endlastfoot

    1.a.1 & mean & checkerboard & 3$D$ filter, support \(M=15\), zero padding\\
    1.a.2 & & & 3$D$ filter, support \(M=15\), nearest value padding\\
    1.a.3 & & & 3$D$ filter, support \(M=15\), periodic padding\\
    1.a.4 & & & 3$D$ filter, support \(M=15\), mirror padding\\
    1.b.1 & & impulse & 2$D$ filter, support \(M=15\), zero padding\\
    \midrule
    2.a & LoG & impulse & 
    \begin{minipage}[t]{\linewidth}
    \begin{itemize}[nosep,after=\strut,leftmargin=*]
        \item zero padding
        \item 3$D$ filter, scale \(\sigma^*=3.0\) mm, filter size cutoff \(4\sigma^*\)
    \end{itemize}
    \end{minipage} \\ \cline{4-4}
    2.b & & checkerboard &
    \begin{minipage}[t]{\linewidth}
    \begin{itemize}[nosep,after=\strut,leftmargin=*]
        \item mirror padding
        \item 3$D$ filter, scale \(\sigma^*=5.0\) mm, filter size cutoff \(4\sigma^*\)
    \end{itemize}
    \end{minipage} \\ \cline{4-4}
    2.c & & checkerboard &
    \begin{minipage}[t]{\linewidth}
    \begin{itemize}[nosep,after=\strut,leftmargin=*]
        \item mirror padding
        \item 2$D$ filter, scale \(\sigma^*=5.0\) mm, filter size cutoff \(4\sigma^*\)
    \end{itemize}
    \end{minipage} \\
    \midrule
    3.a.1 & Laws & impulse & 
    \begin{minipage}[t]{\linewidth}
    \begin{itemize}[nosep,after=\strut,leftmargin=*]
        \item zero padding
        \item 3$D$ filter, E5L5S5 response map
    \end{itemize}
    \end{minipage} \\ \cline{4-4}
    3.a.2 & & & 
    \begin{minipage}[t]{\linewidth}
    \begin{itemize}[nosep,after=\strut,leftmargin=*]
        \item zero padding
        \item 3$D$ filter, E5L5S5 response map
        \item 3$D$ rotation invariance, \texttt{max} pooling
    \end{itemize}
    \end{minipage} \\ \cline{4-4}    
    3.a.3 & & &
    \begin{minipage}[t]{\linewidth}
    \begin{itemize}[nosep,after=\strut,leftmargin=*]
        \item zero padding
        \item 3$D$ filter, E5L5S5 response map
        \item 3$D$ rotation invariance, \texttt{max} pooling 
        \item energy map, distance $\delta = 7$ voxels
    \end{itemize}
    \end{minipage} \\ \cline{4-4}    
    3.b.1 & & checkerboard &
    \begin{minipage}[t]{\linewidth}
    \begin{itemize}[nosep,after=\strut,leftmargin=*]
        \item mirror padding
        \item 3$D$ filter, E3W5R5 response map
    \end{itemize}
    \end{minipage} \\ \cline{4-4}      
    3.b.2 & & &
    \begin{minipage}[t]{\linewidth}
    \begin{itemize}[nosep,after=\strut,leftmargin=*]
        \item mirror padding
        \item 3$D$ filter, E3W5R5 response map
        \item 3$D$ rotation invariance, \texttt{max} pooling
    \end{itemize}
    \end{minipage} \\ \cline{4-4} 
    3.b.3 & & &
    \begin{minipage}[t]{\linewidth}
    \begin{itemize}[nosep,after=\strut,leftmargin=*]
        \item mirror padding
        \item 3$D$ filter, E3W5R5 response map
        \item 3$D$ rotation invariance, \texttt{max} pooling
        \item energy map, distance $\delta = 7$ voxels
    \end{itemize}
    \end{minipage} \\ \cline{4-4}    
    3.c.1 & & checkerboard &
    \begin{minipage}[t]{\linewidth}
    \begin{itemize}[nosep,after=\strut,leftmargin=*]
        \item mirror padding
        \item 2$D$ filter, L5S5 response map
    \end{itemize}
    \end{minipage} \\ \cline{4-4}      
    3.c.2 & & &
    \begin{minipage}[t]{\linewidth}
    \begin{itemize}[nosep,after=\strut,leftmargin=*]
        \item mirror padding
        \item 2$D$ filter, L5S5 response map
        \item 2$D$ rotation invariance, \texttt{max} pooling
    \end{itemize}
    \end{minipage} \\ \cline{4-4} 
    3.c.3 & & &
    \begin{minipage}[t]{\linewidth}
    \begin{itemize}[nosep,after=\strut,leftmargin=*]
        \item mirror padding
        \item 2$D$ filter, L5S5 response map
        \item 2$D$ rotation invariance, \texttt{max} pooling
        \item energy map, distance $\delta = 7$ voxels
    \end{itemize}
    \end{minipage} \\
    \midrule
    4.a.1 & Gabor & impulse &
    \begin{minipage}[t]{\linewidth}
    \begin{itemize}[nosep,after=\strut,leftmargin=*]
        \item zero padding
        \item 2$D$ modulus response map
        \item \(\sigma^*=10.0\) mm, \(\lambda^*=4\) mm, \(\gamma=1/2\)
        \item in-plane orientation \(\theta = \pi/3\)
    \end{itemize}
    \end{minipage} \\ \cline{4-4}
    4.a.2 & & &
    \begin{minipage}[t]{\linewidth}
    \begin{itemize}[nosep,after=\strut,leftmargin=*]
        \item zero padding
        \item 2$D$ modulus response map
        \item \(\sigma^*=10.0\) mm, \(\lambda^*=4\) mm, \(\gamma=1/2\)
        \item 2$D$ rotation invariance, \(\Delta\theta = \pi/4\), \texttt{average} pooling
        \item average 2$D$ responses over orthogonal planes 
    \end{itemize}
    \end{minipage} \\ \cline{4-4}
    4.b.1 & & sphere & 
    \begin{minipage}[t]{\linewidth}
    \begin{itemize}[nosep,after=\strut,leftmargin=*]
        \item mirror padding
        \item 2$D$ modulus response map
        \item \(\sigma^*=20.0\) mm, \(\lambda^*=8\) mm, \(\gamma=5/2\)
        \item in-plane orientation \(\theta = 5\pi/4\)
    \end{itemize}
    \end{minipage} \\ \cline{4-4}    
    4.b.2 & & & 
    \begin{minipage}[t]{\linewidth}
    \begin{itemize}[nosep,after=\strut,leftmargin=*]
        \item mirror padding
        \item 2$D$ modulus response map
        \item \(\sigma^*=20.0\) mm, \(\lambda^*=8\) mm, \(\gamma=5/2\)
        \item 2$D$ rotation invariance, \(\Delta\theta = \pi/8\), \texttt{average} pooling
        \item average 2$D$ responses over orthogonal planes 
    \end{itemize}
    \end{minipage} \\
    \midrule
    5.a.1 & Daubechies 2 & impulse & 
    \begin{minipage}[t]{\linewidth}
    \begin{itemize}[nosep,after=\strut,leftmargin=*]
        \item zero padding
        \item 3$D$ filter, undecimated LHL map -- 1\textsuperscript{st} level
    \end{itemize}
    \end{minipage} \\ \cline{4-4}      
    5.a.2 & & &  
    \begin{minipage}[t]{\linewidth}
    \begin{itemize}[nosep,after=\strut,leftmargin=*]
        \item zero padding
        \item 3$D$ filter, undecimated LHL map -- 1\textsuperscript{st} level
        \item 3$D$ rotation invariance, \texttt{average} pooling
    \end{itemize}
    \end{minipage} \\    
    \midrule
    6.a.1 & Coiflet 1 & sphere & 
    \begin{minipage}[t]{\linewidth}
    \begin{itemize}[nosep,after=\strut,leftmargin=*]
        \item periodic padding
        \item 3$D$ filter, undecimated HHL map -- 1\textsuperscript{st} level
    \end{itemize}
    \end{minipage} \\ \cline{4-4}
    6.a.2 & & & 
    \begin{minipage}[t]{\linewidth}
    \begin{itemize}[nosep,after=\strut,leftmargin=*]
        \item periodic padding
        \item 3$D$ filter, undecimated HHL map -- 1\textsuperscript{st} level
        \item 3$D$ rotation invariance, \texttt{average} pooling
    \end{itemize}
    \end{minipage} \\ 
    \midrule
    7.a.1 & Haar & checkerboard & 
    \begin{minipage}[t]{\linewidth}
    \begin{itemize}[nosep,after=\strut,leftmargin=*]
        \item mirror padding
        \item 3$D$ filter, undecimated LLL map -- 2\textsuperscript{nd} level
        \item 3$D$ rotation invariance, \texttt{average} pooling
    \end{itemize}
    \end{minipage} \\ \cline{4-4}    
    7.a.2 & & & 
    \begin{minipage}[t]{\linewidth}
    \begin{itemize}[nosep,after=\strut,leftmargin=*]
        \item mirror padding
        \item 3$D$ filter, undecimated HHH map -- 2\textsuperscript{nd} level
        \item 3$D$ rotation invariance, \texttt{average} pooling
    \end{itemize}
    \end{minipage} \\    
    \midrule
    8.a.1 & Simoncelli & checkerboard & 
    \begin{minipage}[t]{\linewidth}
    \begin{itemize}[nosep,after=\strut,leftmargin=*]
        \item periodic padding
        \item 3$D$ filter, B map -- 1\textsuperscript{st} level
    \end{itemize}
    \end{minipage} \\ \cline{4-4}     
    8.a.2 & & & 
    \begin{minipage}[t]{\linewidth}
    \begin{itemize}[nosep,after=\strut,leftmargin=*]
        \item periodic padding
        \item 3$D$ filter, B map -- 2\textsuperscript{nd} level
    \end{itemize}
    \end{minipage} \\ \cline{4-4}       
    8.a.3 & & & 
    \begin{minipage}[t]{\linewidth}
    \begin{itemize}[nosep,after=\strut,leftmargin=*]
        \item periodic padding
        \item 3$D$ filter, B map -- 3\textsuperscript{rd} level
    \end{itemize}
    \end{minipage} \\    
    \midrule
    9.a & Riesz-transformed LoG & impulse & 
    \begin{minipage}[t]{\linewidth}
    \begin{itemize}[nosep,after=\strut,leftmargin=*]
        \item zero padding
        \item 3$D$ filter, scale \(\sigma=3.0\) mm, filter size cutoff \(4\sigma\)
        \item \(\boldsymbol{l}=\left(1, 0, 0\right)\)
    \end{itemize}
    \end{minipage} \\ \cline{4-4}    
    9.b.1 & & sphere & 
    \begin{minipage}[t]{\linewidth}
    \begin{itemize}[nosep,after=\strut,leftmargin=*]
        \item zero padding
        \item 3$D$ filter, scale \(\sigma=3.0\) mm, filter size cutoff \(4\sigma\)
        \item \(\boldsymbol{l}=\left(0, 2, 0\right)\)
    \end{itemize}
    \end{minipage} \\ \cline{4-4}      
    9.b.2 & & & 
    \begin{minipage}[t]{\linewidth}
    \begin{itemize}[nosep,after=\strut,leftmargin=*]
        \item zero padding
        \item scale \(\sigma=3.0\) mm, filter size cutoff \(4\sigma\)
        \item 3$D$ filter, \(\boldsymbol{l}=\left(0, 2, 0\right)\)
        \item aligned by structure tensor, $\sigma_{\text{tensor}} = 1 \text{mm}$
    \end{itemize}
    \end{minipage} \\    
    \midrule
    10.a & Riesz-transformed Simoncelli & impulse & 
    \begin{minipage}[t]{\linewidth}
    \begin{itemize}[nosep,after=\strut,leftmargin=*]
        \item zero padding
        \item 3$D$ filter, B map -- 1\textsuperscript{st} level
        \item \(\boldsymbol{l}=\left(1, 0, 0\right)\)
    \end{itemize}
    \end{minipage} \\ \cline{4-4}     
    10.b.1 & & pattern 1 &
    \begin{minipage}[t]{\linewidth}
    \begin{itemize}[nosep,after=\strut,leftmargin=*]
        \item nearest value padding
        \item 3$D$ filter, B map -- 1\textsuperscript{st} level
        \item \(\boldsymbol{l}=\left(0, 2, 0\right)\)
    \end{itemize}
    \end{minipage} \\ \cline{4-4}       
    10.b.2 & & & 
    \begin{minipage}[t]{\linewidth}
    \begin{itemize}[nosep,after=\strut,leftmargin=*]
        \item nearest value padding
        \item 3$D$ filter, B map -- 1\textsuperscript{st} level
        \item \(\boldsymbol{l}=\left(0, 2, 0\right)\)
        \item aligned by structure tensor, $\sigma_{\text{tensor}} = 1 \text{mm}$
    \end{itemize}
    \end{minipage} \\     
    \bottomrule
    \caption{\normalsize Filters, parameters and phantoms for comparing and standardising filter implementations. \textit{Note that 2\(D\) and 3\(D\) rotation invariance for Laws and undecimated wavelet filters (Daubechies 2, Coiflet 1, Haar) is estimated using equivariant right angle rotational representation for separable filters (Appendix \ref{app:separableConvRightAngleEquivariant}). Pooling refers to pooling over response maps obtained under different orientations. 2\(D\) filters are applied in the axial plane by default, except for Gabor filters. For Gabor filters, 2$D$ rotation invariance is estimated by rotating the the filter kernel over multiple orientations in the image plane.}} \label{tab:benchmark_filter_settings}
\end{longtable}
\normalsize
\FloatBarrier
%
%\begin{figure}[h!]
%\centering
%\includegraphics[trim = 0 0 0 0, clip, scale=.7]{PCAconsensus.pdf}
%\hspace{0.5cm}
%\includegraphics[trim = 0 -30 0 0, clip, scale=.7]{BoxplotConsensus.pdf}
%\caption{PCA (left) and boxplot (right) visualisations of a comparison between 120 simulated response maps. The blue and red observations are coming from two distinct uniform distributions. The consensus is marked with a black cross in the scatter plot.}
%\label{fig:responseMapComparison}
%\end{figure}
%
\section{Phase 2: Benchmarking filters using a lung cancer CT image}\label{sec:featureBenchmarkLungCT}
CT images from four patients with non-small-cell lung carcinoma were made available to serve as radiomics phantoms \href{http://dx.doi.org/10.17195/candat.2016.08.1}{(DOI:10.17195/candat.2016.08.1)}. The IBSI previously used the image for the first patient (\texttt{PAT1}) to establish reference values for image features~\cite{ZLV2017}. This image is used here again.

The lung cancer CT image is stored as a stack of slices in \texttt{DICOM} format. The image slices can be identified by the \texttt{DCM\_IMG} prefix. Additionally, the RT structure set (starting \texttt{DCM\_RS}) contains a delineation of the Gross Tumour Volume (GTV) which is used to create a segmentation mask for the ROI. The image and the mask have been converted to the \texttt{NIfTI} format. Note that some import algorithms alter the floating point bit depth of the image and mask. If this occurs, the imported image and mask should be converted back to (at least) 32-bit floating point and rounded prior to any further processing.

Compared to the digital phantoms, the lung cancer CT image offers a more realistic setting to benchmark applications of image filters as part of a radiomics analysis. Image filtering is assessed as part of two different image processing configurations, see Table~\ref{tab:lung_ct_image_processing_configurations}. The main difference between the configurations is that one performs image processing and filtering within each image slice (\textit{i.e.} 2$D$), whereas the other operates in 3$D$. Features are still computed over the entire region of interest, as mentioned in the IBSI 1 reference manual. The filtering configurations are then shown in Table~\ref{tab:lung_ct_image_filter_configurations}.

\marginnote{\footnotesize v5: Updated the 2D-test.}

\begin{table}[h!]
\centering
\small
\begin{tabular}{lcc}
\toprule
\textbf{Parameter} & \textbf{Configuration A} & \textbf{Configuration B}\\
\midrule
slice-wise (2$D$) or as volume (3$D$) & 2$D$ & 3$D$\\
interpolation & no & yes\\
\quad resampled voxel spacing (mm) & --- & $1 \times 1 \times 1$\\
\quad interpolation method & --- & tricubic spline\\
\quad intensity rounding & --- & nearest integer\\
\quad ROI interpolation method & --- & trilinear\\
\quad ROI partial mask volume & --- & $0.5$\\
re-segmentation & & \\
\quad range (HU) & $[-1000, 400]$ & $[-1000, 400]$\\
\quad outlier filtering & none & none \\
image filters & & \\
\quad filters & see Table \ref{tab:lung_ct_image_filter_configurations} & see Table \ref{tab:lung_ct_image_filter_configurations}\\
\quad boundary condition & mirror & mirror\\
% discretisation\textsuperscript{a} & & \\
% \quad intensity histogram & FBN: 16 bins & FBN: 16 bins\\
\bottomrule
\end{tabular}
\normalsize
\caption{Image processing configurations for obtaining reference values for intensity-based statistical features after image filtering in the lung cancer CT image. The different processing steps are described in the IBSI reference manual~\cite{Zwanenburg2020-jt}. Note that the image should not be cropped. Also note the use of a higher order (spline) resampling method instead of linear interpolation. ROI: Region Of Interest; HU: Hounsfield Unit.}
\label{tab:lung_ct_image_processing_configurations}
\end{table}

\begin{table}[h!]
    \centering
    \small
    \begin{tabular}{cp{3cm}ccp{7.5cm}}
    \toprule
    \textbf{ID} & \textbf{Filter} & \textbf{A} & \textbf{B} & \textbf{Filter parameters} \\
    \midrule
    1.A | 1.B & none & \(\times\) & \(\times\) &  \textendash \\ \hline
    2.A | 2.B & mean
    & \(\times\) &  &  
    \begin{minipage}[t]{\linewidth}
    \begin{itemize}[nosep,after=\strut,leftmargin=*]
        \item 2$D$ filter, support \(M=5\) voxels
    \end{itemize}
    \end{minipage} \\ \cline{5-5}    
    & &  & \(\times\) &  
    \begin{minipage}[t]{\linewidth}
    \begin{itemize}[nosep,after=\strut,leftmargin=*]
        \item 3$D$ filter, support \(M=5\) voxels
    \end{itemize}
    \end{minipage} \\  \hline
    3.A | 3.B & LoG
    & \(\times\) &  &  
    \begin{minipage}[t]{\linewidth}
    \begin{itemize}[nosep,after=\strut,leftmargin=*]
        \item 2$D$ filter, scale \(\sigma^*=1.5\) mm, filter size cutoff \(4\sigma\)
    \end{itemize}
    \end{minipage} \\ \cline{5-5}    
    & &  & \(\times\) &  
    \begin{minipage}[t]{\linewidth}
    \begin{itemize}[nosep,after=\strut,leftmargin=*]
        \item 3$D$ filter, scale \(\sigma^*=1.5\) mm, filter size cutoff \(4\sigma\)
    \end{itemize}
    \end{minipage} \\  \hline
    4.A | 4.B & Laws & \(\times\) & & 
    \begin{minipage}[t]{\linewidth}
    \begin{itemize}[nosep,after=\strut,leftmargin=*]
        \item 2$D$ filter, L5E5 energy map, distance $\delta = 7$ voxels
        \item 2$D$ rotation invariance, \texttt{max} pooling
    \end{itemize}
    \end{minipage} \\ \cline{5-5}     
    & & &\(\times\) & 
    \begin{minipage}[t]{\linewidth}
    \begin{itemize}[nosep,after=\strut,leftmargin=*]
        \item 3$D$ filter, L5E5E5 response map
        \item 3$D$ rotation invariance, \texttt{max} pooling
        \item  energy map, distance $\delta = 7$ voxels
    \end{itemize}
    \end{minipage} \\ \hline
    5.A | 5.B & Gabor & \(\times\) &  & 
    \begin{minipage}[t]{\linewidth}
    \begin{itemize}[nosep,after=\strut,leftmargin=*]
        \item 2$D$ modulus response map
        \item \(\sigma^*=5\) mm, \(\lambda^*=2\) mm, \(\gamma=3/2\)
        \item 2$D$ rotation invariance, \(\Delta\theta = \pi/8\), \texttt{average} pooling
    \end{itemize}
    \end{minipage} \\ \cline{5-5}     
    & & &\(\times\) & 
    \begin{minipage}[t]{\linewidth}
    \begin{itemize}[nosep,after=\strut,leftmargin=*]
        \item 2$D$ modulus response map
        \item \(\sigma^*=5\) mm, \(\lambda^*=2\) mm, \(\gamma=3/2\)
        \item 2$D$ rotation invariance, \(\Delta\theta = \pi/8\), \texttt{average} pooling
        \item average 2$D$ responses over orthogonal planes 
    \end{itemize}
    \end{minipage} \\ \hline
    6.A | 6.B & Daubechies 3 & \(\times\) & & 
    \begin{minipage}[t]{\linewidth}
    \begin{itemize}[nosep,after=\strut,leftmargin=*]
        \item 2$D$ filter, undecimated LH map -- 1\textsuperscript{st} level
        \item 2$D$ rotation invariance, \texttt{average} pooling
    \end{itemize}
    \end{minipage} \\ \cline{5-5} 
    & & & \(\times\) & 
    \begin{minipage}[t]{\linewidth}
    \begin{itemize}[nosep,after=\strut,leftmargin=*]
        \item 3$D$ filter, undecimated LLH map -- 1\textsuperscript{st} level
        \item 3$D$ rotation invariance, \texttt{average} pooling
    \end{itemize}
    \end{minipage} \\  \hline
    7.A | 7.B & Daubechies 3 & \(\times\) & & 
    \begin{minipage}[t]{\linewidth}
    \begin{itemize}[nosep,after=\strut,leftmargin=*]
        \item 2$D$ filter, undecimated HH map -- 2\textsuperscript{nd} level
        \item 2$D$ rotation invariance, \texttt{average} pooling
    \end{itemize}
    \end{minipage} \\  \cline{5-5}
    & & & \(\times\) & 
    \begin{minipage}[t]{\linewidth}
    \begin{itemize}[nosep,after=\strut,leftmargin=*]
        \item 3$D$ filter, undecimated HHH map -- 2\textsuperscript{nd} level
        \item 3$D$ rotation invariance, \texttt{average} pooling
    \end{itemize}
    \end{minipage} \\ \hline
    8.A | 8.B & Simoncelli
    & \(\times\) &  &  
    \begin{minipage}[t]{\linewidth}
    \begin{itemize}[nosep,after=\strut,leftmargin=*]
        \item 2$D$ filter, B map -- 1\textsuperscript{st} level
    \end{itemize}
    \end{minipage} \\ \cline{5-5}    
    & &  & \(\times\) &  
    \begin{minipage}[t]{\linewidth}
    \begin{itemize}[nosep,after=\strut,leftmargin=*]
        \item 3$D$ filter, B map -- 1\textsuperscript{st} level
    \end{itemize}
    \end{minipage} \\  \hline
    9.A | 9.B & Simoncelli
    & \(\times\) &  &  
    \begin{minipage}[t]{\linewidth}
    \begin{itemize}[nosep,after=\strut,leftmargin=*]
        \item 2$D$ filter, B map -- 2\textsuperscript{nd} level
    \end{itemize}
    \end{minipage} \\ \cline{5-5}    
    & &  & \(\times\) &  
    \begin{minipage}[t]{\linewidth}
    \begin{itemize}[nosep,after=\strut,leftmargin=*]
        \item 3$D$ filter, B map -- 2\textsuperscript{nd} level
    \end{itemize}
    \end{minipage} \\  \hline
    10.A | 10.B & Riesz-transformed Simoncelli
    & \(\times\) &  &  
    \begin{minipage}[t]{\linewidth}
    \begin{itemize}[nosep,after=\strut,leftmargin=*]
        \item 2$D$ filter, B map -- 1\textsuperscript{st} level
        \item \(\boldsymbol{l}=\left(0, 2\right)\)
    \end{itemize}
    \end{minipage} \\ \cline{5-5}    
    & &  & \(\times\) &  
    \begin{minipage}[t]{\linewidth}
    \begin{itemize}[nosep,after=\strut,leftmargin=*]
        \item 3$D$ filter, B map -- 1\textsuperscript{st} level
        \item \(\boldsymbol{l}=\left(0, 2, 0\right)\)
    \end{itemize}
    \end{minipage} \\  \hline
    11.A | 11.B & Riesz-transformed Simoncelli
    & \(\times\) &  & 
    \begin{minipage}[t]{\linewidth}
    \begin{itemize}[nosep,after=\strut,leftmargin=*]
        \item 2$D$ filter, B map -- 1\textsuperscript{st} level
        \item \(\boldsymbol{l}=\left(0, 2\right)\)
        \item aligned by structure tensor, $\sigma_{\text{tensor}} = 1 \text{mm}$ 
    \end{itemize}
    \end{minipage} \\ \cline{5-5}        
    & & & \(\times\)  & 
    \begin{minipage}[t]{\linewidth}
    \begin{itemize}[nosep,after=\strut,leftmargin=*]
        \item 3$D$ filter, B map -- 1\textsuperscript{st} level
        \item \(\boldsymbol{l}=\left(0, 2, 0\right)\)
        \item aligned by structure tensor, $\sigma_{\text{tensor}} = 1 \text{mm}$ 
    \end{itemize}
    \end{minipage} \\    
    \bottomrule
    \end{tabular}
    \normalsize
    \caption{Filters and parameters for the configurations A (2$D$) and B (3$D$) defined in Table~\ref{tab:lung_ct_image_processing_configurations}. These settings are used to determine reference values for radiomics features computed from filter response maps. Note that 2$D$ and 3$D$ rotation invariance for Laws and undecimated wavelet filters (\emph{e.g.} Daubechies 3) is estimated using equivariant right angle rotational representation for separable filters (Appendix \ref{app:separableConvRightAngleEquivariant}). For Gabor filters, 2$D$ rotation invariance is estimated by rotating the the filter kernel over multiple orientations in the image plane. The padding method can be freely chosen.}
    \label{tab:lung_ct_image_filter_configurations}
\end{table}
\FloatBarrier

Features will be computed based on the response maps of each of the filters applied to the CT image. That is, in contrast to the previous section where the objective is to arrive at a standard filter implementation, here we aim to define reference values for radiomics features. The features that are being assessed are listed below in Section~\ref{sec:radiomicsfeaturesbenchmarking}.

\subsection{Features}\label{sec:radiomicsfeaturesbenchmarking}
Only part of the radiomics features standardised previously need to be reported. This is done for two reasons. First, to facilitate the submission process. And secondly to ensure that sufficient statistical power can be reached as we found that the majority of teams did not implement all features previously~\cite{Zwanenburg2020-jt}.

These features are listed below with their permanent identifiers.

\begin{enumerate}
    \item Mean intensity (\textid{Q4LE})
    \item Intensity variance (\textid{ECT3})
    \item Intensity skewness (\textid{KE2A})
    \item (Excess) intensity kurtosis (\textid{IPH6})
    \item Median intensity (\textid{Y12H})
    \item Minimum intensity (\textid{1GSF})
    \item 10\textsuperscript{th} intensity percentile (\textid{QG58})
    \item 90\textsuperscript{th} intensity percentile (\textid{8DWT})
    \item Maximum intensity (\textid{84IY})
    \item Intensity interquartile range (\textid{SALO})
    \item Intensity range (\textid{2OJQ})
    \item Intensity-based mean absolute deviation (\textid{4FUA})
    \item Intensity-based robust mean absolute deviation (\textid{1128})
    \item Intensity-based median absolute deviation (\textid{N72L})
    \item Intensity-based coefficient of variation (\textid{7TET})
    \item Intensity-based quartile coefficient of dispersion (\textid{9S40})
    \item Intensity-based energy (\textid{N8CA})
    \item Root mean square intensity (\textid{5ZWQ})
\end{enumerate}

In addition some diagnostic features may be reported. Configurations A and B are similar to those used in IBSI 1. However the use of a higher order interpolation method precludes the direct use of diagnostic features from the previous work. Most features are based on the ROI intensity mask after interpolation and re-segmentation, since these features will be different from those previously established:
\begin{enumerate}
    \item Number of voxels in the ROI mask, before interpolation.
    \item Number of voxels in the ROI intensity mask, after interpolation and re-segmentation.
    \item Mean intensity in the ROI intensity mask after interpolation and re-segmentation.
    \item Maximal intensity in the ROI intensity mask after interpolation and re-segmentation.
    \item Minimal intensity in the ROI intensity mask after interpolation and re-segmentation.
\end{enumerate}

\section{Phase 3: Validation}\label{sec:validationPhase}
\marginnote{\footnotesize v6: Introduced section on the validation phase. v7 updated configurations}
The aim of phase 3 is to assess reproducibility of features computed from response maps. We will use a co-registered multi-modality imaging dataset of 51 patients with soft-tissue sarcoma \cite{Vallieres2015-hv,Vallieres2015-tf}, see the \texttt{data\_sets} GitHub repository \footnote{\url{https://github.com/theibsi/data_sets/tree/master/ibsi_validation}}. The images, consisting of CT, 18F-FDG-PET and T1-weighted MR imaging, are preprocessed as follows.

Counts data in PET imaging was converted to body-weight corrected SUV and cropped to 50 mm around the GTV region of interest.

T1-weighted MR images were corrected using the N4-bias field correction algorithm \cite{Tustison2010-zb}. A simple threshold was manually determined to create a patient segmentation mask that excludes air voxels. The N4 algorithm was parameterised to use 3 fitting levels, a maximum of 100 iterations at each level, a convergence threshold of 0.001 and to utilise the patient mask. After bias-field correction, MR images were normalised by mapping intensities to a standard range to improve comparability between different patients. The lower end of the range was set to correspond to air. The upper end of the range was set to the 95\textsuperscript{th} percentile of the intensities within the patient mask, corresponding roughly to subcutaneous fat. This range was then used to linearly map voxel intensities with 0 (air) and 1000 (subcutaneous fat) as anchor points. The images were subsequently cropped to 50 mm around the GTV and stored as integer values for DICOM compliance.

CT images were cropped to 50 mm around the GTV, but did not undergo additional processing.

\subsection{Computing response maps and features for validation}

The validation phase, in concept, mostly follows phase 2, but for more images. Prior to creating the response map, images are processed using the parameters listed in Table \ref{tab:validation_image_processing_configurations}. Response maps are then created using the parameters detailed in Table \ref{tab:validation_image_filter_configurations}. These response maps are then used to compute the same statistical features as listed in Section \ref{sec:radiomicsfeaturesbenchmarking}. Computing diagnostic features is not necessary.

\begin{table}[h!]
\centering
\small
\begin{tabular}{lccc}
\toprule
\textbf{Parameter} & \textbf{CT} & \textbf{MR} & \textbf{PET}\\
\midrule
slice-wise (2$D$) or as volume (3$D$) & 3$D$ & 3$D$ & 3$D$\\
interpolation & yes & yes & yes\\
\quad resampled voxel spacing (mm) & $1 \times 1 \times 1$ & $1 \times 1 \times 1$ & $3 \times 3 \times 3$\\
\quad interpolation method & tricubic spline & tricubic spline & tricubic spline\\
\quad intensity rounding & nearest integer & --- & --- \\
\quad ROI interpolation method & trilinear & trilinear & trilinear\\
\quad ROI partial mask volume & $0.5$ & $0.5$ & $0.5$\\
re-segmentation & & \\
\quad range (HU) & $[-200, 200]$ &  $[0, \infty)$ & $[0, \infty)$\\
\quad outlier filtering & none & none & none\\
image filters & & \\
\quad filters & see Table \ref{tab:validation_image_filter_configurations} & see Table \ref{tab:validation_image_filter_configurations} & see Table \ref{tab:validation_image_filter_configurations}\\
\quad boundary condition & mirror & mirror & mirror\\
\bottomrule
\end{tabular}
\normalsize
\caption{Image processing configurations for the validation phase. The different processing steps are described in the IBSI reference manual~\cite{Zwanenburg2020-jt}. Note that the image should not be cropped. Also note the use of a higher order (spline) resampling method instead of linear interpolation. ROI: Region Of Interest.}
\label{tab:validation_image_processing_configurations}
\end{table}

\begin{table}[h!]
    \centering
    \small
    \begin{tabular}{cp{3cm}p{7.5cm}}
    \toprule
    \textbf{ID} & \textbf{Filter} & \textbf{Filter parameters} \\
    \midrule
    1 & none & \textendash \\ \hline
    2 & mean & 3$D$ filter, support \(M=3\) voxels \\ \hline
    3 & LoG & 3$D$ filter, scale \(\sigma^*=3.0\) mm, filter size cutoff \(4\sigma\) \\ \hline
    4 & Laws & 
    \begin{minipage}[t]{\linewidth}
    \begin{itemize}[nosep,after=\strut,leftmargin=*]
        \item 3$D$ filter, S5E5L5 response map
        \item 3$D$ rotation invariance, \texttt{max} pooling
        \item  energy map, distance $\delta = 5$ voxels
    \end{itemize}
    \end{minipage} \\ \hline
    5 & Gabor & 
    \begin{minipage}[t]{\linewidth}
    \begin{itemize}[nosep,after=\strut,leftmargin=*]
        \item 2$D$ modulus response map
        \item \(\sigma^*=3\) mm, \(\lambda^*=3\) mm, \(\gamma=1\), \(\theta=-5\pi/8\)
    \end{itemize}
    \end{minipage} \\ \hline
    6 & Coiflet 3 & 
    \begin{minipage}[t]{\linewidth}
    \begin{itemize}[nosep,after=\strut,leftmargin=*]
        \item 3$D$ filter, undecimated LHH map -- 1\textsuperscript{st} level
        \item 3$D$ rotation invariance, \texttt{average} pooling
    \end{itemize}
    \end{minipage} \\ \hline
    7 & Coiflet 3 &
    \begin{minipage}[t]{\linewidth}
    \begin{itemize}[nosep,after=\strut,leftmargin=*]
        \item 3$D$ filter, undecimated HHH map -- 2\textsuperscript{nd} level
        \item 3$D$ rotation invariance, \texttt{max} pooling
    \end{itemize}
    \end{minipage} \\ \hline
    8 & Simoncelli &
    \begin{minipage}[t]{\linewidth}
    \begin{itemize}[nosep,after=\strut,leftmargin=*]
        \item 3$D$ filter, B map -- 1\textsuperscript{st} level
    \end{itemize}
    \end{minipage} \\  \hline
    9 & Simoncelli &
    \begin{minipage}[t]{\linewidth}
    \begin{itemize}[nosep,after=\strut,leftmargin=*]
        \item 3$D$ filter, B map -- 2\textsuperscript{nd} level
    \end{itemize}
    \end{minipage} \\
    \bottomrule
    \end{tabular}
    \normalsize
    \caption{Filters and parameters for validation. Note that the \(\lambda\) parameter for the Gabor filter is realistically too small for the PET images. Larger values should be used for actual radiomics studies.}
    \label{tab:validation_image_filter_configurations}
\end{table}

\chapter{Reference values}\label{sec:reference_values}

\section{Reference response maps}

Reference consensus response maps for phase 1 can be downloaded from: \url{https://github.com/theibsi/ibsi_2_reference_data}. Note that no consensus could be found for filter configurations 9 and 10.

To test compliance with the reference standard, a response map should be computed using the parameters listed in Table \ref{tab:benchmark_filter_settings}. Then the voxel-wise differences between the reference response map and the computed response map should be determined. The difference for each voxel should be equal to or less than 1\% of the intensity range of the reference response map.

\section{Reference feature values}

In phase 2, we computed features from response maps created after convolutional filtering. Shown below are their corresponding reference values. These values can also be downloaded in tabular form from: \url{https://github.com/theibsi/ibsi_2_reference_data}.

To test compliance with the reference standard, a response map should be computed using the parameters listed in Tables \ref{tab:lung_ct_image_processing_configurations} and \ref{tab:lung_ct_image_filter_configurations}. Statistical features are then computed from the gross tumour volume (GTV) mask. The computed values are expected to lie within the tolerances indicated in the tables below.

\subsection{Mean intensity}

The reference values for the intensity-based mean intensity feature (\textid{Q4LE}) are shown in Table \ref{tab:mean}.

\subsection{Intensity variance}

The reference values for the intensity-based variance feature (\textid{ECT3}) are shown in Table \ref{tab:variance}.

\begin{minipage}[l]{0.45\textwidth}
\centering
\small
% latex table generated in R 4.2.1 by xtable 1.8-4 package
% Mon Mar 13 17:23:11 2023
\begin{tabular}{cccc}
  \toprule
{\textbf{filter ID}} & {\textbf{value}} & {\textbf{tol.}} & {\textbf{consensus}} \\ 
  \midrule
1.A &  $-$47 &  4.6 & strong \\ 
  1.B & $-$46.4 &  5.9 & very strong \\ 
  2.A & $-$49.9 &  4.4 & strong \\ 
  2.B & $-$49.9 &  5.7 & very strong \\ 
  3.A & $-$2.44 & 0.12 & strong \\ 
  3.B & $-$2.94 &  0.2 & strong \\ 
  4.A &  148 &    3 & strong \\ 
  4.B &  142 &    3 & strong \\ 
  5.A &  103 &    1 & strong \\ 
  5.B & 40.2 &  0.2 & strong \\ 
  6.A & $-$0.185 & 0.025 & strong \\ 
  6.B & $-$0.182 & 0.024 & strong \\ 
  7.A & 0.245 & 0.03 & strong \\ 
  7.B & $-$0.0406 & 0.0051 & strong \\ 
  8.A & 0.248 & 0.047 & moderate \\ 
  8.B & 0.32 & 0.059 & strong \\ 
  9.A & 2.08 & 0.17 & moderate \\ 
  9.B & 2.68 & 0.22 & strong \\ 
  10.A & \textemdash & \textemdash & none \\ 
  10.B & \textemdash & \textemdash & none \\ 
  11.A & \textemdash & \textemdash & none \\ 
  11.B & \textemdash & \textemdash & none \\ 
   \bottomrule
\end{tabular}

\captionof{table}{Reference values for the \textit{mean} feature.}
\label{tab:mean}

\end{minipage}
\quad
\begin{minipage}[r]{0.45\textwidth}
\centering
\small
% latex table generated in R 4.2.1 by xtable 1.8-4 package
% Mon Mar 13 17:23:11 2023
\begin{tabular}{cccc}
  \toprule
{\textbf{filter ID}} & {\textbf{value}} & {\textbf{tol.}} & {\textbf{consensus}} \\ 
  \midrule
1.A & $5.33 \times 10^{4}$ & $2 \times 10^{3}$ & strong \\ 
  1.B & $5.26 \times 10^{4}$ & $2.8 \times 10^{3}$ & very strong \\ 
  2.A & $4.66 \times 10^{4}$ & $1.6 \times 10^{3}$ & strong \\ 
  2.B & $4.44 \times 10^{4}$ & $2.3 \times 10^{3}$ & very strong \\ 
  3.A &  556 &   24 & strong \\ 
  3.B &  720 &   33 & strong \\ 
  4.A & $1.8 \times 10^{4}$ &  500 & strong \\ 
  4.B & $1.11 \times 10^{4}$ &  300 & strong \\ 
  5.A & $1.04 \times 10^{3}$ &   10 & strong \\ 
  5.B &  231 &    2 & strong \\ 
  6.A &  427 &   11 & strong \\ 
  6.B &  250 &    9 & strong \\ 
  7.A & $1.91 \times 10^{3}$ &   40 & strong \\ 
  7.B &  422 &   11 & strong \\ 
  8.A & $1.55 \times 10^{3}$ &   50 & moderate \\ 
  8.B & $1.81 \times 10^{3}$ &   70 & moderate \\ 
  9.A & $4.58 \times 10^{3}$ &  170 & moderate \\ 
  9.B & $5.49 \times 10^{3}$ &  220 & strong \\ 
  10.A & \textemdash & \textemdash & none \\ 
  10.B & \textemdash & \textemdash & none \\ 
  11.A & \textemdash & \textemdash & none \\ 
  11.B & \textemdash & \textemdash & none \\ 
   \bottomrule
\end{tabular}

\captionof{table}{Reference values for the \textit{variance} feature.}
\label{tab:variance}

\end{minipage}
\FloatBarrier

\subsection{Intensity skewness}

The reference values for the intensity-based skewness feature (\textid{KE2A}) are shown in Table \ref{tab:skewness}.

\subsection{Excess intensity kurtosis}

The reference values for the intensity-based excess kurtosis feature (\textid{IPH6}) are shown in Table \ref{tab:kurtosis}.

\begin{minipage}[l]{0.45\textwidth}
\centering
\small
% latex table generated in R 4.2.1 by xtable 1.8-4 package
% Mon Mar 13 17:23:11 2023
\begin{tabular}{cccc}
  \toprule
{\textbf{filter ID}} & {\textbf{value}} & {\textbf{tol.}} & {\textbf{consensus}} \\ 
  \midrule
1.A & $-$2.17 & 0.07 & strong \\ 
  1.B & $-$2.18 & 0.09 & very strong \\ 
  2.A & $-$2.13 & 0.07 & strong \\ 
  2.B & $-$2.13 & 0.09 & very strong \\ 
  3.A & 0.295 & 0.027 & strong \\ 
  3.B & 0.428 & 0.009 & strong \\ 
  4.A & 0.877 & 0.024 & strong \\ 
  4.B & 0.645 & 0.028 & strong \\ 
  5.A & 1.31 & 0.03 & strong \\ 
  5.B & 1.57 & 0.03 & strong \\ 
  6.A & 0.0837 & 0.0188 & strong \\ 
  6.B & 0.157 & 0.018 & strong \\ 
  7.A & 0.0469 & 0.0072 & strong \\ 
  7.B & $-$0.0112 & 0.0027 & strong \\ 
  8.A & $-$0.0473 & 0.0145 & moderate \\ 
  8.B & $-$0.0719 & 0.0163 & moderate \\ 
  9.A & $-$0.0596 & 0.0145 & moderate \\ 
  9.B & $-$0.0858 & 0.0107 & moderate \\ 
  10.A & \textemdash & \textemdash & none \\ 
  10.B & \textemdash & \textemdash & none \\ 
  11.A & \textemdash & \textemdash & none \\ 
  11.B & \textemdash & \textemdash & none \\ 
   \bottomrule
\end{tabular}

\captionof{table}{Reference values for the \textit{skewness} feature.}
\label{tab:skewness}

\end{minipage}
\quad
\begin{minipage}[r]{0.45\textwidth}
\centering
\small
% latex table generated in R 4.2.1 by xtable 1.8-4 package
% Mon Mar 13 17:23:10 2023
\begin{tabular}{cccc}
  \toprule
{\textbf{filter ID}} & {\textbf{value}} & {\textbf{tol.}} & {\textbf{consensus}} \\ 
  \midrule
1.A & 3.65 & 0.34 & strong \\ 
  1.B & 3.71 & 0.47 & very strong \\ 
  2.A & 3.53 & 0.33 & strong \\ 
  2.B & 3.59 & 0.46 & very strong \\ 
  3.A & 7.02 &  0.3 & strong \\ 
  3.B & 6.13 & 0.27 & strong \\ 
  4.A & $-$0.279 & 0.039 & strong \\ 
  4.B & $-$0.711 & 0.044 & strong \\ 
  5.A & 2.84 & 0.14 & strong \\ 
  5.B & 4.34 &  0.2 & strong \\ 
  6.A & 7.72 & 0.24 & strong \\ 
  6.B & 8.98 & 0.35 & strong \\ 
  7.A & 6.22 & 0.16 & strong \\ 
  7.B & 5.45 & 0.09 & strong \\ 
  8.A & 8.19 &  0.3 & moderate \\ 
  8.B & 7.64 & 0.33 & strong \\ 
  9.A & 6.11 & 0.21 & moderate \\ 
  9.B & 5.58 & 0.18 & strong \\ 
  10.A & \textemdash & \textemdash & none \\ 
  10.B & \textemdash & \textemdash & weak \\ 
  11.A & \textemdash & \textemdash & none \\ 
  11.B & \textemdash & \textemdash & none \\ 
   \bottomrule
\end{tabular}

\captionof{table}{Reference values for the \textit{(excess) kurtosis} feature.}
\label{tab:kurtosis}

\end{minipage}
\FloatBarrier

\subsection{Median intensity}

The reference values for the median intensity feature (\textid{Y12H}) are shown in Table \ref{tab:median}.

\subsection{Minimum intensity}
The reference values for the minimum intensity feature (\textid{1GSF}) are shown in Table \ref{tab:minimum}.

\begin{minipage}[l]{0.45\textwidth}
\centering
\small
% latex table generated in R 4.2.1 by xtable 1.8-4 package
% Mon Mar 13 17:23:11 2023
\begin{tabular}{cccc}
  \toprule
{\textbf{filter ID}} & {\textbf{value}} & {\textbf{tol.}} & {\textbf{consensus}} \\ 
  \midrule
1.A &   41 &  0.6 & strong \\ 
  1.B &   41 &  0.7 & very strong \\ 
  2.A & 38.1 &  0.7 & strong \\ 
  2.B & 37.3 &  0.6 & very strong \\ 
  3.A & $-$0.585 & 0.012 & strong \\ 
  3.B & $-$0.919 & 0.024 & strong \\ 
  4.A & 88.7 &    5 & strong \\ 
  4.B &  113 &    4 & strong \\ 
  5.A &   97 &  1.3 & strong \\ 
  5.B & 37.2 &  0.1 & strong \\ 
  6.A & 0.0456 & 0.0041 & strong \\ 
  6.B & 0.0575 & 0.0046 & strong \\ 
  7.A & 0.0675 & 0.0061 & strong \\ 
  7.B & $-$0.0164 & 0.0013 & strong \\ 
  8.A & $-$0.0323 & 0.0073 & moderate \\ 
  8.B & $-$0.00947 & 0.0107 & moderate \\ 
  9.A & 0.14 & 0.028 & moderate \\ 
  9.B & 0.233 & 0.046 & moderate \\ 
  10.A & \textemdash & \textemdash & none \\ 
  10.B & \textemdash & \textemdash & none \\ 
  11.A & \textemdash & \textemdash & none \\ 
  11.B & \textemdash & \textemdash & none \\ 
   \bottomrule
\end{tabular}

\captionof{table}{Reference values for the \textit{median} feature.}
\label{tab:median}

\end{minipage}
\quad
\begin{minipage}[r]{0.45\textwidth}
\centering
\small
% latex table generated in R 4.2.1 by xtable 1.8-4 package
% Mon Mar 13 17:23:11 2023
\begin{tabular}{cccc}
  \toprule
{\textbf{filter ID}} & {\textbf{value}} & {\textbf{tol.}} & {\textbf{consensus}} \\ 
  \midrule
1.A & $-1 \times 10^{3}$ &   10 & strong \\ 
  1.B & $-$997 &    3 & strong \\ 
  2.A & $-$889 &    3 & strong \\ 
  2.B & $-$906 &    5 & very strong \\ 
  3.A & $-$178 &    1 & strong \\ 
  3.B & $-$173 &    5 & strong \\ 
  4.A & 15.5 &  0.1 & strong \\ 
  4.B & 28.5 &  0.1 & strong \\ 
  5.A & 31.4 &  0.5 & strong \\ 
  5.B & 9.53 & 0.11 & strong \\ 
  6.A & $-$245 &    2 & strong \\ 
  6.B & $-$148 &    1 & strong \\ 
  7.A & $-$349 &    7 & strong \\ 
  7.B & $-$203 &    3 & strong \\ 
  8.A & $-$395 &    3 & moderate \\ 
  8.B & $-$411 &    5 & moderate \\ 
  9.A & $-$535 &    1 & moderate \\ 
  9.B & $-$605 &    2 & moderate \\ 
  10.A & \textemdash & \textemdash & none \\ 
  10.B & \textemdash & \textemdash & weak \\ 
  11.A & \textemdash & \textemdash & none \\ 
  11.B & \textemdash & \textemdash & none \\ 
   \bottomrule
\end{tabular}

\captionof{table}{Reference values for the \textit{minimum} feature.}
\label{tab:minimum}

\end{minipage}
\FloatBarrier

\subsection{10\textsuperscript{th} intensity percentile}
The reference values for the 10\textsuperscript{th} intensity percentile feature (\textid{QG58}) are shown in Table \ref{tab:10th_percentile}.

\subsection{90\textsuperscript{th} intensity percentile}
The reference values for the 90\textsuperscript{th} intensity percentile feature (\textid{8DWT}) are shown in Table \ref{tab:90th_percentile}.

\begin{minipage}[l]{0.45\textwidth}
\centering
\small
% latex table generated in R 4.2.1 by xtable 1.8-4 package
% Mon Mar 13 17:23:11 2023
\begin{tabular}{cccc}
  \toprule
{\textbf{filter ID}} & {\textbf{value}} & {\textbf{tol.}} & {\textbf{consensus}} \\ 
  \midrule
1.A & $-$434 &   21 & strong \\ 
  1.B & $-$427 &   29 & very strong \\ 
  2.A & $-$402 &   18 & strong \\ 
  2.B & $-$389 &   25 & very strong \\ 
  3.A & $-$27.1 &  0.5 & strong \\ 
  3.B & $-$32.2 &  0.5 & strong \\ 
  4.A & 26.9 &  0.1 & strong \\ 
  4.B & 35.6 &  0.1 & strong \\ 
  5.A & 68.8 &  0.9 & strong \\ 
  5.B & 24.6 &  0.1 & strong \\ 
  6.A & $-$18.8 &  0.4 & strong \\ 
  6.B & $-$13.8 &  0.5 & strong \\ 
  7.A & $-$39.6 &    1 & strong \\ 
  7.B & $-$20.6 &  0.4 & strong \\ 
  8.A & $-$33.5 &  0.9 & moderate \\ 
  8.B & $-$36.5 &  1.3 & strong \\ 
  9.A & $-$59.2 &  1.8 & moderate \\ 
  9.B & $-$65.9 &  2.2 & strong \\ 
  10.A & \textemdash & \textemdash & none \\ 
  10.B & \textemdash & \textemdash & none \\ 
  11.A & \textemdash & \textemdash & none \\ 
  11.B & \textemdash & \textemdash & none \\ 
   \bottomrule
\end{tabular}

\captionof{table}{Reference values for the \textit{10th percentile} feature.}
\label{tab:10th_percentile}

\end{minipage}
\quad
\begin{minipage}[r]{0.45\textwidth}
\centering
\small
% latex table generated in R 4.2.1 by xtable 1.8-4 package
% Mon Mar 13 17:23:11 2023
\begin{tabular}{cccc}
  \toprule
{\textbf{filter ID}} & {\textbf{value}} & {\textbf{tol.}} & {\textbf{consensus}} \\ 
  \midrule
1.A &   93 &  0.1 & strong \\ 
  1.B &   92 &  0.1 & very strong \\ 
  2.A & 79.8 &  0.1 & strong \\ 
  2.B & 77.2 &  0.1 & strong \\ 
  3.A & 13.2 &  1.3 & strong \\ 
  3.B & 17.4 &  1.9 & strong \\ 
  4.A &  347 &    4 & strong \\ 
  4.B &  293 &    4 & strong \\ 
  5.A &  145 &    2 & strong \\ 
  5.B & 59.3 &  0.3 & strong \\ 
  6.A & 17.7 &  0.4 & strong \\ 
  6.B & 12.1 &  0.4 & strong \\ 
  7.A & 40.5 &  0.9 & strong \\ 
  7.B & 20.4 &  0.4 & strong \\ 
  8.A & 34.4 &    1 & moderate \\ 
  8.B & 38.1 &  1.3 & strong \\ 
  9.A & 71.5 &  1.9 & moderate \\ 
  9.B & 82.8 &  1.8 & moderate \\ 
  10.A & \textemdash & \textemdash & none \\ 
  10.B & \textemdash & \textemdash & none \\ 
  11.A & \textemdash & \textemdash & none \\ 
  11.B & \textemdash & \textemdash & none \\ 
   \bottomrule
\end{tabular}

\captionof{table}{Reference values for the \textit{90th percentile} feature.}
\label{tab:90th_percentile}

\end{minipage}
\FloatBarrier

\subsection{Maximum intensity}
The reference values for the maximum intensity feature (\textid{84IY}) are shown in Table \ref{tab:maximum}.

\subsection{Intensity interquartile range}
The reference values for the intensity-based interquartile range feature (\textid{SALO}) are shown in Table \ref{tab:interquartile_range}.

\begin{minipage}[l]{0.45\textwidth}
\centering
\small
% latex table generated in R 4.2.1 by xtable 1.8-4 package
% Mon Mar 13 17:23:11 2023
\begin{tabular}{cccc}
  \toprule
{\textbf{filter ID}} & {\textbf{value}} & {\textbf{tol.}} & {\textbf{consensus}} \\ 
  \midrule
1.A &  377 &   10 & strong \\ 
  1.B &  377 &   15 & strong \\ 
  2.A &  334 &    7 & strong \\ 
  2.B &  316 &    7 & very strong \\ 
  3.A &  191 &    1 & strong \\ 
  3.B &  204 &    1 & strong \\ 
  4.A &  672 &    4 & strong \\ 
  4.B &  525 &    1 & strong \\ 
  5.A &  354 &    5 & strong \\ 
  5.B &  175 &    3 & strong \\ 
  6.A &  202 &    2 & strong \\ 
  6.B &  155 &    1 & strong \\ 
  7.A &  352 &    4 & strong \\ 
  7.B &  201 &    4 & strong \\ 
  8.A &  408 &    3 & moderate \\ 
  8.B &  374 &    3 & strong \\ 
  9.A &  470 &    6 & moderate \\ 
  9.B &  471 &   13 & moderate \\ 
  10.A & \textemdash & \textemdash & none \\ 
  10.B & \textemdash & \textemdash & weak \\ 
  11.A & \textemdash & \textemdash & none \\ 
  11.B & \textemdash & \textemdash & none \\ 
   \bottomrule
\end{tabular}

\captionof{table}{Reference values for the \textit{maximum} feature.}
\label{tab:maximum}

\end{minipage}
\quad
\begin{minipage}[r]{0.45\textwidth}
\centering
\small
% latex table generated in R 4.2.1 by xtable 1.8-4 package
% Mon Mar 13 17:23:11 2023
\begin{tabular}{cccc}
  \toprule
{\textbf{filter ID}} & {\textbf{value}} & {\textbf{tol.}} & {\textbf{consensus}} \\ 
  \midrule
1.A &   69 &  6.4 & strong \\ 
  1.B &   67 &  9.1 & very strong \\ 
  2.A &   82 & 10.5 & strong \\ 
  2.B & 92.6 & 13.5 & very strong \\ 
  3.A & 8.74 & 0.19 & strong \\ 
  3.B & 11.4 &  0.3 & strong \\ 
  4.A &  219 &    5 & strong \\ 
  4.B &  188 &    4 & strong \\ 
  5.A & 37.2 &  0.6 & strong \\ 
  5.B & 17.4 &  0.1 & strong \\ 
  6.A & 15.8 &  0.1 & strong \\ 
  6.B & 9.35 & 0.15 & strong \\ 
  7.A & 28.8 &  0.4 & strong \\ 
  7.B & 16.3 &  0.2 & strong \\ 
  8.A & 24.8 &  0.3 & moderate \\ 
  8.B & 25.5 &  0.4 & moderate \\ 
  9.A & 34.2 &  0.8 & moderate \\ 
  9.B &   41 &    1 & strong \\ 
  10.A & \textemdash & \textemdash & none \\ 
  10.B & \textemdash & \textemdash & weak \\ 
  11.A & \textemdash & \textemdash & none \\ 
  11.B & \textemdash & \textemdash & none \\ 
   \bottomrule
\end{tabular}

\captionof{table}{Reference values for the \textit{interquartile range} feature.}
\label{tab:interquartile_range}

\end{minipage}
\FloatBarrier

\subsection{Intensity range}
The reference values for the intensity-based interquartile range feature (\textid{2OJQ}) are shown in Table \ref{tab:range}.

\subsection{Intensity-based mean absolute deviation}
The reference values for the intensity-based mean absolute deviation feature (\textid{4FUA}) are shown in Table \ref{tab:mean_absolute_deviation}.

\begin{minipage}[l]{0.45\textwidth}
\centering
\small
% latex table generated in R 4.2.1 by xtable 1.8-4 package
% Mon Mar 13 17:23:11 2023
\begin{tabular}{cccc}
  \toprule
{\textbf{filter ID}} & {\textbf{value}} & {\textbf{tol.}} & {\textbf{consensus}} \\ 
  \midrule
1.A & $1.38 \times 10^{3}$ &   10 & strong \\ 
  1.B & $1.37 \times 10^{3}$ &   20 & very strong \\ 
  2.A & $1.22 \times 10^{3}$ &   10 & strong \\ 
  2.B & $1.22 \times 10^{3}$ &   10 & very strong \\ 
  3.A &  369 &    2 & strong \\ 
  3.B &  377 &    5 & strong \\ 
  4.A &  657 &    4 & strong \\ 
  4.B &  496 &    1 & strong \\ 
  5.A &  322 &    5 & strong \\ 
  5.B &  165 &    3 & strong \\ 
  6.A &  447 &    3 & strong \\ 
  6.B &  303 &    2 & strong \\ 
  7.A &  701 &    9 & strong \\ 
  7.B &  404 &    7 & strong \\ 
  8.A &  803 &    5 & moderate \\ 
  8.B &  785 &    6 & moderate \\ 
  9.A & $1 \times 10^{3}$ &   10 & moderate \\ 
  9.B & $1.08 \times 10^{3}$ &   20 & moderate \\ 
  10.A & \textemdash & \textemdash & none \\ 
  10.B & \textemdash & \textemdash & weak \\ 
  11.A & \textemdash & \textemdash & none \\ 
  11.B & \textemdash & \textemdash & none \\ 
   \bottomrule
\end{tabular}

\captionof{table}{Reference values for the \textit{range} feature.}
\label{tab:range}

\end{minipage}
\quad
\begin{minipage}[r]{0.45\textwidth}
\centering
\small
% latex table generated in R 4.2.1 by xtable 1.8-4 package
% Mon Mar 13 17:23:11 2023
\begin{tabular}{cccc}
  \toprule
{\textbf{filter ID}} & {\textbf{value}} & {\textbf{tol.}} & {\textbf{consensus}} \\ 
  \midrule
1.A &  160 &    5 & strong \\ 
  1.B &  159 &    7 & very strong \\ 
  2.A &  153 &    5 & strong \\ 
  2.B &  149 &    6 & very strong \\ 
  3.A & 13.2 &  0.3 & strong \\ 
  3.B & 15.5 &  0.4 & strong \\ 
  4.A &  116 &    2 & strong \\ 
  4.B & 92.4 &  1.4 & strong \\ 
  5.A & 24.2 &  0.4 & strong \\ 
  5.B & 11.3 &  0.1 & strong \\ 
  6.A &   13 &  0.2 & strong \\ 
  6.B & 9.26 & 0.22 & strong \\ 
  7.A & 26.9 &  0.4 & strong \\ 
  7.B & 13.4 &  0.2 & strong \\ 
  8.A & 23.6 &  0.5 & moderate \\ 
  8.B & 25.3 &  0.6 & moderate \\ 
  9.A & 40.1 &  0.9 & moderate \\ 
  9.B & 45.1 &  1.1 & strong \\ 
  10.A & \textemdash & \textemdash & none \\ 
  10.B & \textemdash & \textemdash & weak \\ 
  11.A & \textemdash & \textemdash & none \\ 
  11.B & \textemdash & \textemdash & none \\ 
   \bottomrule
\end{tabular}

\captionof{table}{Reference values for the \textit{mean absolute deviation} feature.}
\label{tab:mean_absolute_deviation}

\end{minipage}
\FloatBarrier

\subsection{Intensity-based robust mean absolute deviation}
The reference values for the intensity-based robust mean absolute deviation feature (\textid{1128}) are shown in Table \ref{tab:robust_mean_absolute_deviation}.

\subsection{Intensity-based median absolute deviation\id{N72L}}
The reference values for the intensity-based median absolute deviation feature (\textid{N72L}) are shown in Table \ref{tab:median_absolute_deviation}.

\begin{minipage}[l]{0.45\textwidth}
\centering
\small
% latex table generated in R 4.2.1 by xtable 1.8-4 package
% Mon Mar 13 17:23:11 2023
\begin{tabular}{cccc}
  \toprule
{\textbf{filter ID}} & {\textbf{value}} & {\textbf{tol.}} & {\textbf{consensus}} \\ 
  \midrule
1.A & 64.4 &  5.8 & strong \\ 
  1.B & 63.6 &  7.3 & very strong \\ 
  2.A & 67.7 &  5.6 & strong \\ 
  2.B & 68.1 &  6.9 & very strong \\ 
  3.A & 4.98 & 0.16 & strong \\ 
  3.B & 6.37 & 0.19 & strong \\ 
  4.A & 92.9 &  1.8 & strong \\ 
  4.B & 75.9 &  1.4 & strong \\ 
  5.A & 15.8 &  0.2 & strong \\ 
  5.B & 7.31 & 0.06 & strong \\ 
  6.A & 6.82 & 0.08 & strong \\ 
  6.B & 4.21 & 0.09 & strong \\ 
  7.A &   13 &  0.2 & strong \\ 
  7.B &  7.2 &  0.1 & strong \\ 
  8.A & 11.1 &  0.2 & moderate \\ 
  8.B & 11.7 &  0.3 & strong \\ 
  9.A & 17.6 &  0.5 & moderate \\ 
  9.B &   21 &  0.5 & moderate \\ 
  10.A & \textemdash & \textemdash & none \\ 
  10.B & \textemdash & \textemdash & weak \\ 
  11.A & \textemdash & \textemdash & none \\ 
  11.B & \textemdash & \textemdash & none \\ 
   \bottomrule
\end{tabular}

\captionof{table}{Reference values for the \textit{robust mean absolute deviation} feature.}
\label{tab:robust_mean_absolute_deviation}

\end{minipage}
\quad
\begin{minipage}[r]{0.45\textwidth}
\centering
\small
% latex table generated in R 4.2.1 by xtable 1.8-4 package
% Mon Mar 13 17:23:11 2023
\begin{tabular}{cccc}
  \toprule
{\textbf{filter ID}} & {\textbf{value}} & {\textbf{tol.}} & {\textbf{consensus}} \\ 
  \midrule
1.A &  122 &    4 & strong \\ 
  1.B &  121 &    6 & very strong \\ 
  2.A &  116 &    4 & strong \\ 
  2.B &  114 &    5 & very strong \\ 
  3.A & 12.9 &  0.4 & strong \\ 
  3.B & 15.3 &  0.4 & strong \\ 
  4.A &  111 &    2 & strong \\ 
  4.B & 90.8 &  1.6 & strong \\ 
  5.A & 23.7 &  0.4 & strong \\ 
  5.B &   11 &  0.1 & strong \\ 
  6.A &   13 &  0.2 & strong \\ 
  6.B & 9.25 & 0.22 & strong \\ 
  7.A & 26.9 &  0.4 & strong \\ 
  7.B & 13.4 &  0.2 & strong \\ 
  8.A & 23.6 &  0.5 & moderate \\ 
  8.B & 25.3 &  0.6 & moderate \\ 
  9.A &   40 &  0.9 & moderate \\ 
  9.B &   45 &  1.1 & strong \\ 
  10.A & \textemdash & \textemdash & none \\ 
  10.B & \textemdash & \textemdash & weak \\ 
  11.A & \textemdash & \textemdash & none \\ 
  11.B & \textemdash & \textemdash & none \\ 
   \bottomrule
\end{tabular}

\captionof{table}{Reference values for the \textit{median absolute deviation} feature.}
\label{tab:median_absolute_deviation}

\end{minipage}
\FloatBarrier

\subsection{Intensity-based coefficient of variation}
The reference values for the intensity-based coefficient of variation feature (\textid{7TET}) are shown in Table \ref{tab:coefficient_of_variation}.

\subsection{Intensity-based quartile coefficient of dispersion}
The reference values for the intensity-based quartile coefficient of dispersion feature (\textid{9S40}) are shown in Table \ref{tab:quartile_coefficient_of_dispersion}.

\begin{minipage}[l]{0.45\textwidth}
\centering
\small
% latex table generated in R 4.2.1 by xtable 1.8-4 package
% Mon Mar 13 17:23:11 2023
\begin{tabular}{cccc}
  \toprule
{\textbf{filter ID}} & {\textbf{value}} & {\textbf{tol.}} & {\textbf{consensus}} \\ 
  \midrule
1.A & $-$4.92 & 0.42 & strong \\ 
  1.B & $-$4.94 & 0.64 & very strong \\ 
  2.A & $-$4.33 & 0.34 & strong \\ 
  2.B & $-$4.22 & 0.47 & very strong \\ 
  3.A & $-$9.66 & 1.21 & strong \\ 
  3.B & $-$9.12 & 1.63 & strong \\ 
  4.A & 0.904 & 0.007 & strong \\ 
  4.B & 0.743 & 0.005 & strong \\ 
  5.A & 0.313 & 0.003 & strong \\ 
  5.B & 0.377 & 0.004 & strong \\ 
  6.A & $-$112 &   37 & strong \\ 
  6.B & $-$86.9 & 32.6 & strong \\ 
  7.A &  178 &   77 & strong \\ 
  7.B & $-$506 &  149 & strong \\ 
  8.A &  159 &   30 & moderate \\ 
  8.B &  134 &   27 & strong \\ 
  9.A & 32.6 & 24.4 & moderate \\ 
  9.B & 27.7 & 20.4 & strong \\ 
  10.A & \textemdash & \textemdash & none \\ 
  10.B & \textemdash & \textemdash & none \\ 
  11.A & \textemdash & \textemdash & none \\ 
  11.B & \textemdash & \textemdash & none \\ 
   \bottomrule
\end{tabular}

\captionof{table}{Reference values for the \textit{coefficient of variation} feature.}
\label{tab:coefficient_of_variation}

\end{minipage}
\quad
\begin{minipage}[r]{0.45\textwidth}
\centering
\small
% latex table generated in R 4.2.1 by xtable 1.8-4 package
% Mon Mar 13 17:23:11 2023
\begin{tabular}{cccc}
  \toprule
{\textbf{filter ID}} & {\textbf{value}} & {\textbf{tol.}} & {\textbf{consensus}} \\ 
  \midrule
1.A &    1 & 0.85 & strong \\ 
  1.B & 0.944 & 0.925 & very strong \\ 
  2.A & 1.85 & 0.66 & strong \\ 
  2.B & 2.97 & 0.58 & very strong \\ 
  3.A & $-$2.93 & 0.06 & strong \\ 
  3.B & $-$2.34 & 0.07 & strong \\ 
  4.A & 0.773 & 0.003 & strong \\ 
  4.B & 0.699 & 0.003 & strong \\ 
  5.A & 0.187 & 0.001 & strong \\ 
  5.B & 0.226 & 0.002 & strong \\ 
  6.A &  491 &   10 & strong \\ 
  6.B & $-$162 &   27 & strong \\ 
  7.A &  104 &   66 & strong \\ 
  7.B & $-$684 &  130 & strong \\ 
  8.A & $-$441 &   29 & moderate \\ 
  8.B & \textemdash & \textemdash & none \\ 
  9.A &   47 & 15.2 & moderate \\ 
  9.B & 47.4 & 20.7 & strong \\ 
  10.A & \textemdash & \textemdash & none \\ 
  10.B & \textemdash & \textemdash & none \\ 
  11.A & \textemdash & \textemdash & none \\ 
  11.B & \textemdash & \textemdash & none \\ 
   \bottomrule
\end{tabular}

\captionof{table}{Reference values for the \textit{quartile coefficient of dispersion} feature.}
\label{tab:quartile_coefficient_of_dispersion}

\end{minipage}
\FloatBarrier

\subsection{Intensity-based energy}
The reference values for the intensity-based energy feature (\textid{N8CA}) are shown in Table \ref{tab:energy}.

\subsection{Root mean square intensity}
The reference values for the root mean square intensity feature (\textid{5ZWQ}) are shown in Table \ref{tab:root_mean_square}.

\begin{minipage}[l]{0.45\textwidth}
\centering
\small
% latex table generated in R 4.2.1 by xtable 1.8-4 package
% Mon Mar 13 17:23:11 2023
\begin{tabular}{cccc}
  \toprule
{\textbf{filter ID}} & {\textbf{value}} & {\textbf{tol.}} & {\textbf{consensus}} \\ 
  \midrule
1.A & $6.96 \times 10^{9}$ & $4.5 \times 10^{8}$ & strong \\ 
  1.B & $1.96 \times 10^{10}$ & $1.9 \times 10^{9}$ & very strong \\ 
  2.A & $6.15 \times 10^{9}$ & $3.9 \times 10^{8}$ & strong \\ 
  2.B & $1.68 \times 10^{10}$ & $1.6 \times 10^{9}$ & very strong \\ 
  3.A & $7.03 \times 10^{7}$ & $4.2 \times 10^{6}$ & strong \\ 
  3.B & $2.61 \times 10^{8}$ & $1.9 \times 10^{7}$ & strong \\ 
  4.A & $5 \times 10^{9}$ & $2.7 \times 10^{8}$ & strong \\ 
  4.B & $1.12 \times 10^{10}$ & $7 \times 10^{8}$ & strong \\ 
  5.A & $1.46 \times 10^{9}$ & $2 \times 10^{7}$ & strong \\ 
  5.B & $6.62 \times 10^{8}$ & $9 \times 10^{6}$ & strong \\ 
  6.A & $5.35 \times 10^{7}$ & $2 \times 10^{6}$ & strong \\ 
  6.B & $8.96 \times 10^{7}$ & $5.3 \times 10^{6}$ & strong \\ 
  7.A & $2.39 \times 10^{8}$ & $9 \times 10^{6}$ & strong \\ 
  7.B & $1.51 \times 10^{8}$ & $7 \times 10^{6}$ & strong \\ 
  8.A & $1.94 \times 10^{8}$ & $9 \times 10^{6}$ & moderate \\ 
  8.B & $6.48 \times 10^{8}$ & $3.9 \times 10^{7}$ & strong \\ 
  9.A & $5.74 \times 10^{8}$ & $3.1 \times 10^{7}$ & moderate \\ 
  9.B & $1.97 \times 10^{9}$ & $1.4 \times 10^{8}$ & strong \\ 
  10.A & \textemdash & \textemdash & none \\ 
  10.B & \textemdash & \textemdash & none \\ 
  11.A & \textemdash & \textemdash & none \\ 
  11.B & \textemdash & \textemdash & none \\ 
   \bottomrule
\end{tabular}

\captionof{table}{Reference values for the \textit{energy} feature.}
\label{tab:energy}

\end{minipage}
\quad
\begin{minipage}[r]{0.45\textwidth}
\centering
\small
% latex table generated in R 4.2.1 by xtable 1.8-4 package
% Mon Mar 13 17:23:11 2023
\begin{tabular}{cccc}
  \toprule
{\textbf{filter ID}} & {\textbf{value}} & {\textbf{tol.}} & {\textbf{consensus}} \\ 
  \midrule
1.A &  236 &    5 & strong \\ 
  1.B &  234 &    7 & very strong \\ 
  2.A &  222 &    5 & strong \\ 
  2.B &  217 &    7 & very strong \\ 
  3.A & 23.7 &  0.5 & strong \\ 
  3.B &   27 &  0.6 & strong \\ 
  4.A &  200 &    4 & strong \\ 
  4.B &  177 &    3 & strong \\ 
  5.A &  108 &    2 & strong \\ 
  5.B &   43 &  0.2 & strong \\ 
  6.A & 20.7 &  0.3 & strong \\ 
  6.B & 15.8 &  0.3 & strong \\ 
  7.A & 43.7 &  0.6 & strong \\ 
  7.B & 20.6 &  0.3 & strong \\ 
  8.A & 39.3 &  0.7 & moderate \\ 
  8.B & 42.5 &  0.9 & moderate \\ 
  9.A & 67.7 &  1.3 & moderate \\ 
  9.B & 74.1 &  1.6 & strong \\ 
  10.A & \textemdash & \textemdash & none \\ 
  10.B & \textemdash & \textemdash & none \\ 
  11.A & \textemdash & \textemdash & none \\ 
  11.B & \textemdash & \textemdash & none \\ 
   \bottomrule
\end{tabular}

\captionof{table}{Reference values for the \textit{root mean square} feature.}
\label{tab:root_mean_square}

\end{minipage}
\FloatBarrier

\chapter{Conclusion}
In this document we detail various convolutional image filters to standardise their implementation in radiomics software. Such filters are used to convert medical images into response maps that emphasise characteristics such as edges, blobs or directional structures. Quantifying such characteristics using standardised radiomics software will open the way to their reproducible assessment and validation, and their translation from bench to bedside.

\chapter*{Acknowledgments}
This work was partially supported by the Swiss National Science Foundation (grant 205320\_179069), the Swiss Personalized Health Network (IMAGINE and QA4IQI projects) and the Canada CIFAR AI Chairs program.

\addcontentsline{toc}{chapter}{Bibliography}
\bibliography{./refs}

\appendix
\chapter{Equivariant right angle rotational representation for separable filters} \label{app:separableConvRightAngleEquivariant}
Right-angle rotations of separable filters can be efficiently obtained using permutations and unidimensional filter flipping.
The flipping of a 1$D$ vector $g_1 \in \mathbb{Z}^n$ can be obtained by multiplication with an $n\times n$ exchange matrix $\mathrm{J}$, where 
$$\mathrm{J}[i,j] = \begin{cases}
1\quad \text{if}\quad j=n-i+1,\\
0\quad \text{otherwise}.
\end{cases}$$

Note that a $0$ should be appended to any even-sized filter kernel, to allow for equivalence with methods where rotational invariance is reached through rotating the image, see Sec.~\ref{sec:combiningDSandRotInv}. For instance, $g_1 = (1,2,3,4)$ is first extended to $g_1^{*}=(1,2,3,4,0)$, and then flipped by computing $Jg_1^{*}=(0,4,3,2,1)$.

\section{2$D$ case}
In the 2$D$ case, four versions of a filter can be obtained by right-angle rotations (\textit{i.e.} $0$, $\frac{\pi}{2}$, $\pi$ and $\frac{3\pi}{2}$). The permutations and flips required to rotate the filter $g[\boldsymbol{k}]= g_{1}[k]\otimes g_{2}[k]$ are listed in the following, where $g^{\theta}$ is a rotation $\mathrm{R}_{\theta}$ by $\theta$  of the filter $g$ so that $g^{\theta}=g(\mathrm{R}_{\theta}\cdot)$.

$g^0= g_{1}\otimes g_{2},$

$g^{\frac{\pi}{2}}= \mathrm{J} g_{2}\otimes g_{1},$

$g^{\pi}= \mathrm{J} g_{1}\otimes \mathrm{J} g_{2},$

$g^{\frac{3\pi}{2}}= g_{2}\otimes \mathrm{J} g_{1}.$

\section{3$D$ case}
When considering 3$D$ filters, 24 right angle rotations are possible (cube symmetries). 
We use a $(k_1,k_2,k_3)$ (\textit{i.e.} $x,y,z$) extrinsic Euler notation $(\alpha,\beta,\gamma)$ representing a first rotation by $\alpha$ on the $k_3$ axis, followed by a rotation by $\beta$ on the $k_2$ axis and a last rotation by $\gamma$ on the $k_1$ axis.
We denote $g^{(\alpha,\beta,\gamma)}$ a 3$D$ filter $g$ rotated by such angles so that $g^{(\alpha,\beta,\gamma)} = g(\mathrm{R}_{\alpha}\mathrm{R}_{\beta}\mathrm{R}_{\gamma}\cdot)$.
The flips and permutations of the 1$D$ vectors required to obtain these 24 rotations of a separable filter $g[\boldsymbol{k}]= g_{1}[k]\otimes g_{2}[k]\otimes g_{3}[k]$ are listed in the following.
%$g[\boldsymbol{k}]=g[k_1,k_2,k_3]= g_{1}[k_1] g_{2}[k_2] g_{3}[k_3]$ are listed in the following.

% 1
$g^{(0,0,0)}= g_{1}\otimes g_{2}\otimes g_{3},$

% 2
$g^{(0,\frac{\pi}{2},0)}= \mathrm{J}g_{3}\otimes g_{2}\otimes g_{1},$

% 3
$g^{(0,\pi,0)}= \mathrm{J}g_{1}\otimes g_{2}\otimes \mathrm{J}g_{3},$

% 4
$g^{(0,\frac{3\pi}{2},0)}= g_{3}\otimes g_{2}\otimes \mathrm{J}g_{1},$

% 5
$g^{(\frac{\pi}{2},0,\frac{\pi}{2})}= g_{2}\otimes g_{3}\otimes g_{1},$

% 6
$g^{(\frac{\pi}{2},0,\frac{3\pi}{2})}= g_{2}\otimes \mathrm{J}g_{3}\otimes \mathrm{J}g_{1},$

% 7
$g^{(\frac{\pi}{2},0,0)}= g_{2}\otimes \mathrm{J}g_{1}\otimes g_{3},$

% 8
$g^{(\pi,0,0)}= \mathrm{J}g_{1} \otimes \mathrm{J}g_{2}\otimes g_{3},$

% 9
$g^{(\frac{3\pi}{2},0,0)}= \mathrm{J}g_{2}\otimes g_{1}\otimes g_{3},$

% 10
$g^{(0,\frac{\pi}{2},\frac{3\pi}{2})}= \mathrm{J}g_{3}\otimes \mathrm{J}g_{1}\otimes g_{2},$

% 11
$g^{(0,\frac{\pi}{2},\pi)}= \mathrm{J}g_{3}\otimes \mathrm{J}g_{2}\otimes \mathrm{J}g_{1},$

% 12
$g^{(0,\frac{\pi}{2},\frac{\pi}{2})}= \mathrm{J}g_{3}\otimes g_{1}\otimes \mathrm{J}g_{2},$

% 13
$g^{(\frac{\pi}{2},\pi,0)}= \mathrm{J}g_{2}\otimes \mathrm{J}g_{1}\otimes \mathrm{J}g_{3},$

% 14
$g^{(\pi,\pi,0)}= g_{1}\otimes \mathrm{J}g_{2}\otimes \mathrm{J}g_{3},$

% 15
$g^{(\frac{3\pi}{2},\pi,0)}= g_{2}\otimes g_{1}\otimes \mathrm{J}g_{3},$

% 16
$g^{(0,\frac{3\pi}{2},\frac{\pi}{2})}= g_{3}\otimes \mathrm{J}g_{1}\otimes \mathrm{J}g_{2},$

% 17
$g^{(0,\frac{3\pi}{2},\pi)}= g_{3}\otimes \mathrm{J}g_{2}\otimes g_{1},$

% 18
$g^{(0,\frac{3\pi}{2},\frac{3\pi}{2})}= g_{3}\otimes g_{1}\otimes g_{2},$

% 19,
$g^{(\pi,0,\frac{\pi}{2})}= \mathrm{J}g_{1}\otimes g_{3}\otimes g_{2}$

% 20
$g^{(\frac{3\pi}{2},0,\frac{\pi}{2})}= \mathrm{J}g_{2}\otimes g_{3}\otimes \mathrm{J}g_{1},$

% 21
$g^{(0,0,\frac{\pi}{2})}= g_{1}\otimes g_{3}\otimes \mathrm{J}g_{2},$

% 22
$g^{(\pi,0,\frac{3\pi}{2})}= \mathrm{J}g_{1}\otimes \mathrm{J}g_{3}\otimes \mathrm{J}g_{2},$

% 23
$g^{(\frac{3\pi}{2},0,\frac{3\pi}{2})}= \mathrm{J}g_{2}\otimes \mathrm{J}g_{3}\otimes g_{1},$

% 24
$g^{(0,0,\frac{3\pi}{2})}= g_{1}\otimes \mathrm{J}g_{3}\otimes g_{2}.$

\end{document}